\documentclass{emulateapj}
\font\boldsym=cmmib10      
                                                                                                                                                                                                                             
\usepackage{natbib}
\usepackage{epstopdf} 
\def \bea {\begin{eqnarray}}
\def \ena {\end{eqnarray}}                  

\def    \ba     {\bf  a}

\def	\Angstrom	{\,{\rm \AA}}		
\def	\ba	{{\bf a}}
\def	\be	{{\bf e}}

\def	\B	{{\rm B}}

\def	\bJ	{{\bf J}}

\def    \bmu    {{\hbox{\boldsym\char'026}}}	

\def    \bomega {{\hbox{\boldsym\char'041}}}	
\def    \bepsilon {{\hbox{\boldsym\char'017}}}	

\def	\C	{{\rm C}}

\def	\cm	{\,{\rm cm}}
\def	\d	{{\rm d}}
\def    \disk {{\rm disk}}
\def	\sd	{{\rm sd}}

\def	\D	{{\rm D}}
\def	\eff	{{\rm eff}}
\def	\ed	{{\rm ed}}

\def	\gas	{\,{\rm gas}}

\def	\GHz	{\,{\rm GHz}}

\def	\H	{{\rm H}}

\def	\VRE	{{\rm VRE}}

\def	\IR	{{\rm IR}}

\def	\K	{{\rm K}}
\def    \kB    {k_{\rm B}}

\def	\mtxt	{{\rm m}}
\def	\ntxt	{{\rm n}}

\def	\pc	{\,{\rm pc}}
\def     \peak  {{\rm peak}}

\def	\s	{\,{\rm s}}
\def	\sp	{{\rm sp}}
\def	\sr	{\,{\rm sr}}
\def    \ln  {\,{\rm ln}}
\def	\rot	{{\rm rot}}
\def	\vib	{{\rm vib}}
\def	\WIM	{{\rm WIM}}
\def	\CNM	{{\rm CNM}}

\def	\xhat	{\hat{\bf x}}
\def	\yhat	{\hat{\bf y}}
\def	\zhat	{\hat{\bf z}}

\begin{document}
\title{Spinning Dust Emission: Effects of irregular grain shape, transient heating 
and comparison with WMAP results}
\author{Thiem Hoang \altaffilmark{1}, A. Lazarian\altaffilmark{1}, and 
B. T. Draine\altaffilmark{2}}

\altaffiltext{1}{Astronomy Department, University of Wisconsin, Madison, WI 53706, USA}
\altaffiltext{2}{Department of Astrophysical Sciences, Princeton University, Princeton, NJ 08544, USA}

\begin{abstract}
{\em Planck}  is expected to answer crucial questions on the early Universe, but
it also provides further understanding on anomalous microwave emission. Electric dipole
 emission from spinning dust grains continues to be the favored interpretation 
 of anomalous microwave emission. 
In this paper, we present a method to calculate  the rotational emission
from small grains of irregular shape with moments of inertia $I_{1}\ge I_{2}\ge 
I_{3}$.
 We show that a 
torque-free rotating irregular grain with a given angular momentum 
radiates at multiple frequency modes. The resulting spinning dust spectrum
 has peak frequency and emissivity increasing with the degree of grain shape
 irregularity,
which is defined by $I_{1}:I_{2}:I_{3}$. We discuss how the 
orientation of dipole moment $\bmu$  in body coordinates affects the spinning 
dust spectrum for different 
regimes of internal thermal fluctuations. We show that the spinning dust 
emissivity for the case of strong thermal fluctuations  
is less sensitive to the orientation of $\bmu$ than in
the case of weak thermal fluctuations. 
 We calculate spinning dust spectra for a range of gas density and
 dipole moment. The effect of compressible turbulence on spinning dust emission 
 is investigated. 
We show that the emission in a turbulent medium 
 increases by a factor from $1.2$--$1.4$ relative to that in a uniform medium,
as sonic Mach number $M_{\s}$ increases 
from $2$--$7$. Finally, spinning dust parameters are constrained by fitting our 
improved model to five-year 
{\it Wilkinson Microwave Anisotropy Probe}
 cross-correlation foreground spectra, for both the H$\alpha$-correlated and 
100 $\mu$m-correlated emission spectra.
 \end{abstract}

\keywords{ISM: dust, extinction ---
          ISM: microwave emission ---
	  galaxies: ISM ---
          infrared: galaxies
	  }
\section{Introduction}

Cosmic Microwave Background (CMB) experiments 
(see Bouchet et al. 1999; Tegmark et al. 2000; Efstathiou 2003; Bennet et al. 2003)
 are of great importance for studying the early universe and its subsequent 
expansion. Precision cosmology with {\it Wilkinson Microwave 
Anisotropy Probe} 
({\it WMAP}) and Planck satellite requires a good model of the microwave 
foreground emission to allow the subtraction of microwave foreground contamination 
from the CMB radiation. 

Three well-known components of the diffuse microwave Galactic 
foreground consist of synchrotron emission, free-free emission from plasma 
(thermal bremsstrahlung) and thermal emission from dust. However, in the 
microwave range of frequency from $10$--$100$ GHz, anomalous emission which was 
difficult to reconcile with the components above was reported (Kogut et al. 1996a, 1996b).

An explanation for the anomalous emission was proposed by Draine \& Lazarian 
(1998ab, hereafter DL98 model), where it was identified as electric dipole emission 
from rapidly spinning dust grains (hereafter spinning dust emission). 
Although spinning dust emission had
 been discussed previously (see Erickson 1957; Ferrara \& Dettmar 1994), Draine \& Lazarian
were the first to include the variety of excitation and damping processes that are 
relevant for very small grains.

While the DL98 model appears to be in general agreement with observations 
(see Lazarian \& Finkbeiner 2003; Finkbeiner 2004), it did not 
account for a number of physical effects, namely, the non-sphericity of grain 
shapes, internal relaxation, and transient spin-up due to ion collisions. 

The Planck Collaboration 2011b has reported new observations of spinning dust 
emission in new environments. Spinning dust now provides a potential diagnostic 
tool for interstellar dust properties (shape, size distribution, dipole moment).
A comprehensive model of spinning dust for different grain shapes is 
required.

Recent studies showed that the correspondence of the DL98 model to 
observations can be improved by adjusting the parameters of the model. 
For instance, the five-year  ({\it WMAP}) data showed a broad 
bump at $\sim40$ GHz in the H$\alpha$-correlated emission
(Dobler \& Finkbeiner 2008; Dobler, Draine \& Finkbeiner 2009). The $\sim 40$ GHz frequency 
of the peak is higher than the value predicted ($\sim 23$ GHz)
by the DL98 model for standard parameters of the warm ionized medium (WIM). 
Dobler et al. (2009) showed that the bump is consistent with the DL98 model modified so that 
the characteristic dipole moment 
of grains is decreased and gas density of the WIM is increased,
 relative to the typical spinning dust parameters in Draine \& Lazarian 
 (1998b, hereafter DL98b). 

Ali-Ha\"imoud, Hirata \& Dickinson (2009) revisited the spinning dust model and presented an 
analytic solution to the Fokker-Planck (FP) equation that describes the rotational
 excitation of a spherical grain if the discrete nature of the impulses can
 be neglected.

Hoang, Draine \& Lazarian (2010, hereafter HDL10) improved the 
DL98 model by accounting for a number of physical effects.
The main modifications in their improved model of spinning dust
emission are as follows.

(i) Disk-like grains rotate with the grain symmetry axis $\ba_{1}$ not perfectly aligned with 
the angular momentum $\bJ$. The disaligned rotation of 
$\bomega$ with $\ba_{1}$ causes 
wobbling of the grain principal axes with respect to $\bomega$ and $\bJ$ due to 
internal thermal fluctuations.

(ii) Distribution functions for grain angular momentum and velocity 
are obtained exactly using the Langevin equation (LE) for
 the evolution of grain angular momentum in an inertial coordinate system. 

(iii) The limiting cases of fast internal relaxation and no internal relaxation
are both considered for calculation of the angular momentum distributions
 and emissivity of spinning dust.

(iv) Infrequent collisions of single ions which deposit an angular 
momentum larger than the grain angular momentum prior to collision
 are treated as Poisson-distributed events.

The wobbling disk-like grain has anisotropic rotational damping and  excitation.
 Such an anisotropy can increase the peak emissivity by a factor $\sim 2$, and
 increases the peak frequency by a factor $1.4-1.8$, 
 compared to the results from the DL98 model. 

The effect of the grain wobbling on electric dipole emission was independently
studied in Silsbee et al. (2011) using a Fokker-Planck (FP) equation approach,
but they disregarded the effect of internal relaxation and transient spin-up
 by infrequent ion collisions. 

Earlier models of spinning dust emission dealt with axisymmetric grains having
moments of inertia $I_{1}>I_{2}=I_{3}$ (see HDL10;
 Silsbee et al. 2011). An axisymmetric grain of an angular momentum $\bJ$ radiates 
 in general at four frequency modes 
$\dot\phi\pm \dot\psi, \dot\phi$ and $\dot\psi$, where $\dot\phi=J/I_{2}$  is
 the angular frequency of precession of the grain symmetry axis $\ba_{1}$ about $\bJ$, 
 and $\dot\psi=J\cos\theta(1/I_{1}-1/I_{2})$ with $\theta$ being the angle between $\bJ$ 
 and $\ba_{1}$ is the angular frequency of the grain rotation about its symmetry 
axis $\ba_{1}$. For an ``irregular'' (i.e. triaxial body) grain with $I_{1}>I_{2}>I_{3}$, in addition to the 
precession of the grain axis of maximum moment of inertia $\ba_{1}$ 
(hereafter {\it axis of major inertia}) or of the grain axis of minimum moment 
of inertia $\ba_{3}$ (hereafter {\it axis of minor inertia})
 around the angular momentum $\bJ$, 
the grain principal axes 
wobble with respect to $\bJ$. This wobbling, that is different from the wobbling
 due to the internal thermal fluctuations studied in HDL10 for the disk-like grain,
 can result in a more complex 
 electric dipole emission spectrum, so that it modifies the spinning dust emission.
 We will quantify this effect in \S 3.  

Very small grains (smaller than $20\Angstrom$) are important for spinning dust 
emission. 
But also within this range of grain size, the grain dust temperature has strong 
fluctuations 
 due to absorption of UV photons (Greenberg 1968;
 Draine \& Anderson 1985). As a result, they can not be characterized by
a single equilibrium temperature, and are described by
a temperature distribution function (Draine \& Li 2001). The temperature fluctuations 
induce wobbling of the grain axes with respect to the angular momentum 
and modify the spinning dust spectrum. This issue is addressed in \S 5.1.

We investigate also spinning dust emission in the presence of compressible 
turbulence. 
An increase of the spinning dust emission arises as a result of the 
non-linear dependence of emissivity on gas density.

{The seven-year {\it WMAP} data (Gold et al. 2011) and Planck early results
 (Planck Collaboration 2011ab) appear to support spinning dust emission as a source of the 
anomalous microwave emission in new regions with a wide range of physical parameters.
The {\it WMAP} thermal dust-correlated 
spectra in Gold et al. (2011) for two regions within the Galactic plane
indicate that spinning dust emission should peak around $22$ GHz. However, the improved 
model of spinning dust developed by HDL10 predicted a peak frequency above $\sim$ 30 GHz, 
for all media using the same physical parameters as in DL98b. We explore
parameter space for the HDL10 spinning dust model, and characterize 
spinning dust parameters by fitting to the latest observation data.}

The structure of the paper is as follows. In \S 2 we present
the assumptions and notations adopted throughout this paper. In \S 3, we describe
the torque-free motion of an irregular grain and the exchange of vibrational-rotational
 energy. The power spectrum and spinning dust emissivity for irregular grains
  are presented in \S 4.  
 In \S 5, we investigate the influence of grain temperature fluctuations, dipole moment orientation, and explore the parameter space of gas density and magnitude of dipole moment  for spinning dust emissivity.
  \S 6 is devoted to 
the effects of fluctuations of gas density due to compressible turbulence on
 spinning dust emission. 
Constraints on spinning dust parameters obtained by fitting theoretical model to
 the H$\alpha$-correlated and thermal-dust-correlated emission spectra are presented in \S 7. 
Discussion and summary of principal results are given in \S 8 and 9, respectively.

\section{Assumptions and notations}
\subsection{Assumptions}
In the present paper, we follow the same assumptions for 
grain size distribution and rotation dynamics as in HDL10.
The grain size $a$ is
defined as the radius of a sphere of equivalent volume.\footnote{Denote $V$ 
be the volume of the grain, then $a$ is defined as $V=4\pi a^{3}/3$.} Unless 
stated otherwise, we assume throughout the paper that grains smaller than 
$a_{2}=6\times 10^{-8}$ cm 
are planar, and grains larger than $a_{2}$ are approximately spherical. 
We adopt rotational damping and excitation coefficients
due to gas-grain interaction, infrared emission, and electric dipole
damping from HDL10. Our notation is summarized in Table 
\ref{tab:notation}. 

\subsection{Grain electric dipole moment}
The electric dipole moment of a grain arises from the intrinsic dipole moment of 
molecules within the grain and from the asymmetric distribution of the grain 
charge. The former is shown to be dominant (see DL98b). 

The grain dipole
moment $\bmu$ can be written as 
\bea
{\bmu}={\bmu}_i + {\bepsilon}Ze a_x~~,~~~
\ena
where $\bmu_i$ is the intrinsic dipole moment of an uncharged grain,
$Ze$ is the grain charge, $a_{x}$ is the excitation-equivalent radius of the 
grain, and the vector ${\bepsilon}a_x$ is
the displacement between the grain center of mass and the charge centroid.

Following DL98b, the magnitude of the dipole moment is given by  
\bea
\mu^{2}=23\left[\left(\frac{a_{x}}{a}\right)^2 \langle
  Z^2\rangle+3.8\left(\frac{\beta}{0.4~\D}\right)^2a_{-7}\right]a_{-7}^{2}
~\D^{2},~~~
\label{mu2}
\ena
where $\langle Z^{2}\rangle$ is the
mean square grain charge, $\beta$ is the dipole moment per atom of the
grain, and $a_{-7}=a/10^{-7}$cm. Above, $\epsilon \sim 0.1$ is assumed.

\begin{table}[htb]
\begin{center}
\caption{\label{tab:notation}
         Notations and Meanings}
\begin{tabular}{l c l}
\hline\hline
Symbol & Meaning \cr
\hline
$a$ & grain size \cr
$\nu$    & frequency \cr
$n_{\H}$ & H nucleus density \cr 
$x_{\rm H}$ & H ionization fraction\cr
$x_{\rm M}$ & metal ionization fraction\cr
$T_{\rm gas}$& gas temperature\cr
$T_{\rm vib}$ & dust vibrational temperature\cr
$T_{\rm d}$ & decoupling temperature for \cr
& vibrational-rotational energy exchange\cr
$\tau_{\H}$ & gas damping time\cr
$\tau_{\rm ed}$ & electric dipole damping time\cr
$\beta_{0}$ & characteristic dipole moment\cr
$\bmu$& electric dipole moment\cr
$\omega_{\rm T}$ & thermal angular velocity\cr
$\bJ$ & angular momentum\cr
$\ba_{1},\ba_{2},\ba_{3}$ & basis vectors of grain principal axes \cr
$I_{1},I_{2},I_{3}$& eigenvalues of moment of inertia tensor, \cr
$\xhat,\yhat,\zhat$ & coordinate systems associated to $\bJ\|\zhat$\cr
$\be_{1},\be_{2},\be_{3}$& inertial coordinate system fixed to the lab \cr
$\theta,~\phi,~\psi$ & Euler angles of grain axes in external \cr
& system $\xhat\yhat\zhat$\cr
$q$& $2I_{1}E_{\rm rot}/J^{2}$ with $E_{\rm rot}$ rotational energy\cr
$P_{\ed}$& emission power\cr
$j_{\nu}$& emissivity at frequency $\nu$\cr
${\rm case}~ 1$& $\bmu$ oriented so that $\mu_{1}^{2}=\mu_{2}^{2}=\mu_{3}^{2}$\cr
${\rm case}~ 2$& $\bmu$ perpendicular to $\ba_{1}$\cr
${\rm model}~ A$& 25$\%$ of grains having $\beta=\beta_{0}/2$, \cr
&$50\%$ of grains having $\beta_{0}$\cr
& and $25\%$ grains having $2\beta_{0}$\cr
\cr
\hline\hline\\
\end{tabular}
\end{center}
\end{table}

\begin{table}[htb]
\begin{center}
\caption{Idealized Environments For 
 Interstellar Matter}\label{ISM}
\begin{tabular}{llll} \hline\hline\\
\multicolumn{1}{c}{\it Parameters} & \multicolumn{1}{c}{CNM}& 
{WNM} &{WIM}\\[1mm]
\hline\\
$n_{\H}$~(cm$^{-3}$) &30 &0.4 &0.1 \\[1mm]
$T_{\gas}$~(K)& 100 & 6000 &8000 \\[1mm]
$\chi$ &1 &1 &1 \\[1mm]
$x_{\H}$ &0.0012 &0.1 &0.99 \\[1mm]
$x_{\rm M}$ &0.0003 &0.0003 &0.001 \\[1mm]
$y=2n({\H}_{2})/n_{\H}$& 0. & 0. & 0. \cr
\cr
\hline\hline\\
\end{tabular}
\end{center}
\end{table}

The electric dipole moment, assumed to be fixed in the grain body system, 
 can be decomposed into three components along the grain principal axes:
\bea
\bmu=\mu_{1}\ba_{1}+\mu_{2}\ba_{2}+\mu_{3}\ba_{3},\label{mu_eq}
\ena
where $\ba_{1},\ba_{2}$ and $\ba_{3}$ are the basis vectors of 
principal axes of the grain.

Because the orientation of $\bmu$ in the grain body is not well constrained, 
in the present paper, we study two limiting cases.
 In case 1, $\bmu$ is fixed in the grain body 
 such that $\mu_{1}^{2}=\mu_{2}^{2}=\mu_{3}^{2}=\mu^{2}/3$. 
In case 2, $\bmu$ is perpendicular to
 the grain axis of major inertia $\ba_{1}$, so that $\mu_{1}^{2}=0$
 and $\mu_{2}^{2}=\mu_{3}^{2}=\mu_{\perp}^{2}/2=\mu^{2}/2$,
 where $\mu_{\perp}^{2}=\mu_{2}^{2}+\mu_{3}^{2}$. 

Table \ref{ISM} presents typical physical parameters for various phases 
of the interstellar medium (ISM), including the cold neutral medium (CNM),
 warm neutral medium (WNM) and warm ionized medium (WIM).

\section{Rotational Dynamics of Grain of Irregular Shape}

Very small dust grains (also PAHs) are expected to be planar, while larger grains 
are likely to be more-or-less spherical. Even though planar, small PAHs are not
 expected to be perfectly symmetric. In general, for both PAHs and larger 
particles, we expect the eigenvalues $I_{i}$ of the moment of inertia tensor to 
be nondegenerate: $I_{1}>I_{2}>I_{3}$. We will refer to grains 
with nondegenerate eigenvalues as ``irregular''.

\subsection{Torque-free rotation and flip states}\label{torq-free}
Let us consider an irregular grain characterized by a triaxial body 
 having moments of inertia 
$I_{1}>I_{2}>I_{3}$ along the principal axes $\ba_{1}, \ba_{2}$ and $\ba_{3}$.
\begin{figure}
\includegraphics[width=0.4\textwidth]{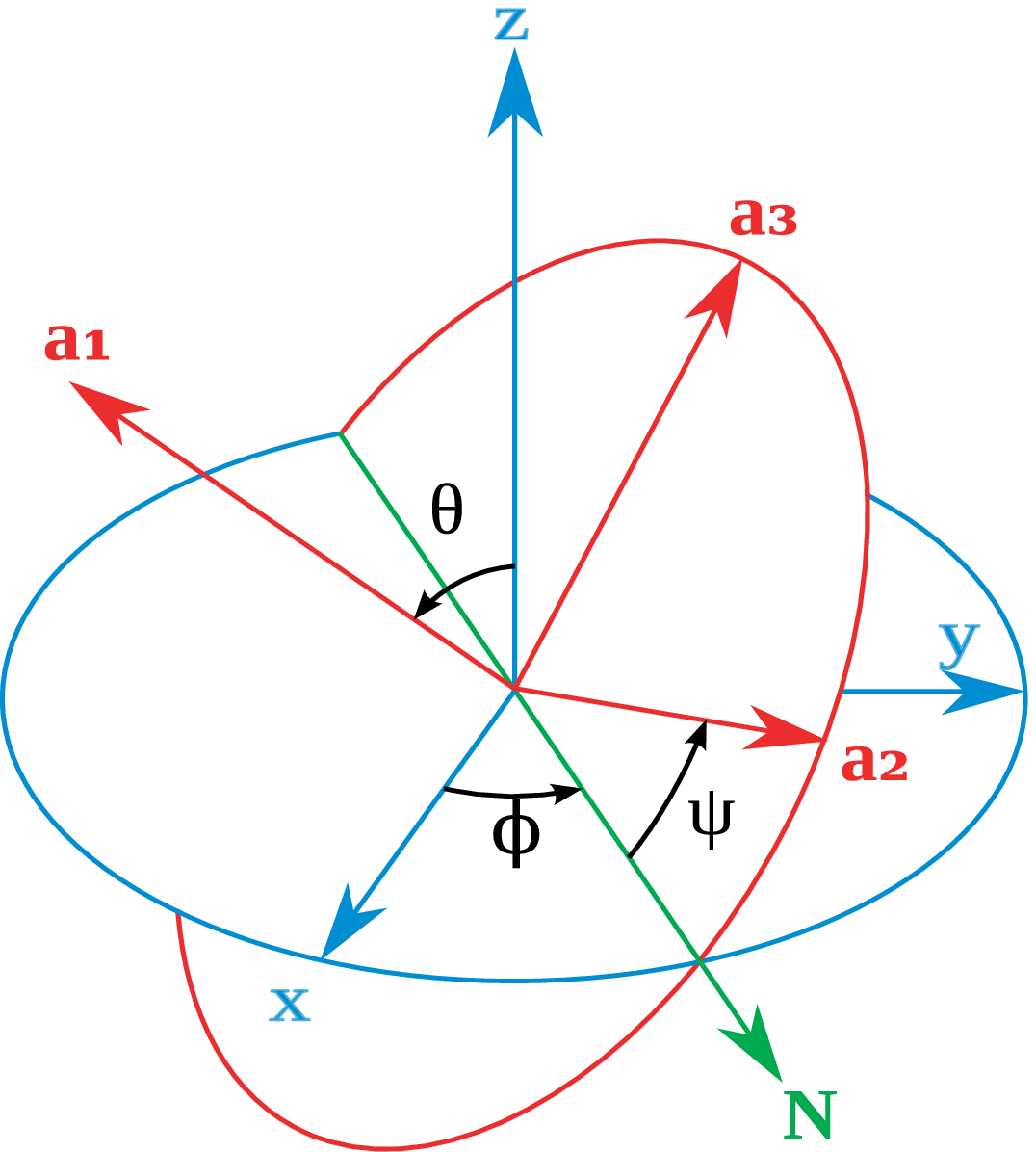}
\caption{Torque-free motion of a grain having three principal axes $\ba_{1}, \ba_{2}$
 and $\ba_{3}$ in an inertial coordinate system with 
$\zhat \| \bJ$ and $\xhat,~\yhat\perp \bJ$ described by three Euler angles 
$\theta, \phi$ and $\psi$.}
\label{euler}
\end{figure}

The dynamics of the irregular grain is more complicated than that of a disk-like
grain. In addition to the precession of the axis of major inertia 
 $\ba_{1}$
 about $\bJ$ as in the disk-like grain, the axis $\ba_{1}$  wobbles rapidly,
resulting in variations of the angle $\theta$ between $\ba_{1}$ and $\bJ$ at 
a rate $\dot\theta$. 
Therefore, we expect the wobbling of the irregular grain can 
produce more frequency modes than those observed in the disk-like grain with 
$I_{2}=I_{3}$ (see HDL10).

To describe the torque-free motion of an irregular grain having a rotational energy
$E_{\rm rot}$, we use conserved quantities, including the angular momentum $\bJ$,
and a dimensionless parameter  that characterizes the deviation of the grain 
rotational energy from its minimum value,
\bea
q=\frac{2I_{1}E_{\rm rot}}{J^{2}}.\label{q_eq}
\ena

 Following Weingartner \& Draine (2003), we define the total number of states $s$ 
in phase space for $q$ ranging from $1$ to $q$ as 
\bea
s\equiv 1-\frac{2}{\pi}\int_{0}^{\psi_{1}}d\psi \left[\frac{I_{3}(I_{1}-I_{2}q)+
I_{1}(I_{2}-I_{3})\cos^{2}\psi}{I_{3}(I_{1}-I_{2})+I_{1}(I_{2}-I_{3})
\cos^{2}\psi}\right]^{1/2},~~~~\label{sq}
\ena
where
\bea
\psi_{1}=\cos^{-1}\left[\frac{I_{3}(I_{2}q-I_{1})}{I_{1}(I_{2}-I_{3})}\right]^{1/2},
\ena
for $q>q_{\rm sp}$ and $\psi_{1}=\pi/2$ for $q\leq q_{\rm sp}$, with 
$q_{\rm sp}\equiv I_{1}/I_{2}$ being the separatrix between the two regimes.

The instantaneous orientation of the grain in an inertial coordinate system is given 
by three Euler angles $\psi, \phi$ and $\theta$ (see Figure \ref{euler}). 
The angular velocities are obtained by solving the Euler equations of motion.
The torque-free motion of an irregular grain is treated in detail in Appendix A.

Briefly, the Euler equations give rise to two sets of solutions corresponding 
to $+$ and $-$ signs 
for $\omega_{i}$ with $i=1, 2$ and $3$ (see Eqs \ref{omep1}-\ref{omep3}).
It can be seen that for $q\le q_{\sp}$, two rotation states 
$\pm$ correspond to 
$\omega_{1}>0$ and $\omega_{1}<0$, i.e., $\ba_{1}\cdot\bJ>0$ and $\ba_{1}\cdot\bJ<0$.
 Following Weingartner \& Draine (2003), we refer to these rotation states as 
the {\it positive} flip state and {\it negative} flip state with 
respect to $\ba_{1}$ (see also Hoang \& Lazarian 2009a). For 
$q>q_{\sp}$, similarly, two rotation states correspond to $\omega_{3}>0$ and 
$\omega_{3}<0$, i.e., $\ba_{3}\cdot\bJ<0$ and $\ba_{3}\cdot\bJ>0$. These rotation
 sates are referred to as the {\it positive} and {\it negative} flip states with 
respect to $\ba_{3}$. 

\subsection{Internal relaxation, thermal fluctuations and thermal flipping}
\label{sec:inter}

\subsubsection{Internal relaxation and thermal fluctuations}

Internal relaxation (e.g. imperfect elasticity, Barnett relaxation; Purcell 1979;
Lazarian \& Efroimsky 1999)
 arises from the transfer of grain rotational energy to vibrational modes. For
cold grains, this process tends to result in nearly perfect alignment of the grain axis of major
 inertia with the angular 
momentum.\footnote{This rotation configuration has minimum rotational energy 
or highest entropy.} Naturally, if the grain has nonzero vibrational energy, 
 energy can also be transferred from the vibrational modes into
grain rotational energy (Jones \& Spitzer 1969). 

For an isolated grain, a small amount of energy gained
 from the vibrational modes results in fluctuations of the rotational energy 
$E_{\rot}$ when the grain angular momentum is conserved. Such fluctuations in $E_{\rot}$ result in fluctuations of $q$ and of the angle 
 $\theta$ between $\ba_{1}$ and $\bJ$ for an axisymmetric
 grain. 
For an irregular grain, the fluctuations in $E_{\rot}$
are described by fluctuations in $s(q)$ (see Eq. \ref{sq}).
Over time, the fluctuations in $E_{\rot}$ establish a local thermal equilibrium
 (LTE) at a {\it rotational energy equilibrium temperature} $T_{\rm rot}$. 

\subsubsection{Exchange of Vibrational-Rotational Energy}

The Intramolecular 
Vibrational-Rotational Energy Transfer process (IVRET) due to imperfect elasticity
 occurs on a timescale $10^{-2}$ s,
 for a grain of a few Angstroms (Purcell 1979),
 which is shorter than the IR emission time. So, when the vibrational energy 
decreases due to IR emission, as long as the Vibrational-Rotational (V-R)
energy exchange exists, interactions between vibrational
 and rotational systems maintain a thermal equilibrium, i.e.,
$T_{\rm rot}\approx T_{\rm vib}$. As a result, the LTE distribution
 function of rotational energy reads (hereafter VRE regime;
 see Lazarian \& Roberge 1997):
\bea
f_{\VRE}(s,J)\propto {\rm exp}\left(-\frac{E_{\rm rot}}{k_{\B}T_{\rot}}\right)
\approx{\rm exp}\left(-\frac{E_{\rm rot}}{k_{\B}T_{\vib}}\right).\label{eq:fs}
\ena

The existence of V-R energy exchange depends on the 
rotational modes.  In principle,
 the rotational energy can change in increments of $\hbar \omega$, with $\omega$
 being the angular frequency of rotational modes. However, at low vibrational energy $E_{\vib}$, the 
vibrational mode spectrum is sparse, and there may not be available transitions
 with $\Delta E_{\vib}=\hbar\omega_{\rot}$. For the case $\ba_{1}$ parallel to 
$\bJ$, the V-R energy exchange occurs only when the grain is hot just after absorption
 of a UV photon, because the grain rotating with $\ba_{1}\|\bJ$ has only one 
rotational mode $\omega=\omega_{\rot}\equiv J/I_{1}$.  
As the grain cools down further, there is no available vibrational transition
 corresponding to an energy change $\hbar\omega_{\rot}$, and the V-R coupling 
ceases.

For disk-like or irregular grain, multiple rotational frequency
 modes  are observed, and some modes correspond to energy 
$\hbar \omega > \hbar\omega_{\rot}$ (see Eq. \ref{ome_m} or Fig. \ref{fft}). This
 may  allow V-R energy exchange even when the vibrational 
energy is relatively low. Nevertheless, the energy separation between the vibrational
 ground state and the first vibrationally-excited state is large enough that V-R
 energy exchange is unlikely to take place after the PAH cools to the vibrational
 ground sate. The key, then, is to know the vibrational ``temperature'' of a cooling
 PAH at the time when V-R energy exchange ceases to be effective. Sironi \& Draine
 (2009) estimated a {\it decoupling} temperature $T_{\d}=65$ K for a PAH with 200 C 
atoms.

In regions with the average starlight background, the mean time interval between 
absorptions of starlight photon will be $\sim 10^{6}$ s for a PAH containing
 200 C atoms. Because the PAH will cool below $65$ K in $\sim 10^{3}$ s (see,
 e.g., Fig. 9 of Draine \& Li 2001), V-R energy exchange will be suppressed most 
 of the time, taking place only during $\sim 10^{3}$ s intervals following
 absorption of starlight photons. However, the $10^{6}$ s interval between 
photon absorptions is short compared to the timescale over which $J$ will change 
 appreciably, and we will therefore approximate the V-R 
energy exchange process as continuous, even though it is episodic.

Substituting $E_{\rot}$ as a function of $J$ and $q$ from Equation (\ref{q_eq})
 into Equation (\ref{eq:fs}),  
the distribution function for the rotational energy becomes
\bea
f_{\VRE}(s,J)= A{\rm exp}\left(-\frac{q(s)J^{2}}{2I_{1}k_{\B}T_{\vib}}\right),
\label{fs}
\ena
where $A$ is a normalization constant such that $\int_{0}^{1} f_{\VRE}(s,J)ds=1$.

Define a rotational
 temperature $T_{J}=J^{2}/2I_{1}k_{\B}$.  Thus, the distribution for $q$
becomes $f_{\VRE}\propto {\rm exp}\left(-q(s)T_{J}/T_{\vib}\right)$. 
When $T_{J}\gg T_{\vib}$, the grain is suprathermally rotating, and 
$\ba_{1}$ is aligned with $\bJ$. 
In contrast, when $T_{J} \ll T_{\vib}$, the grain rotates subthermally,
 and thermal fluctuations randomize the grain axes with respect to $\bJ$. 
For fixed angular momentum, the thermal fluctuations increase with $T_{\vib}$.

\subsubsection{Thermal flipping}
Transfer of energy between the vibrational and rotational modes due to dissipation
 or thermal fluctuations will cause $E_{\rot}$, and therefore $q$, to vary with 
time. If $q$ crosses the separatrix state $q=q_{\sp}$, 
the new flip state is effectively chosen randomly. Changes
 in $q$ can also result from external forcing (e.g., collisions with gas atoms
 or absorption/emission of radiation), that may change both $J$ and $E_{\rot}$.
 If the external torques are not ``chiral'', then the changes in $q$ over
 time will result in equal probabilities of the two flip states. However, the 
chirality of the grain may lead to $q$-changing processes that depend not 
only on $q$, but also on which flip state the grain is in. The present treatment
 neglects any grain chirality.

\section{Electric Dipole Emission from Grains of Irregular Shape}
 Below we investigate electric dipole emission arising from irregular grains. 
 We first calculate emission power spectra arising from a torque-free rotating 
irregular grain. Then, we compute the spinning dust emissivity and compare 
 the results with those from disk-like grains. 

\subsection{Power Spectrum}

A torque-free rotating  irregular grain can rotate either in the positive or negative 
flip state, and the treatment  in this section is applicable for a general  
flip state.
To find the electric dipole emission spectrum from such an isolated irregular grain, 
we follow the approach in HDL10. We first represent the dipole moment $\bmu$
 in an inertial coordinate system, and then compute its second derivative. 
We obtain
\bea
\ddot{\bmu}=\sum_{i=1}^{3}\mu_{i}\ddot{\ba}_{i},\label{dotmu}
\ena
where $\mu_{i}$ are components of $\bmu$ along principal axes $\ba_{i}$,
$\ddot{\ba}_{i}$ are second 
derivative of $\ba_{i}$ with respect to time, and $i=1, 2$ and $3$. 

The instantaneous emission power of the dipole moment is defined by 
\bea
P_{\rm ed}(J,q,t)=\frac{2}{3c^{3}}\ddot\bmu^{2}.\label{power}
\ena

The power spectrum is then obtained from the Fourier 
transform (FT) for the components of $\ddot\bmu$. 
For example, the amplitude of $\ddot\mu_{x}$ at the frequency $\nu_{k}$ is 
defined as
\bea
\ddot\mu_{x,k}=\int_{-\infty}^{+\infty} \ddot\mu_{x}(t) 
{\rm exp}\left(-i2\pi\nu_{k}t\right)dt,
\ena
where $k$ denotes the frequency mode. The emission power at the positive 
frequency $\nu_{k}$ is given by
\bea
P_{\ed,k}(J,q)=\frac{4}{3c^{3}}(\ddot\mu_{x,k}^{2}+\ddot\mu_{z,k}^2+
\ddot\mu_{z,k}^2),
\ena
where the factor $2$ arises from the positive/negative frequency symmetry 
of the Fourier spectrum. To reduce the spectral leakage in the FT, we convolve
the time function $\ddot\bmu$ with a Blackman-Harris window function (Harris 1978). 
The power
 spectrum needs to be corrected for the power loss due to the window function.

The total emission power from all frequency modes for a given $J$ and $q$ is then
\bea
P_{\ed}(J,q)=\sum_{k}P_{\ed,k}(J,q)
\equiv\frac{1}{T}\int_{0}^{T}dt\left(\frac{2}{3c^{3}}\ddot\bmu^{2}\right),~~~\label{ptot}
\ena
where $T$ is the integration time. \footnote{This is the result of Parseval's
Theorem.}

Here we compute power spectra for case 1 ($\mu_{1}=\mu/\sqrt{3}$) of $\bmu$ orientation,
 but our approach can be used straightforward for any orientation of $\bmu$.
 
\begin{figure*}
\includegraphics[width=0.33\textwidth]{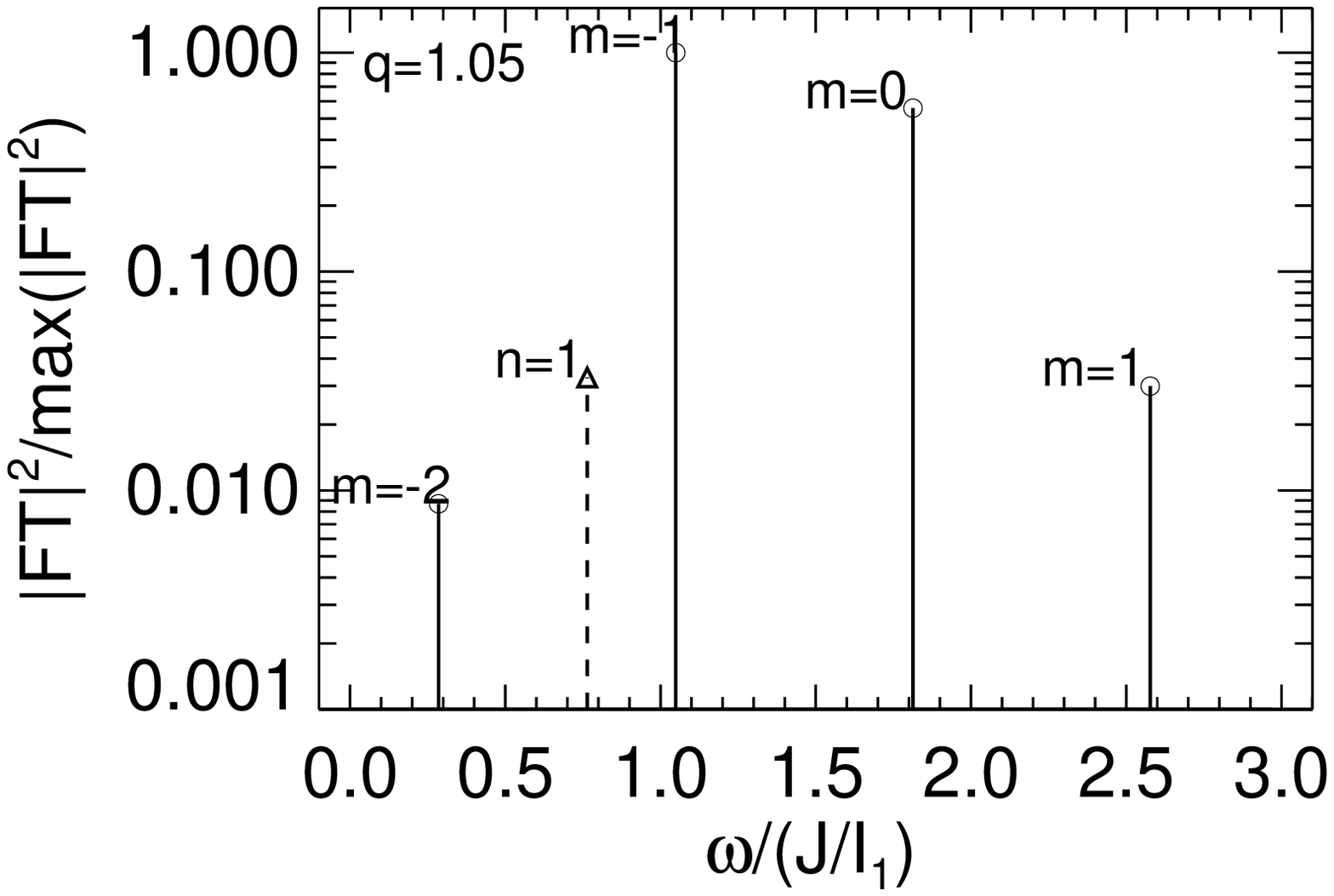}
\includegraphics[width=0.33\textwidth]{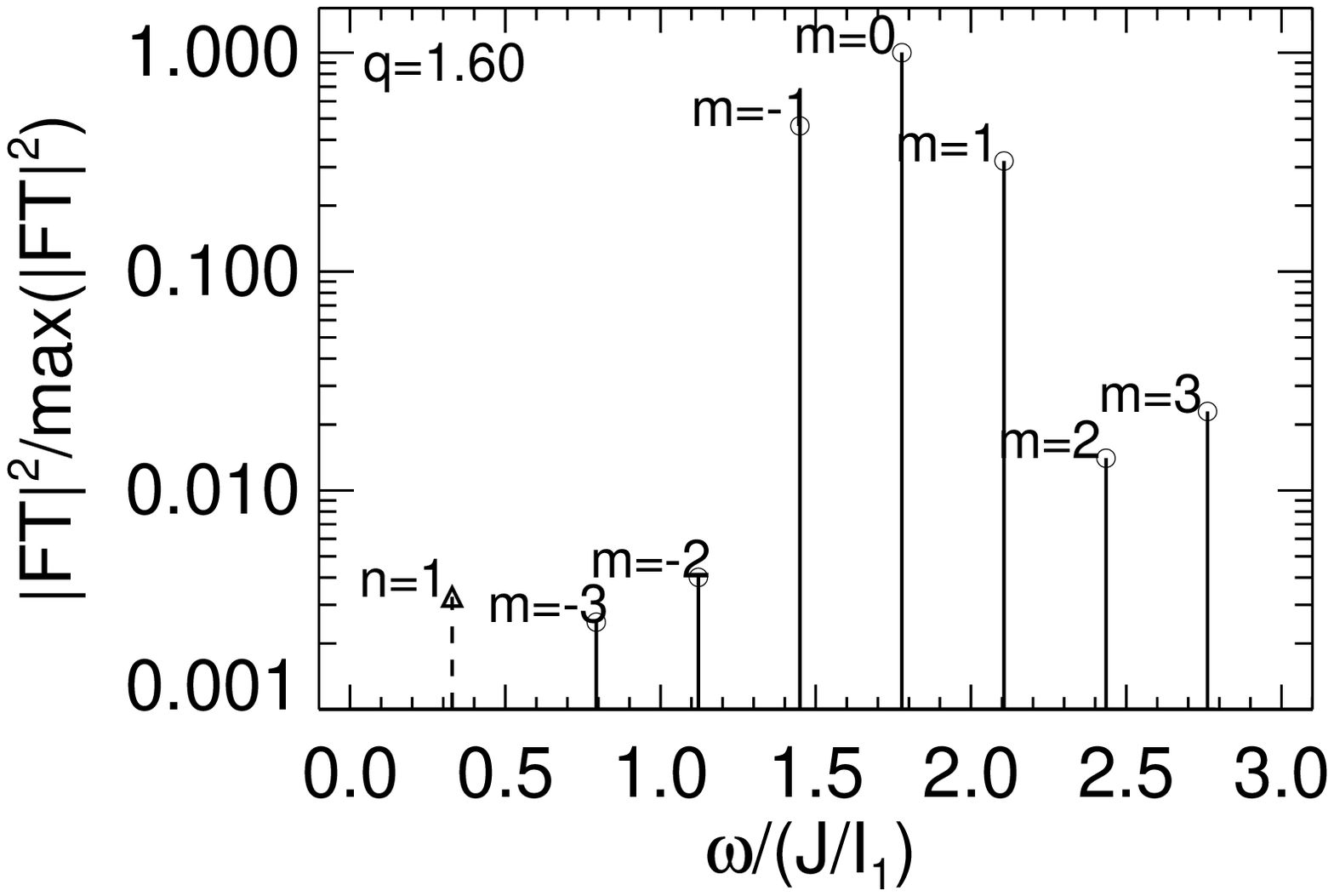}
\includegraphics[width=0.33\textwidth]{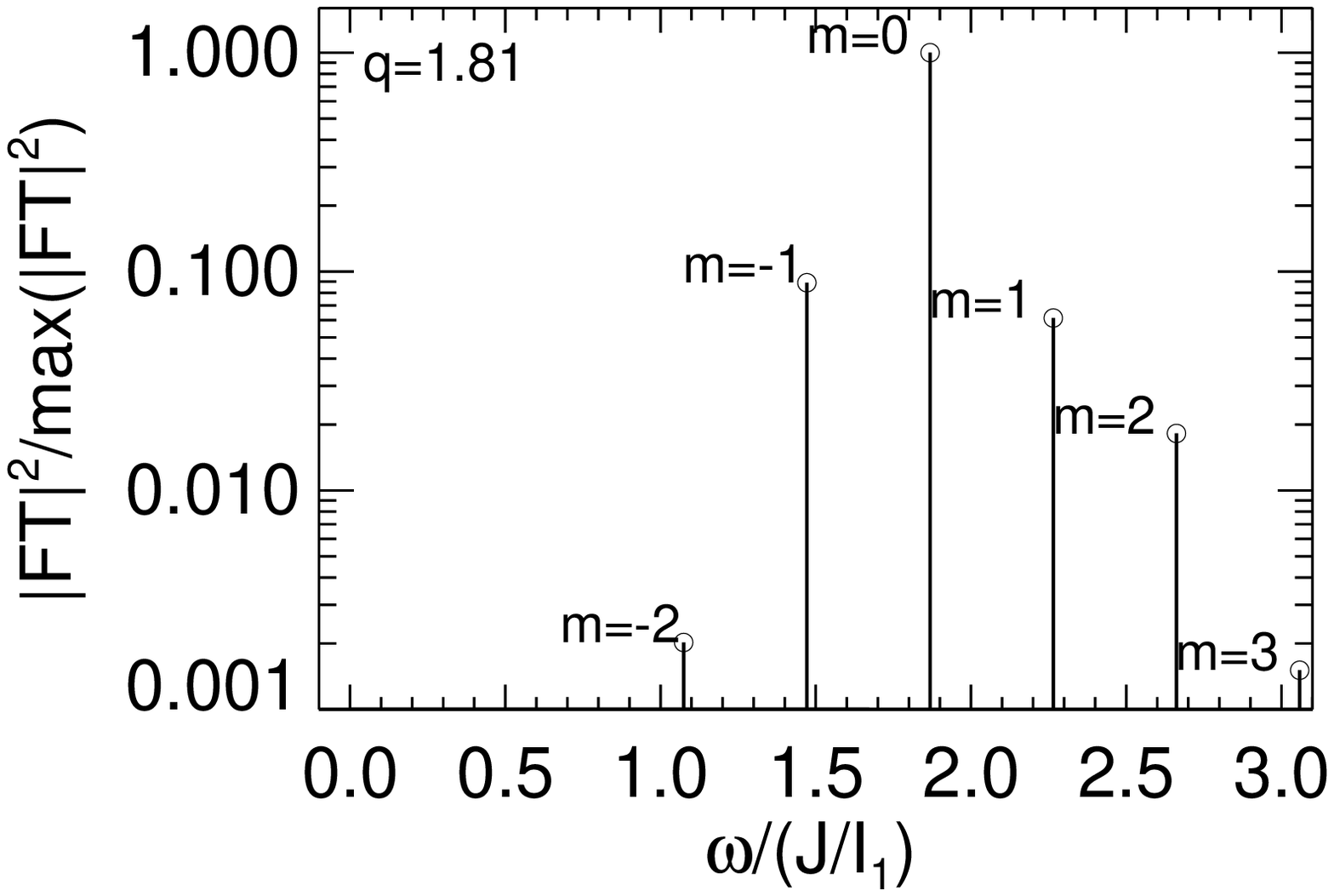}
\caption{Normalized power spectrum  of a torque-free rotating irregular grain
with $I_{1}:I_{2}:I_{3}=1:0.6:0.5$ 
for different values of $q=1.05, 1.60$ 
(i.e. $q<q_{\rm sp}\equiv I_{1}/I_{2}$) and
 $q=1.81>q_{\rm sp}$.
 The components of $|{\rm FT}(\ddot{\mu}_x)|^2/\max(|{\rm FT}(\ddot{\mu}_x)|^2))$
 (or $|{\rm FT}(\ddot{\mu}_y)|^2/\max(|{\rm FT}(\ddot{\mu}_x)|^2)$)
 are indicated by circles, while the components of $|{\rm FT}(\ddot{\mu}_z)|^2/
\max(|{\rm FT}(\ddot{\mu}_x)|^2)$
 are indicated by triangles. Orders of in-plane modes $m$  and out-of-plane modes 
 $n$ are indicated, and case 1 ($\mu_{1}=\mu/\sqrt{3}$) of $\bmu$ orientation is assumed.}
\label{fft}
\end{figure*}

Figure \ref{fft} presents normalized power spectra (squared amplitude of
Fourier transforms), $|{\rm FT}(\mu_{x,y})|^{2}/\max(|{\rm FT}(\mu_{x})|^{2})$ and 
$|{\rm FT}(\mu_{z})|^{2}/\max(|{\rm FT}(\mu_{x})|^{2})$
for the components $\ddot\mu_{x}$ (or $\ddot\mu_{y}$) and $\ddot{\mu}_{z}$, for 
a torque-free rotating irregular grain having the ratio of moments of inertia 
$I_{1}:I_{2}:I_{3}=1:0.6:0.5$ in the positive flip state with various $q$. 
Circles and triangles denote peaks
of the power spectrum for components of $\ddot\mu_{x}$ (or $\ddot\mu_{y}$) and
$\ddot\mu_{z}$, respectively. 

 Multiple frequency modes  are observed in the power spectra
  of the irregular grain, but in Figure \ref{fft} we  show only 
the modes with power no less than $10^{-3}$ the maximum value. 
Both positive and negative flip states have the same frequency modes. 

For $q<I_{1}/I_{2}$, we found that power spectra for $\ddot\mu_{x}$ 
(or $\ddot\mu_{y}$) have angular frequency modes 
\bea
\omega_{m}\approx \langle \dot\phi\rangle+ 
m\langle|\dot\psi|\rangle,\label{ome_m}
\ena
 where the bracket denotes the averaging value over
 time, and $m=0,\pm 1,\pm 2...$ denote the order of the mode. 
The frequency modes for $\ddot\mu_{z}$  are given by 
\bea
\omega_{n}=n \langle|\dot{\psi}|\rangle,\label{ome_n}
\ena
 where $n$ is integer and $n\ge 1$. 
 
The above frequency modes are equivalent to results obtained from a quantum mechanical treatment. For the limiting case of an axisymmetric grain with $I_{1}>I_{2}=I_{3}$,
 the grain rotational Hamiltonian is $H=J^{2}/2I_{2}-J_{z}^{2}(I_{1}-I_{2})/(2I_{1}I_{2})$ where $J_{z}=J\cos\theta$
 is the projection of $\bJ$ on the symmetry axis. From a quantitization, $J^{2}\rightarrow J(J+1)\hbar^{2}$ and $J_{z}\rightarrow K\hbar$ where $J$ and $K$ are quantum numbers for the angular momentum and 
 its projection on the grain symmetry axis, the rotational energy is written as $E_{J,K}=J(J+1)\hbar^{2}/2I_{2}-K^{2}\hbar^{2}(I_{1}-I_{2})/(2I_{1}I_{2})$.
 For PAHs, $J$ is usually very large, so an emission mode arising from the transition from 
 the energy level $E_{J,K}$ to $E_{J',K'}$ have frequency given by 
 \bea
 \omega_{JJ',KK'}=\frac{1}{\hbar}\left(\frac{\partial E_{J,K}}{\partial J}\Delta J+\frac{\partial E_{J,K}}{\partial K}\Delta K\right),
 \ena
 where $\partial E_{J,K}/\partial J\approx (J\hbar^{2}/I_{2})=\hbar\dot\phi$ and $\partial E_{J,K}/\partial K=(K\hbar^{2}/I_{1}I_{2})(I_{1}-I_{2})=\hbar\dot\psi$. The selection rules
for electric dipole transitions correspond to $\Delta J=0,~\pm 1$, and $\Delta K=0,~\pm 1$, so the total emission modes are $\omega_{JJ',KK'}= \dot\phi,~\dot\phi\pm \dot\psi$
  and $\dot\psi$, as found in HDL10.  
The rotational energy levels of asymmetric rotors are specified
     by three quantum numbers -- the total angular momentum $J$ and
     two additional quantum numbers $K_{-1}$ and $K_{+1}$.  Electric
     dipole transitions obey selection rules $\Delta J=0, \pm1$,
     $\Delta K_{-1}=\pm1, \pm3$, and $\Delta K_{+1}=\pm1, \pm3$
     (see, e.g., Townes \& Schawlow 1955) \footnote{The rotational levels
     of a particular asymmetric rotor, H$_2$O, are shown in Draine
(2011).} .
     In the large $J$ limit, these selection rules will correspond to
     the modes of Equations (\ref{ome_m}) and (\ref{ome_n}).

For $q>I_{1}/I_{2}$,
only the mode $m=0$ has angular frequency given by Equation (\ref{ome_m}).

In the following, the emission modes induced by the oscillation of $\mu_{x}$ or 
$\mu_{y}$, which lie in the $\xhat\yhat$ plane, perpendicular 
to $\bJ$, are called {\it in-plane} modes, and those induced by the oscillation of 
$\mu_{z}$ in the direction perpendicular to the $\xhat\yhat$ plane, are called 
{\it out-of-plane} modes. The order of mode is denoted by
$m$ and $n$, respectively.

Figure \ref{fft} also shows that the emission power for out-of-plane modes 
$\omega_{n}$ is negligible compared to the power emitted by in-plane modes $\omega_{m}$.

\begin{figure*}
\includegraphics[width=0.33\textwidth]{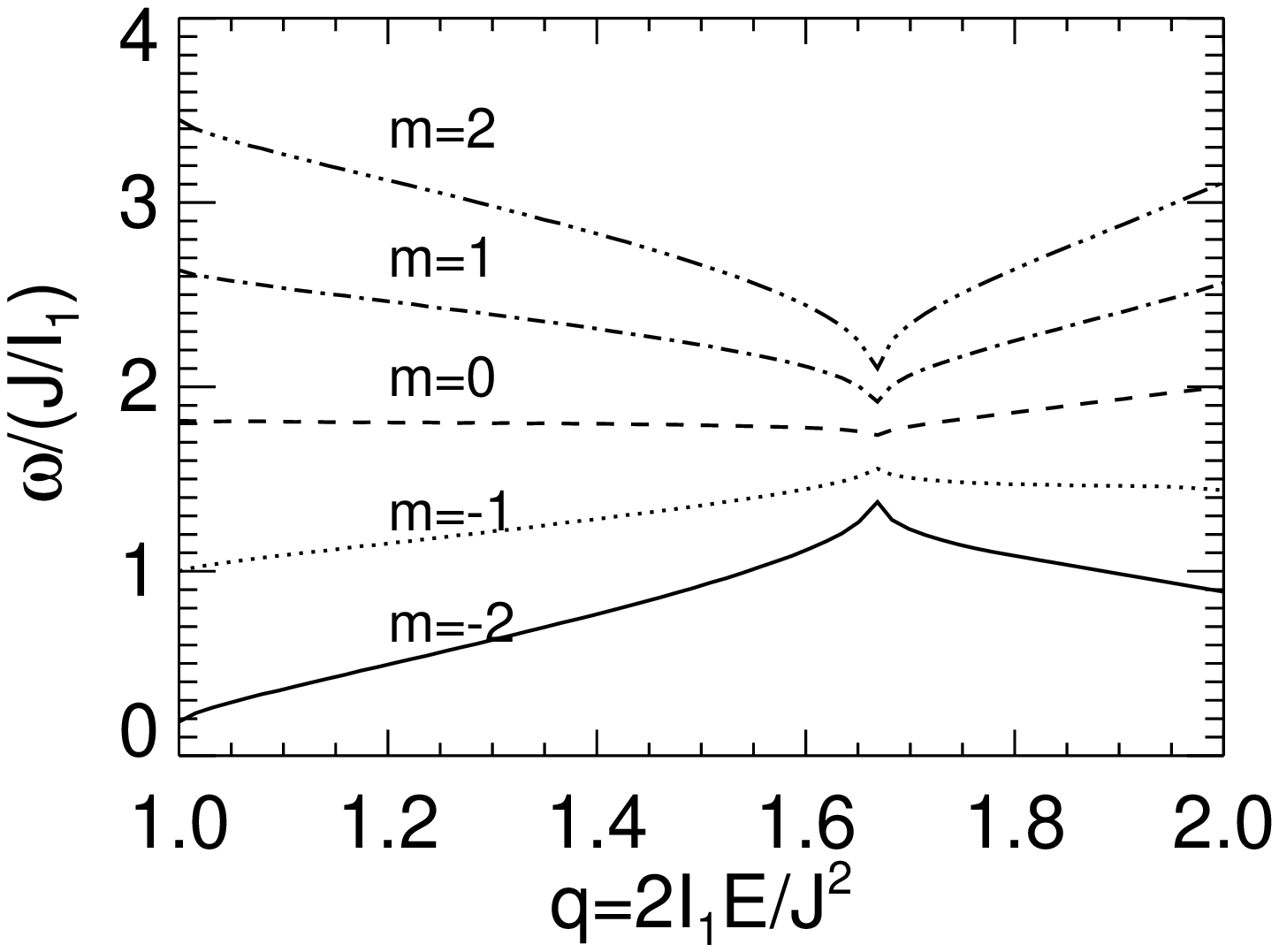}
\includegraphics[width=0.33\textwidth]{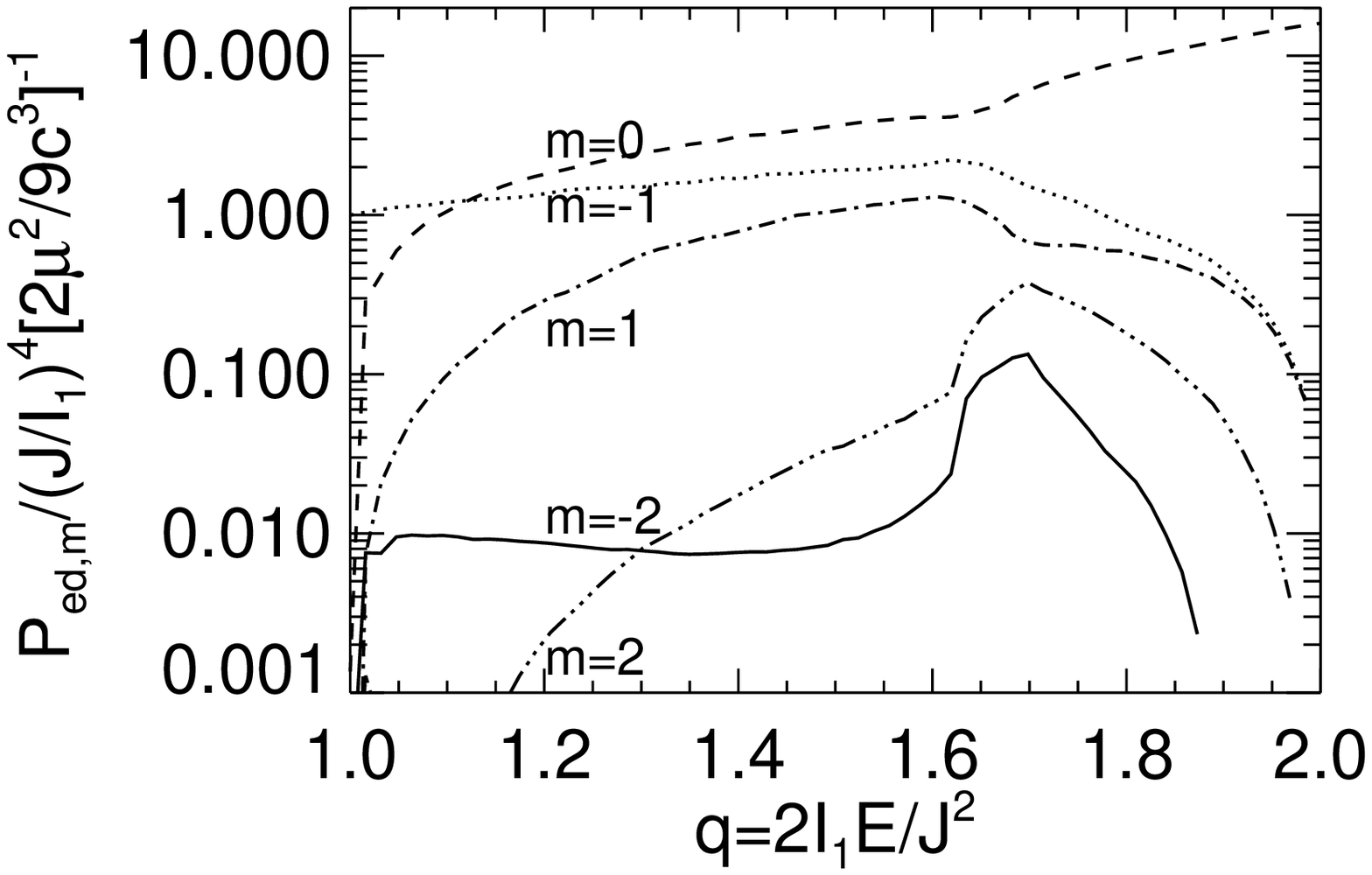}
\includegraphics[width=0.33\textwidth]{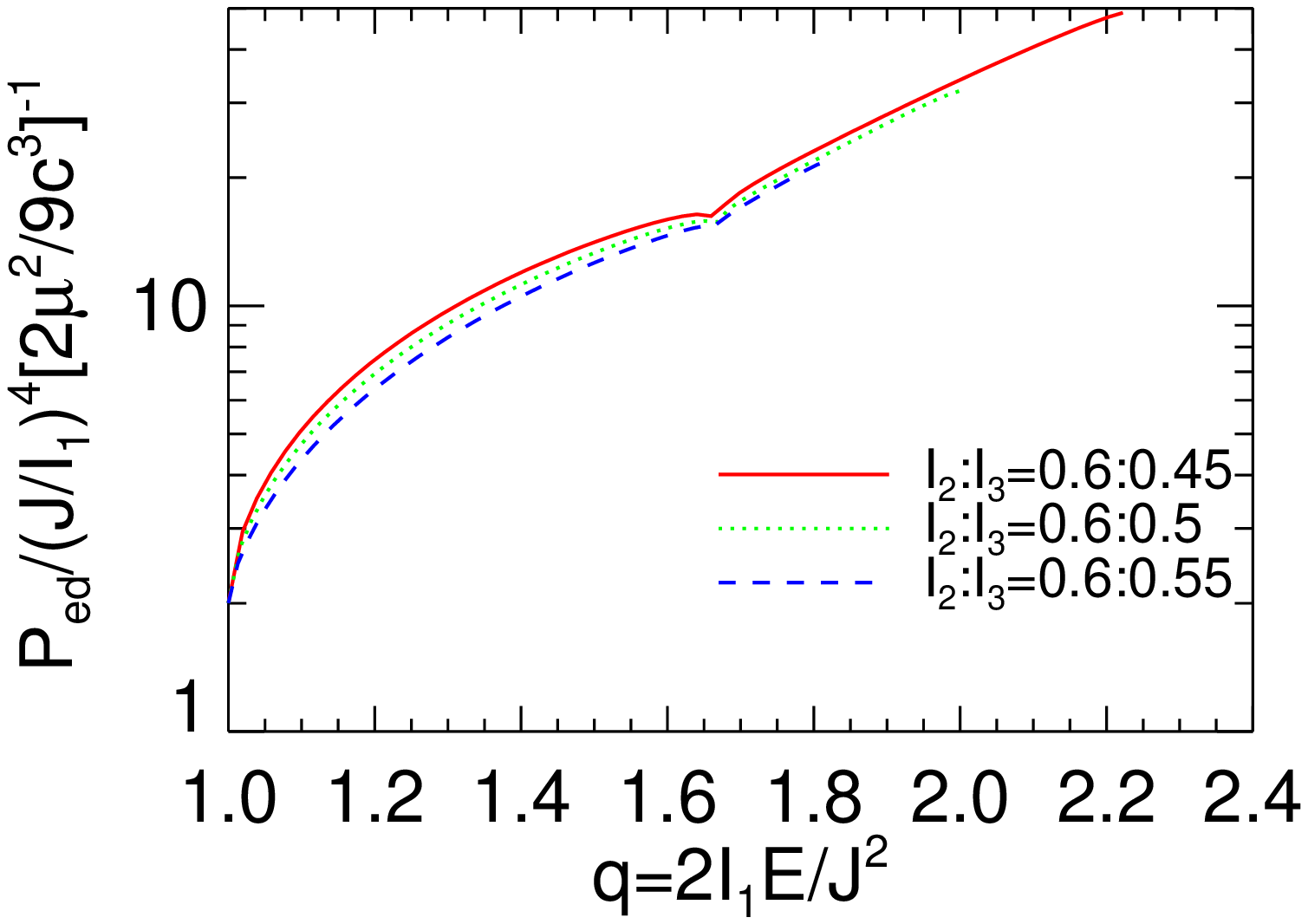}
\caption{{\it Left}: Angular frequency of in-plane modes as a function of $q$ 
for an irregular grain with $I_{1}:I_{2}:I_{3}=1:0.6:0.5$. 
The kink at $q=q_{\rm sp}\equiv I_{1}/I_{2}$ corresponds
 to the change from rotation around $\ba_{1}$ to rotation around $\ba_{3}$.
 {\it Middle}: Emission power radiating at
frequency modes $\omega_{m}$ as a function of $q$.
{\it Right:} Total emission power from all frequency modes as a function of $q$ for 
various ratios $I_{2}:I_{3}$ while $I_{1}:I_{2}=1:0.6$ is 
kept constant. Case 1 ($\mu_{1}=\mu/\sqrt{3}$) of $\bmu$ orientation is considered.}
\label{ome_q}
\end{figure*}

The upper panel in Figure \ref{ome_q} shows the variation of the angular frequency
of the dominant modes $m=0,\pm 1, \pm 2 $ with $q$. The kink at 
$q=q_{\rm sp}=I_{1}/I_{2}$ in $\omega_{m}$
arises from the change of grain rotation state. For $q< q_{\rm sp}$, the frequency 
$\omega_{m=0}=\langle \dot\phi\rangle$ is almost constant, whereas it increases 
with $q$ to $\omega_{m=0}=J/I_{3}$ at $q=I_{1}/I_{3}$. The frequency of the 
mode $m=1$ also increases with $q$ for $q>q_{\rm sp}$. The increase of frequency for the two 
dominant modes $m=0$ and $1$ for $q>q_{\rm sp}$
 can result in an increase of peak frequency of 
spinning dust spectrum. This issue will be addressed in the next section.

The middle panel in Figure \ref{ome_q} shows the emission power
at the dominant frequency modes $|m|\le 2$. For $q\sim 1$, most of emission 
comes from the mode $m=-1$. As $q$ increases, the mode $m=0$ becomes dominant.

The total emission power from all modes for irregular grains of different 
$I_{2}:I_{3}$ and $I_{1}:I_{2}=1:0.6$ is shown in the right panel.

We compute emission power spectra to infer $\omega_{k}$ and $P_{\ed, k}$, 
as functions of $J$ and $q$, for various ratio of moments of inertia $I_{1}:I_{2}:I_{3}$. 
The obtained data will be used later to compute spinning dust emissivity.

\subsection{Spinning Dust Emissivity}\label{sec:jnu_irreg}
To calculate the spinning dust emissivity, the first step is to find the distribution
 function for grain angular momentum.

\subsubsection{Angular Momentum Distribution Function}

Following HDL10, to find the distribution function for the grain angular momentum
$\bJ$, we solve Langevin equations describing the evolution of $\bJ$ in time 
in an inertial coordinate system. They read 
\bea
dJ_{i}=A_{i}dt+\sqrt{B_{ii}}dq_{i},\mbox{~for~} i=\mbox{~x,~y,~z},\label{le3d}
\ena
where $dq_{i}$ are random Gaussian variables with $\langle dq_{i}^{2}\rangle=dt$, 
$A_{i}=\langle {\Delta J_{i}}/{\Delta t}\rangle$ 
and $B_{ii}=\langle \left({\Delta J_{i}}\right)^{2}/{\Delta t}\rangle$ are 
damping and
diffusion coefficients defined in the inertial coordinate system.

For an irregular grain, to simplify calculations, we adopt the $A_{i}$ and 
$B_{ii}$ for a disk-like grain obtained in HDL10. Following DL98b and HDL10, the disk-like
grain has radius $R$ and thickness $L=3.35\Angstrom$, and the ratio of moments of inertia along 
and perpendicular to the grain symmetry axis $h=I_{\|}/I_{\perp}$. 
Thereby, the effect of nonaxisymmetry on $A_{i}$ and $B_{ii}$ is ignored, 
and we examine only the effect the wobbling resulting from the irregular shape 
has on the spinning dust emissivity. 

In dimensionless units, $\bJ'\equiv \bJ/I_{\|}\omega_{\rm T,\|}$ with 
$\omega_{\rm T,\|}\equiv \left(2k_{\B}T_{\gas}/I_{\|}\right)^{1/2}$ being the thermal
 angular velocity of the grain along the grain symmetry axis,
and $t'\equiv t/\tau_{\rm H,\|}$, Equation (\ref{le3d}) becomes 
\bea
dJ'_{i}=A'_{i}dt'+\sqrt{B'_{ii}}dq'_{i},\label{le3d2}
\ena
where $\langle dq_{i}^{'2}\rangle=dt'$ and
\bea
A'_{i}&=&-\frac{J'_{i}}{\tau'_{\gas,{\rm eff}}}-\frac{2}{3}\frac{J_{i}^{'3}}
{\tau'_{\ed,{\rm eff}}},~~~~\\
B'_{ii}&=&\frac{B_{ii}}{2I_{\|}\kB T_{\gas}}\tau_{\H,\|},
\ena
where 
\bea
\tau'_{\gas,{\rm eff}}&=& \frac{\tau_{\gas,{\rm eff}}}{\tau_{\H,\|}}=
\frac{F_{\rm tot,\|}^{-1}}{\cos^{2}\theta+
\gamma_{\H}\sin^{2}\theta},\\
\gamma_{\rm H}&=&\frac{F_{\rm tot,\perp}\tau_{\rm H,\|}}{F_{\rm tot,\|}
\tau_{\rm H,\perp}},\\
\tau'_{\ed,{\rm eff}}&=&\frac{\tau_{\ed,{\rm eff}}}{\tau_{\rm H,\|}},~~~
\ena
where $\tau_{\H,\|}$ and $\tau_{\H,\perp}$ are rotational damping times due to gas of 
purely hydrogen atom for rotation along parallel and perpendicular direction
 to the grain symmetry axis $\ba_{1}$, 
$\tau_{\ed,{\rm eff}}$
 is the effective damping time due to electric dipole emission (see Appendix B),
 $\theta$ is the angle between $\ba_{1}$ and $\bJ$, and $F_{\rm tot,\|}$ and 
 $F_{\rm tot,\perp}$ are total damping coefficients parallel and perpendicular 
to $\ba_{1}$ (see HDL10). 

At each time step, the angular momentum $J_{i}$ obtained from LEs is recorded and
 later used to find the distribution function $f_{J}$ with normalization
$\int_{0}^{\infty} f_{J}dJ=1$.

\subsubsection{Emissivity from Irregular Grains}

As discussed in \S 4.1, an irregular grain rotating with a given 
angular momentum $J$ radiates at frequency 
modes $\omega_{k}\equiv \omega_{m}$ with $m=0,\pm 1, \pm 2...$ 
and $\omega_{k}\equiv \omega_{n}$ with $n=1,2,3...$ (see Eqs \ref{ome_m} and 
\ref{ome_n}). For simplicity, we define
 the former
as $\omega_{\mtxt_{i}}$ and the latter as $\omega_{\ntxt_{i}}$ where 
 $i$ indicates the value for $m$ and $n$. These frequency modes depend on 
 the parameter $q(s)$, which is determined by the internal thermal fluctuations within the grain. 

To find the emissivity at an observation frequency 
$\nu=\omega/2\pi$, we need to know how much emission each mode $\omega_{k}$ contributes to the observation frequency $\nu$. 

Consider an irregular grain rotating with the angular momentum $J$, the probability 
of finding the emission at the angular frequency $\omega$ depends on the 
probability of finding the value $\omega$ such that
\bea
pdf(\omega|J)d\omega=f_{\VRE}(s,J)ds= A{\rm exp}\left(-\frac{q(s)J^{2}}
{2I_{1}k_{\B}T_{\vib}}\right)ds,~~~~~\label{eq:pdf}
\ena
where we assumed the VRE regime with $f_{\VRE}$ given by Equation (\ref{fs}).

For the mode $\omega\equiv\omega_{k}(s)$, from Equation (\ref{eq:pdf}) 
we can derive 
\bea
pdf_{k}(\omega|J)=\left(\frac{\partial \omega_{k}}{\partial s}\right)^{-1}
f_{\VRE}(s,J).
\ena

The emissivity from the mode $k$ is calculated as
\bea
j_{\nu,k}^{a}&=&\frac{1}{4\pi}\int_{J_{l}}^{J_{u}}P_{\ed,k}(J,q_{\le})f_{J}(J)
  pdf_{k}(\omega|J)2\pi~ dJ\nonumber\\
&&+\frac{1}{4\pi}\int_{J_{l}}^{J_{u}}P_{\ed,k}(J,q_{>})f_{J}(J) 
pdf_{k}(\omega|J)2\pi~ dJ,~~~~~\label{jnu_nua}
\ena
where $q_{\le}$ and $q_{>}$ denote $q\le q_{\sp}$ and $q>q_{\sp}$, respectively;
 $J_{l}$ and $J_{u}$ are lower and upper limits for $J$ 
corresponding to a given angular frequency $\omega_{k}(J,q)=\omega$, and 
$2\pi$ appears due to the change of variable from $\nu$ to $\omega$.

Emissivity by a grain of size $a$ at the observation frequency $\nu$ arising
 from all emission modes is then
\bea
j_{\nu}^{a}&\equiv&\sum_{k}j_{\nu,k}^{a}.
\ena

Consider for example the emission mode $k\equiv \mtxt_{0}$. For the case 
$I_{2}$ slightly larger than $I_{3}$, this mode has the angular frequency 
$\omega_{\mtxt_{0}}=\langle \dot\phi\rangle=(J/I_{1})
q_{0}$ with $q_{0}$ obtained from  calculation of $\omega_{\mtxt_{0}}$, is 
independent on $q$ for  $q<q_{\sp}$.\footnote{$q_{0}$ approaches $I_{1}/I_{2}$ 
as $I_{3} \rightarrow I_{2}$, i.e., when 
irregular shape becomes spheroid.}  As a result 
\bea
pdf_{\mtxt_{0}}(\omega|J)=\delta\left(\omega-(J/I_{1})q_{0}\right).
\ena
Thus, the first term of Equation 
(\ref{jnu_nua}), denoted by $j_{\nu,\mtxt_{0},\le}^{a}$, is rewritten as
\bea
j_{\nu,\mtxt_{0},\le}^{a}&=&\frac{1}{2}\int_{J_{l}}^{J_{u}}P_{\ed,\mtxt_{0}}
(J,q_{\le})f_{J}(J)
 \delta \left(\omega-(J/I_{1}) q_{0}\right)dJ,\nonumber\\
&=&\frac{1}{2}\frac{I_{1}f_{J}(J_{0})}{q_{0}} P_{\ed,\mtxt_{0}}(J_{0},q(s)),
\label{jnu_les}
\ena
where $J_{0}=I_{1}\omega/q_{0}$, and the value of $q(s)$ remains to be determined.

For $q>q_{\sp}$, $\langle \dot\phi\rangle$ is a function of $q$. Hence, 
the emissivity (\ref{jnu_nua}) for the mode $k\equiv \mtxt_{0}$ becomes
 \bea
j_{\nu,\mtxt_{0}}^{a}&=&\frac{1}{2}\frac{I_{1}f_{J}(J_{0})}{q_{0}}
\int_{0}^{s_{\sp}} ds P_{\ed,\mtxt_{0}}(J_{0},q(s))f_{\VRE}(J_{0},s)\nonumber\\
&&+\frac{1}{2}\int_{J_{l}}^{J_{u}}P_{\ed,\mtxt_{0}}(J,q_{>})f_{J}(J) 
pdf_{\mtxt_{0}}(\omega|J)dJ,~~~~~\label{jnua1}
\ena
where $s_{\sp}$ is the value of $s$ corresponding to $q=q_{\sp}$, and the term $P_{\ed,\mtxt_{0}}(J_{0},q(s))$ in Equation (\ref{jnu_les})
has been replaced by its average value over the internal thermal distribution $f_{\VRE}$.

The emissivity per H is obtained by integrating $j_{\nu}^{a}$ over the grain 
size distribution:
\bea
{j_\nu\over n_\H} = 
{1\over n_\H}
\int_{a_{\rm min}}^{a_{\rm max}} da {dn\over da}j_{\nu}^{a}~,\label{jnued}
\ena
where $j_{\nu}^{a}$ is given by Equation (\ref{jnu_nua}).

\subsubsection{Emissivity from Disk-like Grains}\label{jnu-disk}

HDL10 showed that a disk-like grain with an angular momentum $\bJ$ radiates at 
frequency modes 
\bea
\omega_{\mtxt_{i}}&\equiv&\dot\phi+i \dot\psi=\frac{J}{I_{\|}}
\left[h+i(1-h)\cos\theta\right],\label{eq:omem}\\
\omega_{n_{1}}&\equiv&\dot\psi=\frac{J}{I_{\|}}(1-h)\cos\theta,\label{eq:omen}
\ena
where $i=0$ and $\pm 1$.

 The emission power of these modes are given by following analytical forms 
(HDL10; Silsbee et al. 2011):
\bea
P_{\omega_{\mtxt_{0}}}&=&\frac{2\mu_{\|}^{2}}{3c^{3}}\omega_{\mtxt_{0}}^{4}
\sin^{2}\theta,
\label{pedm0}\\
P_{\omega_{\mtxt_{\pm1}}}&=&\frac{\mu_{\perp}^{2}}{6c^{3}}\omega_{\mtxt_{\pm1}}^{4}
(1\pm\cos\theta)^{2},\label{pedm1}\\
P_{\omega_{\ntxt_{1}}}&=&\frac{2\mu_{\perp}^{2}}{3c^{3}}\omega_{\ntxt_{1}}^{4}
\sin^{2}\theta.
\label{pedn1}
\ena

For the disk-like grain, from Equation (\ref{sq}), we obtain the number of state
in phase space $s$ for $q$ spanning from $1-q$:
\bea
s=1-\left(\frac{h-q}{h-1}\right)^{1/2}=1-\cos\theta,\label{s_theta}
\ena
where $q=1+(h-1)\sin^{2}\theta$ has been used.
Thus, for an arbitrary mode with frequency $\omega_{k}$, we obtain
\bea
pdf_{k}(\omega|J)d\omega=f_{\VRE}(s,J)ds=f_{\VRE}(\theta,J)\sin\theta d\theta.~~~
\ena

Taking use of $\omega=\omega_{k}(J,\theta)$, we derive
\bea
pdf_{k}(\omega|J)=f_{\VRE}(\theta,J)\left(\frac{\partial \omega_{k}}
{\partial \theta}\right)^{-1}\sin\theta.\label{pdf_theta}
\ena

Therefore, by substituting Equations (\ref{pedm0})-(\ref{pedn1}) in Equation 
(\ref{jnu_nua}), the emissivity at the observation frequency $\nu=\omega/(2\pi)$ from a 
disk-like grain of size $a$ is now given by
\bea
j_{\nu}^{a}&\equiv&\frac{1}{2} \frac{f_{J}(I_{\|}\omega/h)}{h}
\frac{2\mu_{\|}^{2}}{3c^{3}}\omega^{4}\langle\sin^{2}\theta\rangle\nonumber\\
&&+\frac{1}{2} \frac{\mu_{\perp}^{2}}{6c^{3}}\omega^{4}\int_{J_{l}}^{J_{u}}
pdf_{\mtxt_{1}}(\omega|J)
 f_{J}(J)dJ\nonumber\\
&&+\frac{1}{2} \frac{\mu_{\perp}^{2}}{6c^{3}}\omega^{4}\int_{J_{l}}^{J_{u}}
pdf_{\mtxt_{-1}}(\omega|J)
 f_{J}(J)dJ\nonumber\\
&&+\frac{1}{2} \frac{\mu_{\perp}^{2}}{3c^{3}}\omega^{4}\int_{J_{l}}^{J_{u}}
pdf_{\ntxt_{1}}(\omega|J)
 f_{J}(J)dJ,\label{jnua_disk}
\ena
where $pdf_{\mtxt_{\pm1}}$ and $pdf_{\ntxt_{1}}$ are easily derived by using 
Equation (\ref{pdf_theta}) for $\omega_{\mtxt_{\pm1}}$ and $\omega_{\ntxt_{1}}$,
 and $J_{l}=I_{\|}\omega/(2h-1)$ and $J_{u}=I_{\|}\omega$ for $m_{\pm 1}$ mode, 
and $J_{l}=I_{\|}\omega/(h-1)$ and $J_{u}=\infty$ for $n_{1}$ mode (see Eqs 
\ref{eq:omem} and \ref{eq:omen}) . 

As in HDL10, we also consider the regime without internal relaxation. Then,
the distribution function $f_{\VRE}(\theta,J)$ is replaced by a Maxwellian
distribution (hereafter Mw regime):
\bea
f_{\rm Mw}(\theta)=\frac{h}{4\pi}\frac{1}{\left(\cos^{2}\theta+h\sin^{2}
\theta\right)^{3/2}}.\label{f_Mw}
\ena
For symmetric rotors ($h=1$), the distribution function for $\theta$ is isotropic:
\bea
f_{\rm iso}(\theta)=\frac{1}{4\pi},\label{f_iso}
\ena
satisfying the normalization condition 
$\int_{0}^{\pi}f_{\rm iso}(\theta)2\pi\sin\theta d\theta=1$

\subsection{Results}
In this section, we present results for spinning dust emissivity  from 
grains of irregular shape obtained using the method in Section \ref{sec:jnu_irreg}.

\subsubsection{A Simple Model of Irregular Grain}

We assume that smallest grains of size $a \le a_{2}=6\Angstrom$, 
have planar shape, and larger grains are spherical. To compare with emission from
the disk-like grains, we consider the simplest case of the irregular shape in which
the circular cross-section of the disk-like grain is changed to elliptical 
cross-section. We are interested in the emission of two PAHs with the same
mass $M$ and thickness $L$, therefore, the length of  the semi-axes of elliptical disk is constrained by
\bea
M=\pi R^{2}L=\pi b_{2}b_{3}L,\label{Md}
\ena
where $R=(4a^{3}/3L)^{1/2}$ is the radius of the disk-like grain,
 $b_{2}$ and $b_{3}$ 
are the length of semi-axes $\ba_{2}$ and $\ba_{3}$, and $b_{1}=L$
is kept constant. Assuming that the circular disk is 
compressed by a factor $\alpha\le 1$ along $\ba_{2}$, then Equation (\ref{Md})
yields
\bea
b_{2}=\alpha R,~~b_{3}=\alpha^{-1}R.
\ena
Denote the parameter $\eta\equiv b_{3}/b_{2}=\alpha^{-2}$, then the degree of grain 
shape irregularity is completely characterized by $\eta$.

The moments of inertia along the principal axes for the irregular grain are
\bea
I_{1}=\frac{M}{4}\left(b_{2}^{2}+b_{3}^{2}\right)=\frac{I_{\|}}{2}\left(\eta+\eta^{-1}\right),
\ena
where $I_{\|}$ is the moment of inertia for the disk-like grain. The ratios of
moment of inertia read
\bea
\frac{I_{1}}{I_{2}}=\frac{3\left(\eta+\eta^{-1}\right)}{(L/R)^{2}+3\eta},\label{I1_I2}\\
\frac{I_{2}}{I_{3}}=\frac{(L/R)^{2}+3\eta}{(L/R)^{2}+3\eta^{-1}},\label{I2_I3}
\ena

For each grain size $a$, the parameter $\eta$ is increased from $\eta=1$ 
to $\eta=\eta_{\max}$. However, $\eta_{\max}$ is constrained by the fact that
the shortest axis $\ba_{2}$ should not be shorter than the grain thickness $L$.
We below assume conservatively $\eta_{\max}\sim 3/2$.
 
\subsubsection{Rotational Emission Spectrum}

Although irregular 
grains radiate in large number of frequency modes, we take into account only the dominant 
modes with order $|m| \le 2$, and disregard the modes that contribute less than $\sim 0.5\%$
to the total emission.
We also assume that grains smaller than $a_{2}$ have 
a fixed vibrational temperature $T_{\vib}$, and that for the instantaneous
 value of $J$ the rotational energy has a probability distribution $f_{\VRE}$
(i.e. VRE regime).

 We adopt the grain size distribution $dn/da$ from Draine \&
Li (2007) with
the total to selective extinction $R_{\rm V}=3.1$ and the total carbon
abundance per hydrogen nucleus $b_{\rm C}=5.5\times 10^{-5}$ in carbonaceous 
grains with $a_{\min}=3.55\Angstrom$ and $a_{\max}=100\Angstrom$.

The  spinning dust emissivity is calculated for a so-called model A (similar
 to DL98b; HDL10), in which $25\%$ of 
grains have  the electric dipole moment
parameter $\beta=2\beta_{0}$, $50\%$ have $\beta=\beta_{0}$ and $25\%$ have 
$\beta=0.5\beta_{0}$ with $\beta_{0}=0.4$ D. In the rest of the paper, the 
notation {\it model} A is omitted, unless stated otherwise.

\begin{figure*}
\includegraphics[width=0.33\textwidth]{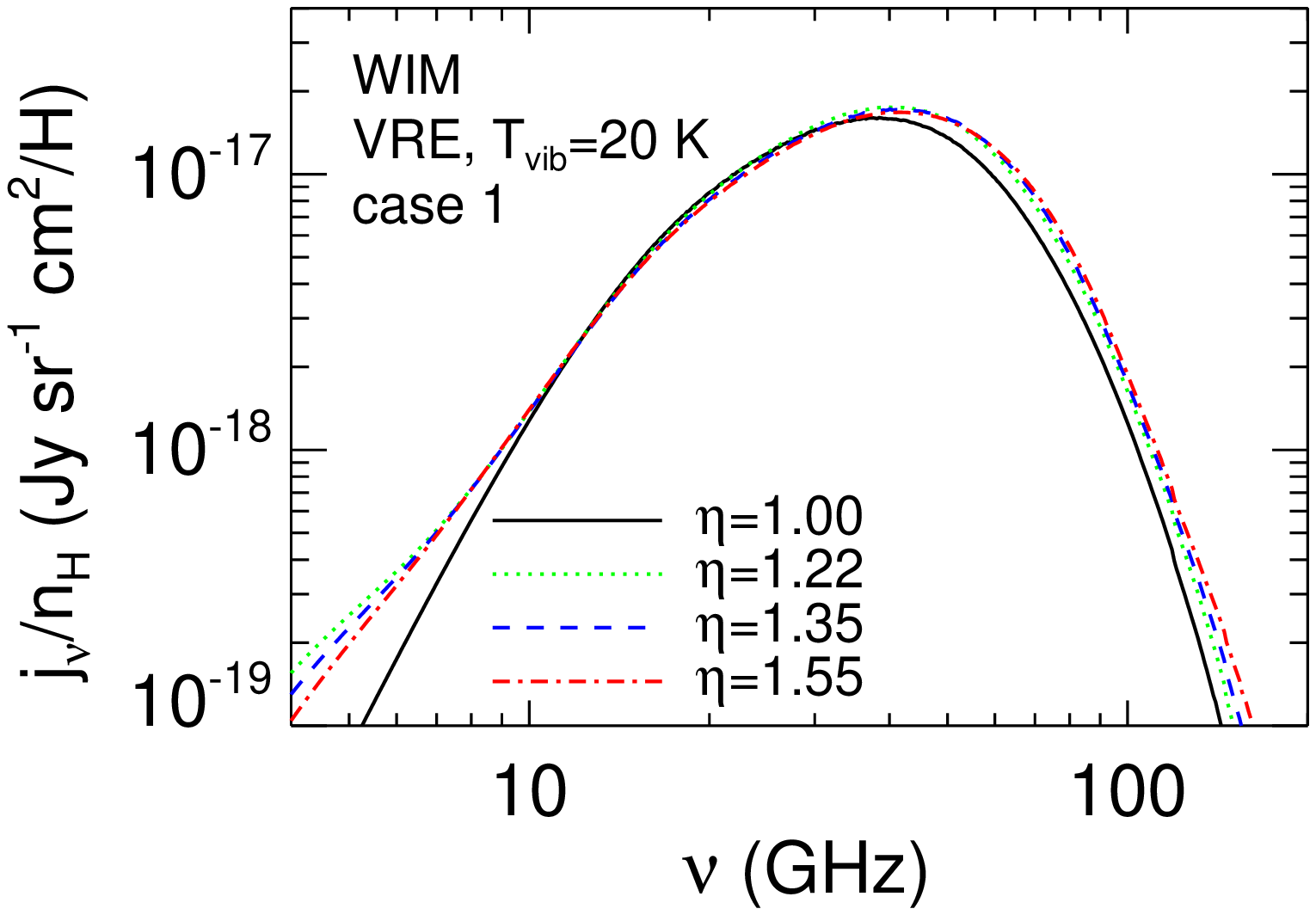}
\includegraphics[width=0.33\textwidth]{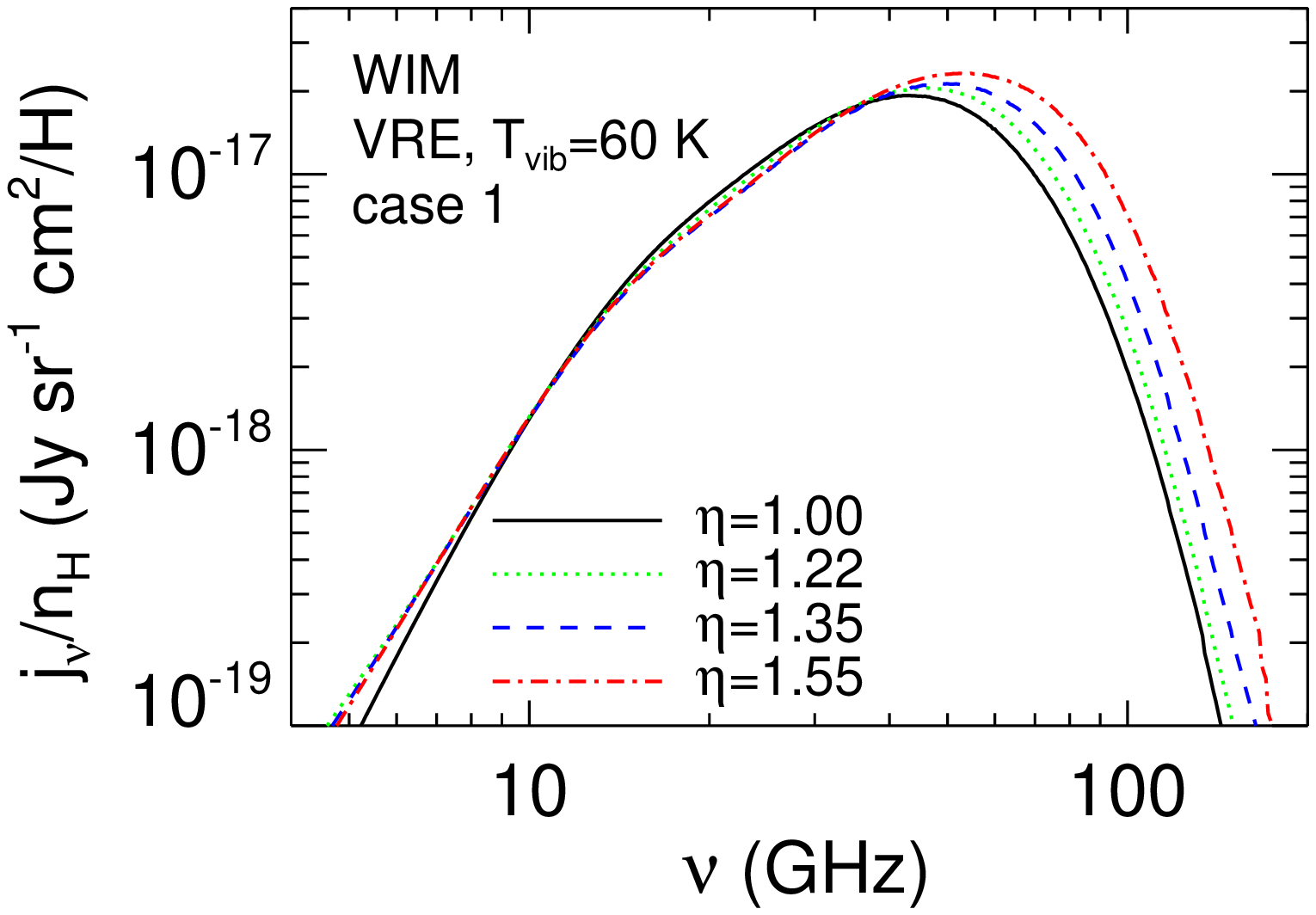}
\includegraphics[width=0.33\textwidth]{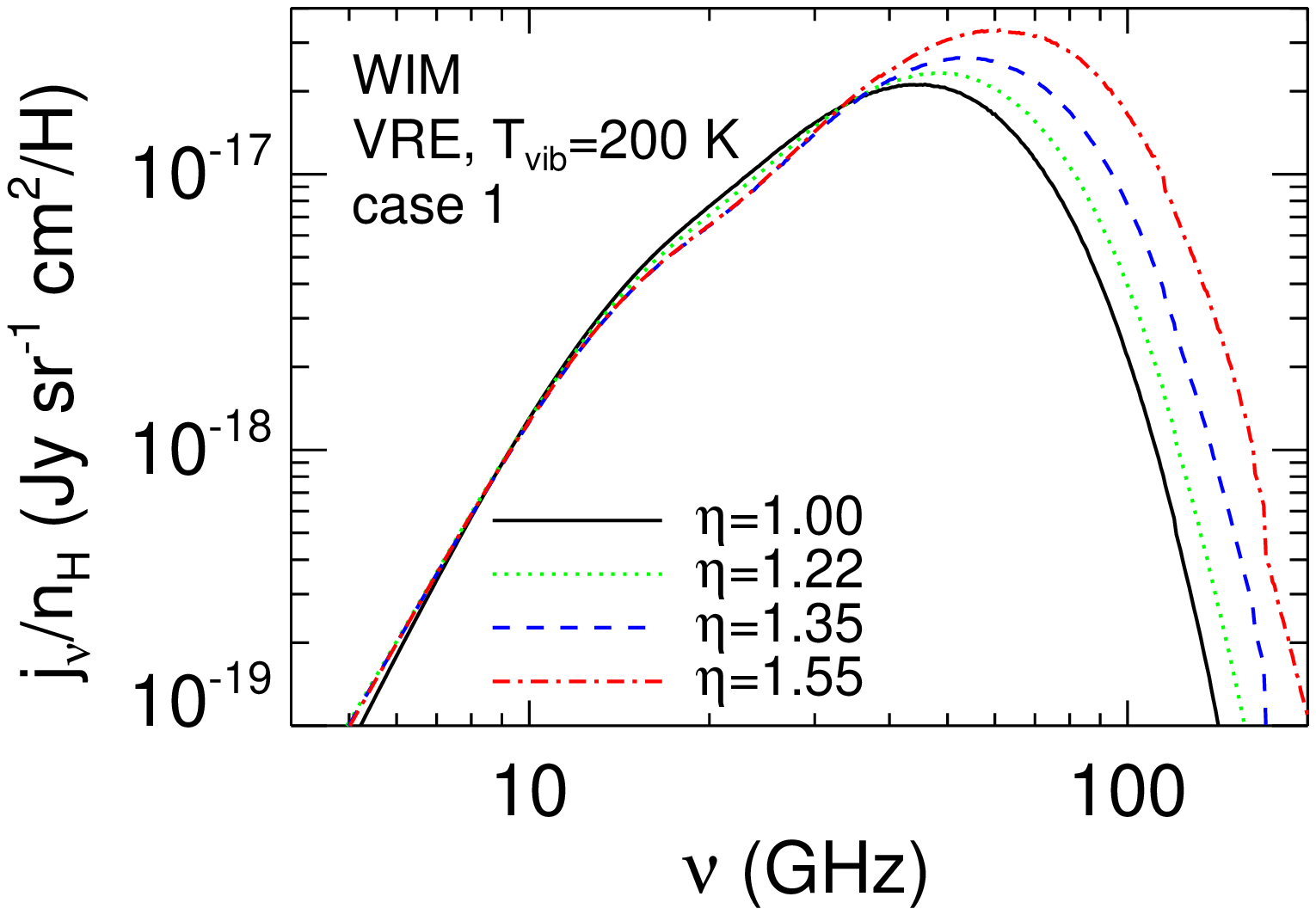}
\caption{Emissivity per H from irregular grains of different degree of 
irregularity $\eta=b_{3}/b_{2}$  with $T_{\vib}=20$,
 $60$ and $200$ K for case 1 ($\mu_{1}=\mu/\sqrt{3}$) of dipole moment orientation in the WIM. 
 The emission spectrum shifts to higher frequency as $\eta$ decreases. 
 The grain mass is held fixed as $\eta$ changes.}
\label{jnu_irr}
\end{figure*}

 Figure \ref{jnu_irr} shows  the spinning dust spectra  for different degree of
 irregularity $\eta$ and various dust temperature $T_{\vib}$ in the WIM. 
 The emission spectrum for a given $T_{\vib}$ shifts to higher frequency as 
  $\eta$ decreases (i.e. the degree of  grain irregularity  increases), 
  but their spectral profiles remain similar. 
  
\begin{figure*}
\includegraphics[width=0.5\textwidth]{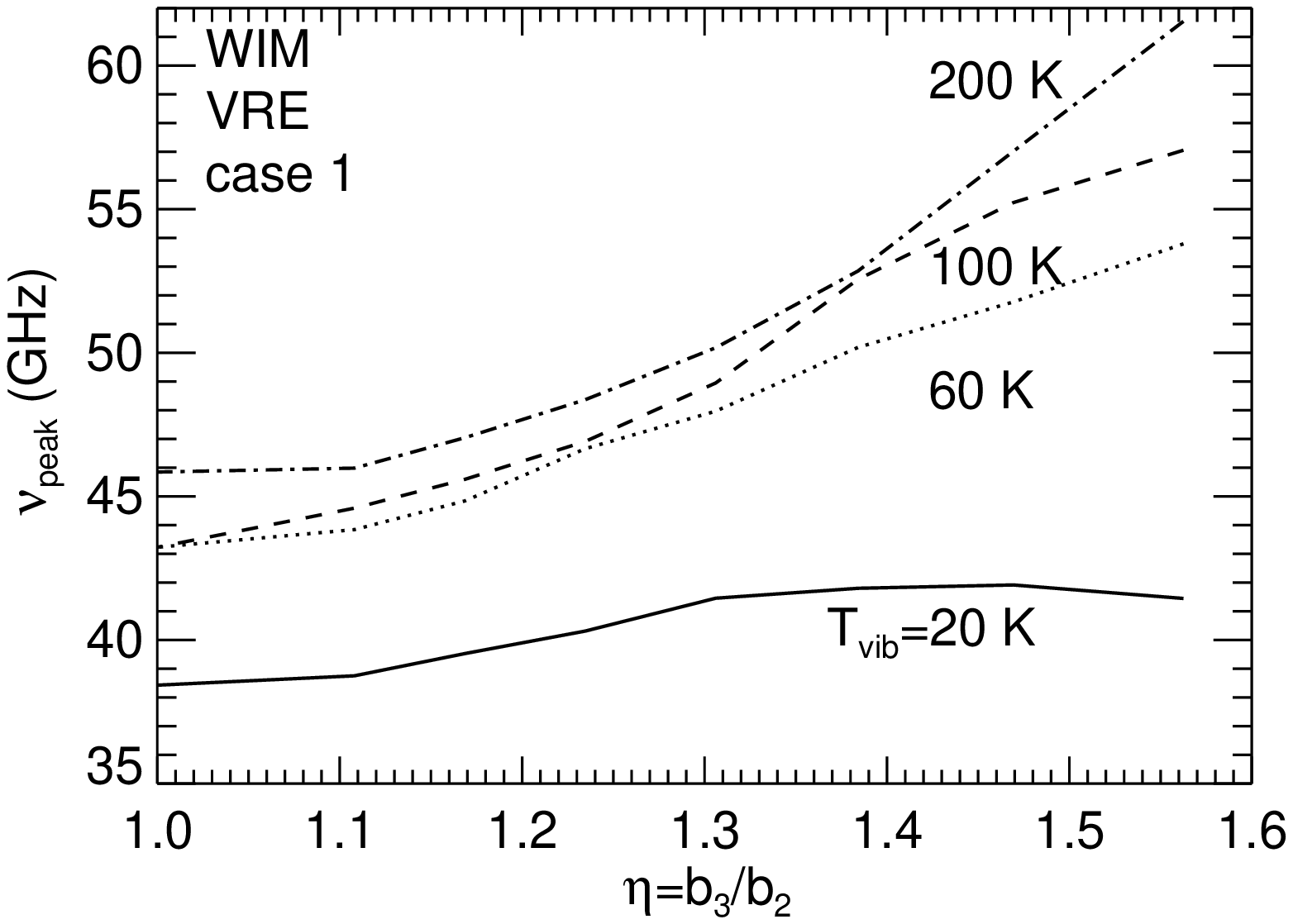}
\includegraphics[width=0.5\textwidth]{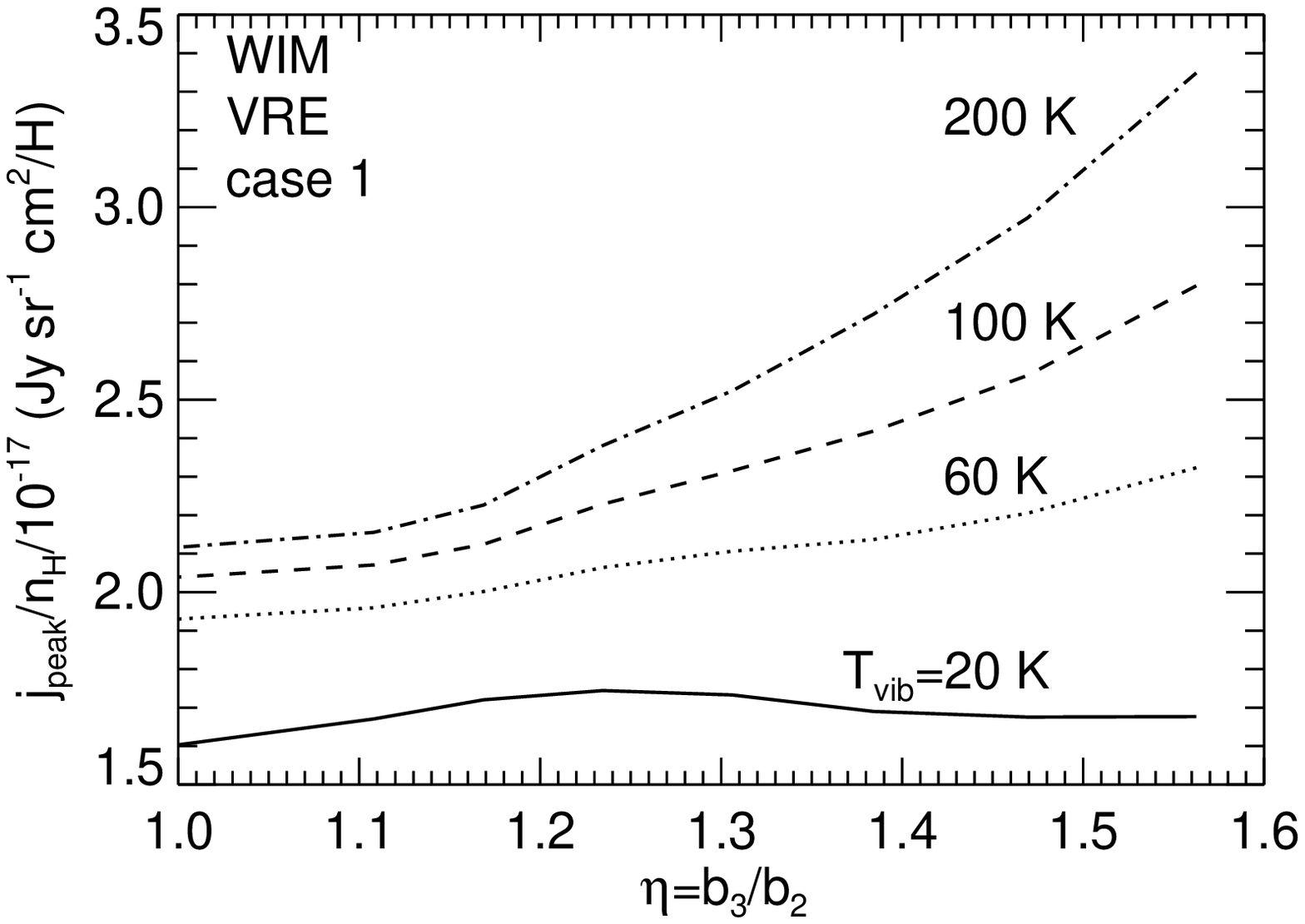}
\caption{The variation of peak frequency ({\it left}) and peak emissivity ({\it right})  
as a function of the length ratio of semi-axes of elliptical disk, $\eta=b_{3}/b_{2}$ 
for various $T_{\vib}$ and case 1 ($\mu_{1}=\mu/\sqrt{3}$) of $\bmu$ orientation 
in the WIM conditions. Both $\nu_{\peak}$ and $j_{\peak}$ increase fast with 
$\eta$ for $T_{\vib}\ge 60~\K$, but they changes slightly at  $T_{\vib}=20~\K$}.\label{jpeak_irr}
\end{figure*}

\begin{figure*}
\includegraphics[width=0.5\textwidth]{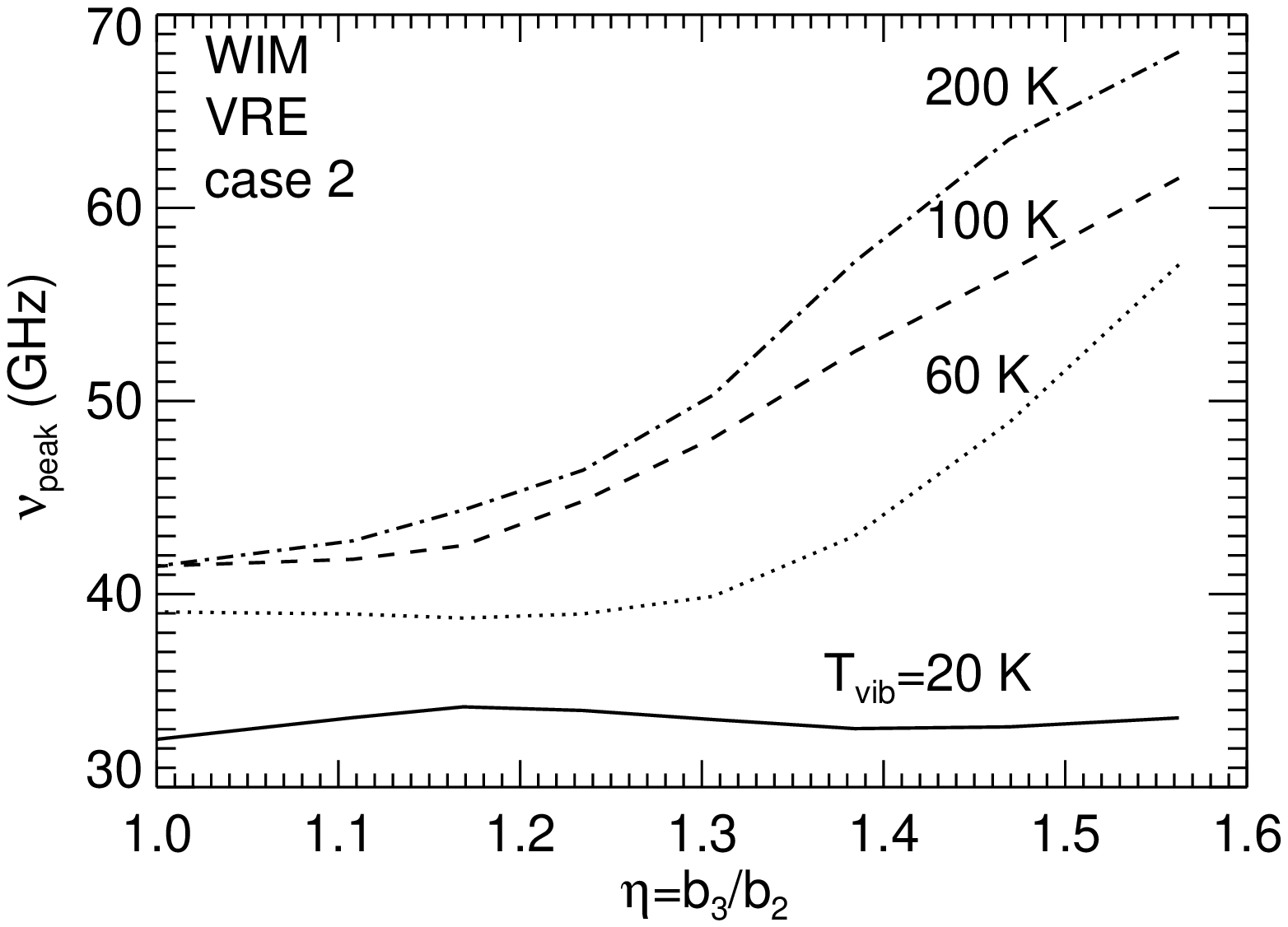}
\includegraphics[width=0.5\textwidth]{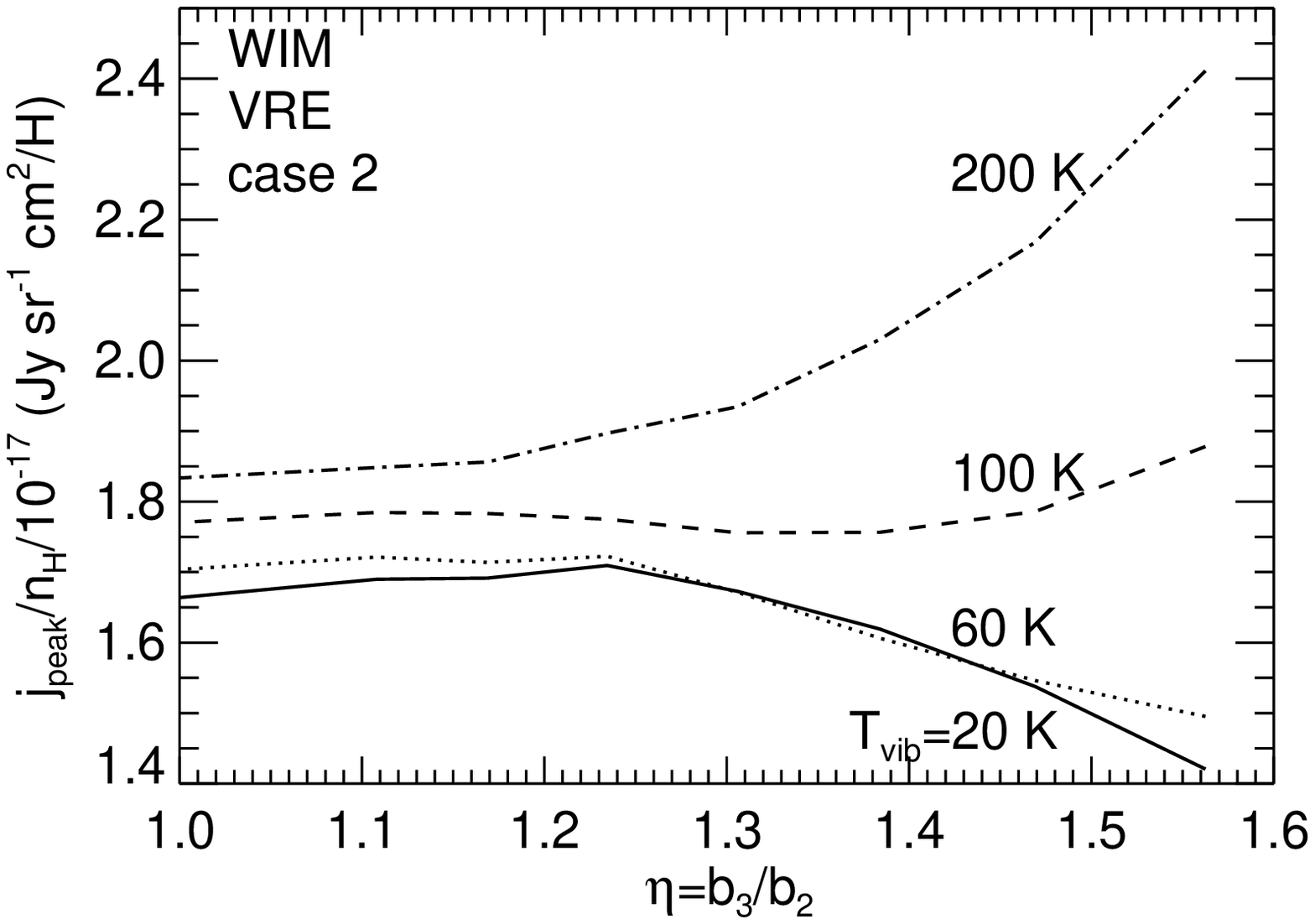}
\caption{Similar to Fig. \ref{jpeak_irr}, but for case 2 ($\mu_{1}=0$) of $\bmu$ orientation. $\nu_{\peak}$ 
increases with $\eta$ for $T_{\vib}>20~\K$. $j_{\peak}$ increase with $\eta$ for 
$T_{\vib}\ge 100~\K$. For $T_{\vib} < 60~ K$, $j_{\peak}$
increases slightly to $\eta \sim 1.23$ and then drops.}
\label{jpeak_irr2}
\end{figure*}

To see more clearly the effect of grain shape irregularity,
in Figure \ref{jpeak_irr}, we show the variation of peak frequency
$\nu_{\peak}$ and peak emissivity $j_{\peak}$ as a function of $\eta$ for different $T_{\vib}$ 
and for case 1 ($\mu_{1}=\mu/\sqrt{3}$) of $\bmu$ orientation for the WIM.

For low $T_{\vib}$ (i.e. $T_{\vib}=20~\K$), $\nu_{\peak}$ 
increases slowly with $\eta$, while $j_{\peak}$ varies slightly. As
$T_{\vib}$ increases, both $\nu_{\peak}$ and $j_{\peak}$ increase rapidly with $\eta$,
but the former increases faster.

 Figure \ref{jpeak_irr2}  shows the results for case 2 ($\mu_{1}=0$) of $\bmu$ orientation.
Similar to case 1, $\nu_{\peak}$ increases with $\eta$ for $T_{\vib}\ge 60~\K$, but
changes rather little for $T_{\vib}=20~\K$. 
The variation of $j_{\peak}$ with $\eta$  is more complicated. For $T_{\vib}\le 60~\K$,
it increases first to $\eta\sim 1.23$, and then drops as $\eta$ increases.

\section{Spinning Dust Spectrum: exploring parameter space}

Below we explore  parameter space by varying dust temperature, the lower grain size cutoff, 
the grain dipole moment, and the gas density. Our results from previous section indicate that the spinning dust spectrum 
 from grains of irregular shape is shifted in general to higher frequency, 
 relative to that from  disk-like grains, but their spectral profile remains similar.
 Therefore,  it is sufficient to explore parameter space
 of spinning dust for disk-like grains, extrapolating to irregular grains 
using our results from the previous section.  
Spinning dust emissivity is then calculated using Equation (\ref{jnued}) 
in which $j_{\nu}^{a}$ is given by Equation (\ref{jnua_disk}).

\subsection{Effect of internal thermal fluctuations}
  In this section, we account for the fluctuations of the dust temperature 
in smallest grains. 

Let $T_{\d}$ be the decoupling temperature for the 
V-R energy exchange. For $T_{\vib}>T_{\d}$,
the V-R energy exchange is efficient, and $T_{\rot} \approx T_{\vib}$. For the instantaneous
value of $J$ the rotational energy has a probability distribution (\ref{fs}) 
determined by V-R energy exchange (the VRE regime). For $T_{\vib}<T_{\d}$, no V-R energy
 exchange exists.
The rotational emissivity averaged over temperature fluctuations for a fixed 
grain size reads
\bea
\langle j_{\nu}^{a}\rangle_{T}&=&\int j_{\nu}^{a}(T_{\vib}> T_{\d})\frac{dP}{dT_{\vib}}dT_{\vib}\nonumber\\
&&+
\int  j_{\nu}^{a}(T_{\rot}\le T_{\d})P_{0}\delta(T_{\rot}-T_{\d})
dT_{\rot},\label{jnu_Td}
\ena
where 
\bea
P_{0}=\int_0^{T_{\d}}\frac{dP}{dT_{\vib}}dT_{\vib},
\ena
 is the probability of the grain having $T_{\vib}\le T_{\d}$, and $dP/dT$
 is the probability for the grain having temperature in $T, T+dT$.

Figure \ref{jnu_Td} shows spinning dust spectra averaged over temperature 
distribution, $\langle j_{\nu}(T_{\d})\rangle_{T}$, for various decoupling temperature 
$T_{\d}$ compared to spectra
 for grains having a single vibrational temperature $T_{\vib}=T_{\d}$,
denoted by $j_{\nu}(T_{\vib}=T_{\d})$. 
 For low $T_{\d}\le 70~\K$, the averaged emissivity $\langle j_{\nu}(T_{\d})\rangle_{T}$
 is larger than 
 $j_{\nu}(T_{\vib}=T_{\d})$, but the difference between $j_{\nu}(T_{\vib}=T_{\d})$
 and $\langle j_{\nu}(T_{\d})\rangle_{T}$ decreases as  $T_{\d}$ increases. 

For  $T_{\d}= 60$ K, Figure \ref{jnu_Td} shows that $\langle j_{\nu}(T_{\d})\rangle_{T}$
is approximate to $j_{\nu}(T_{\vib}=T_{\d})$ with the difference less 
than a few percent. In the rest of the paper, we adopt 
a conservative value $T_{\d}= 60$ K and calculate the spinning dust emissivity 
assuming that all grains have a single temperature $T_{\d}$.
 
\begin{figure}
\includegraphics[width=0.5\textwidth]{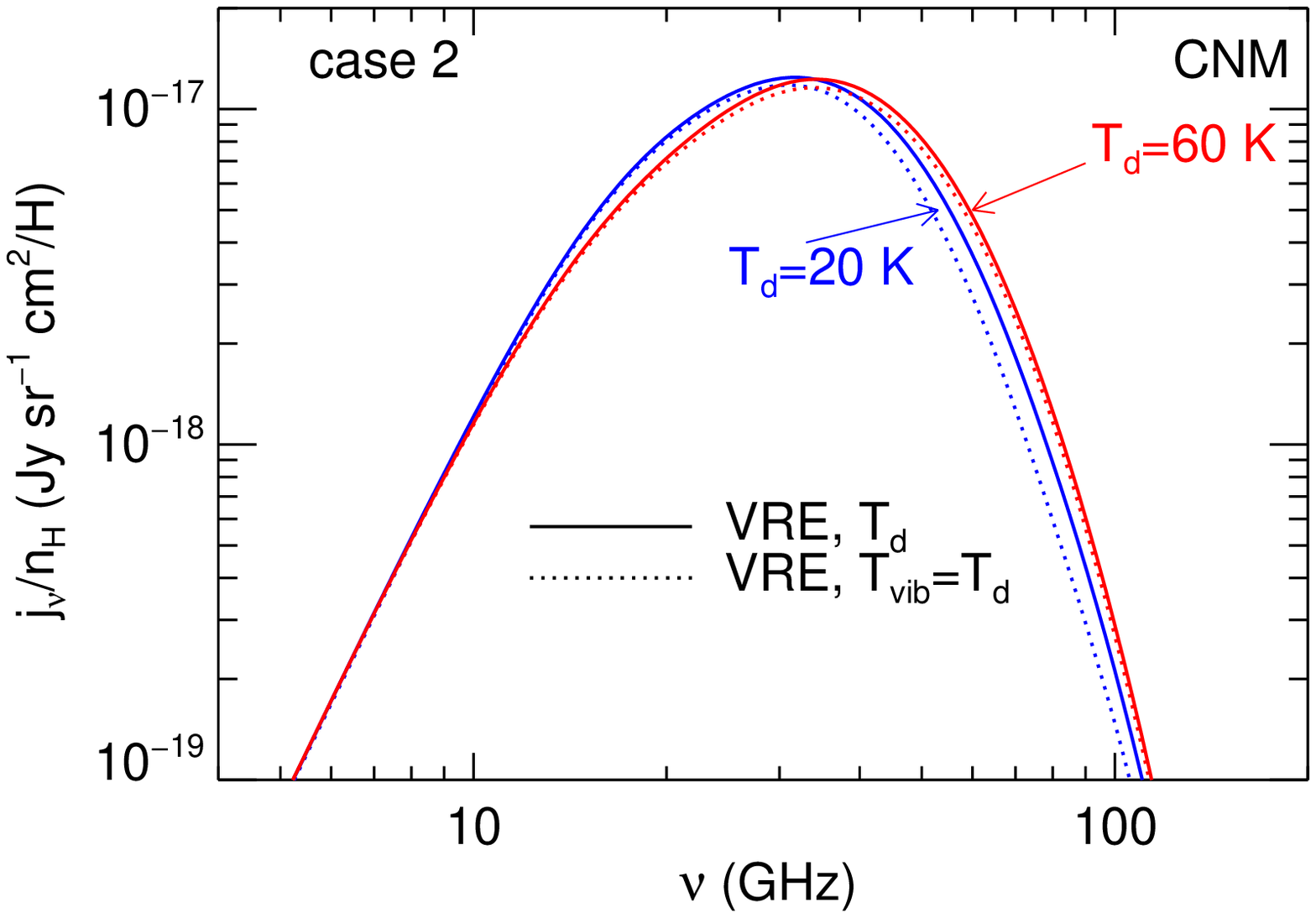}
\includegraphics[width=0.5\textwidth]{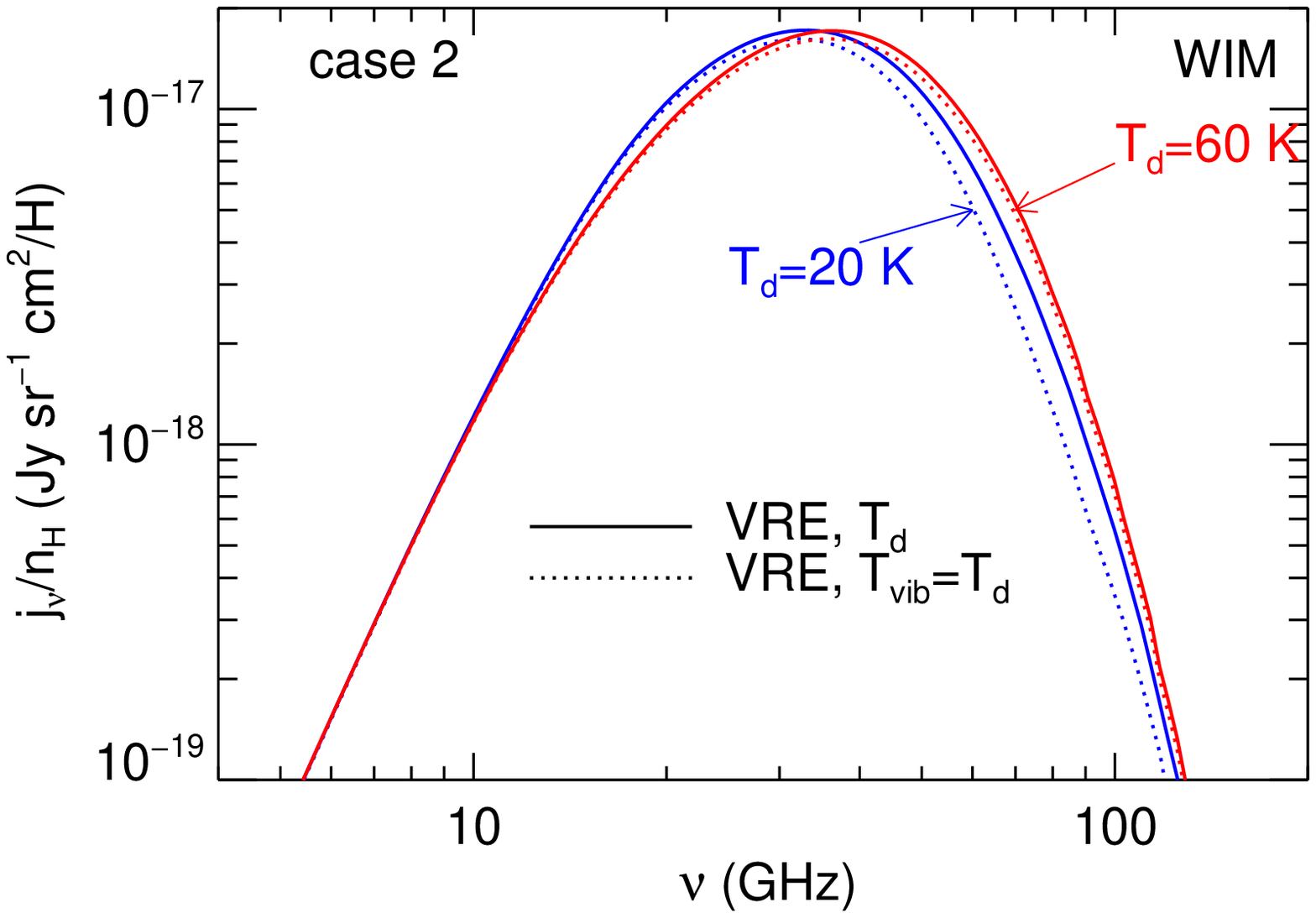}
\caption{Emissivity per H for case 2 ($\mu_{1}=0$) of the $\bmu$ orientation and for 
the CNM and WIM. Solid and dotted lines are emissivity spectra averaged over temperature
 distribution, $\langle j_{\nu}(T_{\d})\rangle_{T}$, and spectra for grains having
 $T_{\vib}=T_{\d}$, $j_{\nu}(T_{\vib}=T_{\d})$, respectively. Two values $T_{\d}=20$ and $60~\K$ are chosen
 for illustration.}
\label{jnu_Td}
\end{figure}

\subsection{Minimum size $a_{\min}$}\label{sec:amin}

The spinning dust emission spectrum is sensitive to the population of smallest 
dust grains, and its peak frequency is mostly determined by 
the smallest grains. When $a_{\min}$ is increased, the peak 
frequency $\nu_{\peak}$ decreases accordingly. 

\begin{figure}
\includegraphics[width=0.5\textwidth]{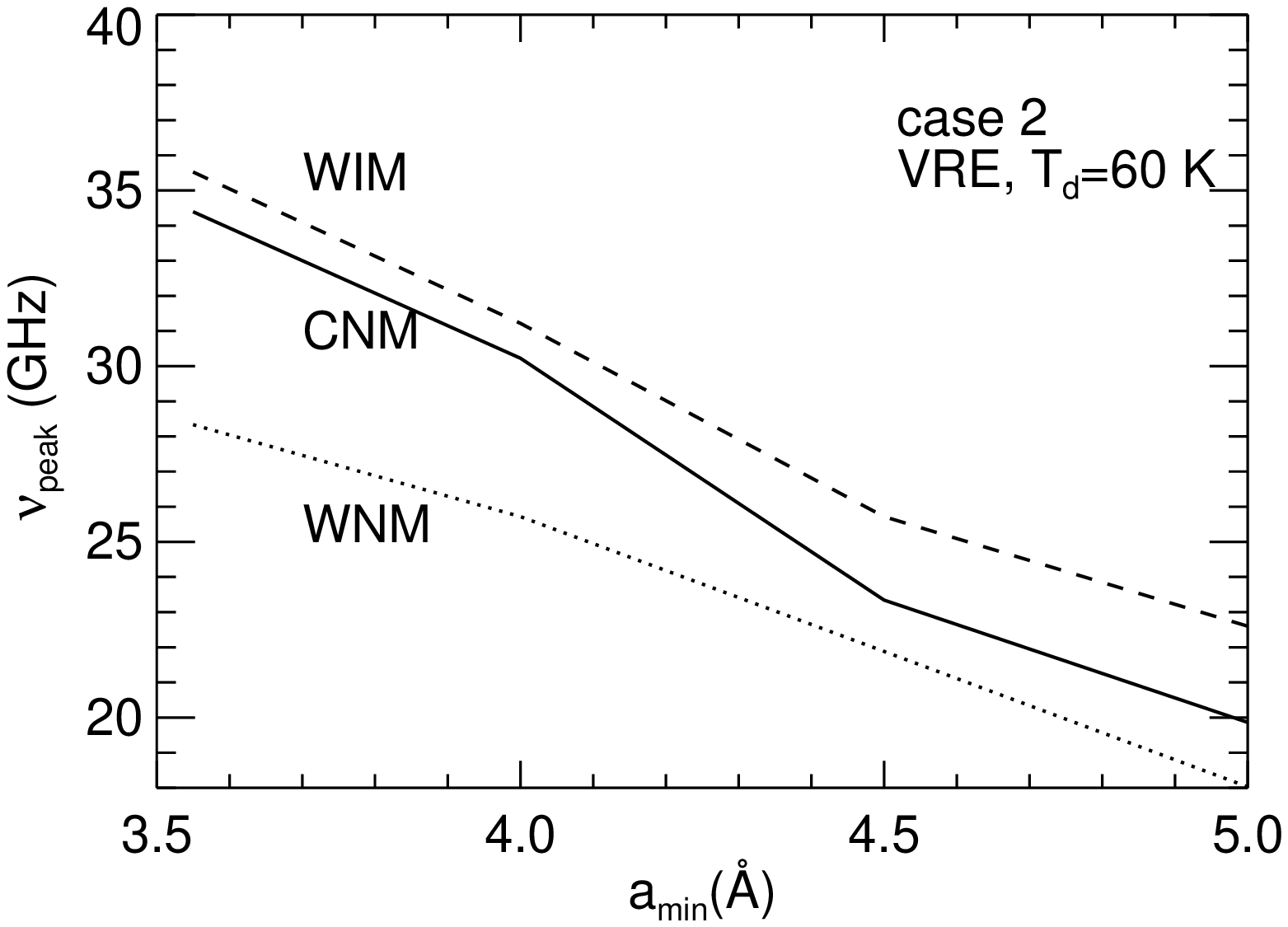}
\caption{Decrease of the peak frequency $\nu_{\peak}$ of spinning dust spectrum 
with the lower cutoff of the grain size distribution $a_{\min}$ for
various environment conditions.}
\label{nupk_amin}
\end{figure}

Figure \ref{nupk_amin} shows the variation of $\nu_{\peak}$ as 
a function of $a_{\min}$ for various environments, and 
case 2 ($\mu_{1}=0$) of $\bmu$ orientation and VRE regime ($T_{\d}=60~\K$). 
 As expected, $\nu_{\peak}$ decreases generically with  $a_{\min}$ increasing.

\subsection{Effect of electric dipole moment}
\subsubsection{Variation of orientation of electric dipole moment}

Silsbee et al. (2011) found that the total emission power $P_{\ed}$ and the peak
 frequency are rather insensitive to the orientation of the 
electric dipole moment. That stems from their calculations
 assuming no internal relaxation (i.e. $T_{\vib}\rightarrow\infty$), so that
 the angle $\theta$ is drawn from a isotropic distribution function $f_{\rm iso}$. Below
 we study the variation of emissivity $j_{\nu}$ with the orientation of dipole accounting
 for the internal relaxation.

To investigate the variation of $j_{\nu}$ with the orientation of $\bmu$, 
in addition to  case 1 ($\mu_{1}=\mu/\sqrt{3}$) and case 2 ($\mu_{1}=0$) of $\bmu$ orientation, 
we consider values of $\mu_{\perp}^{2}/\mu^{2}$ between 2/3 (case 1) and 1 (case 2).  
We also consider the regime of fast internal relaxation (VRE)
 and without internal relaxation (Mw and iso).  Model A for the
distribution of electric dipole moment is adopted.

\begin{figure}
\includegraphics[width=0.5\textwidth]{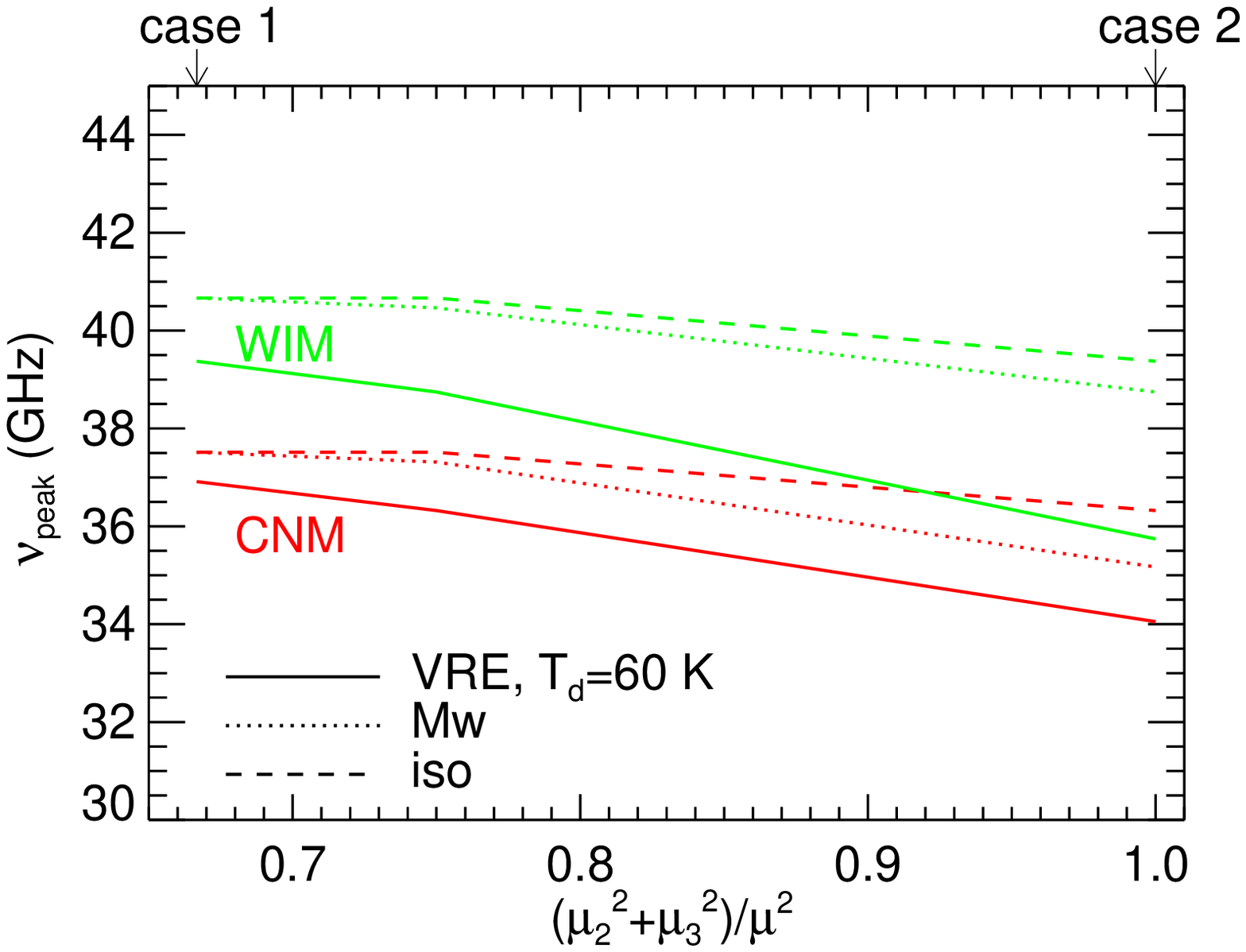}
\includegraphics[width=0.5\textwidth]{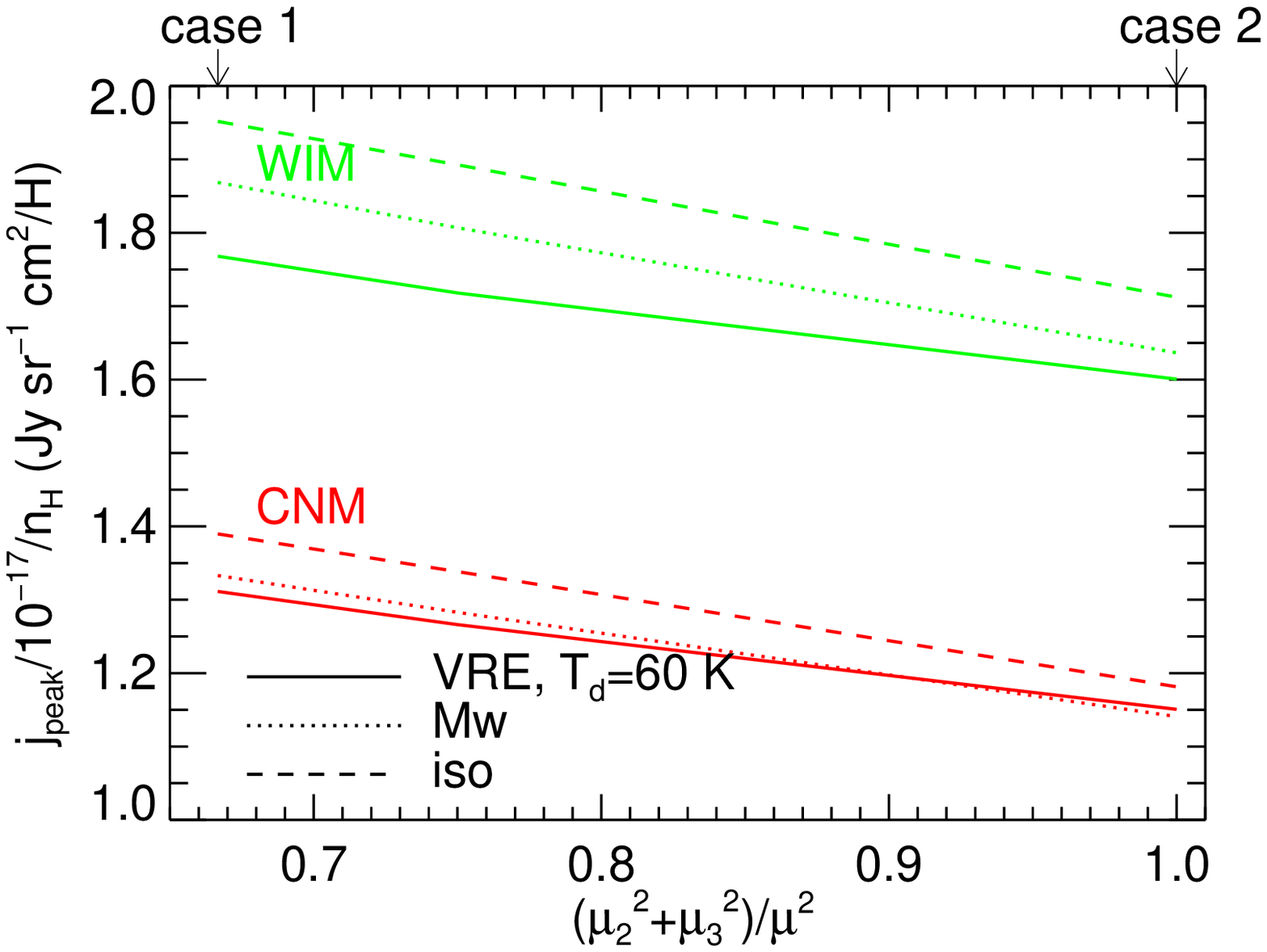}
\caption{ Variation
 of the peak frequency $\nu_{\peak}$ ({\it upper}) and peak emissivity ({\it lower})
with $\bmu$ orientation ($(\mu_{2}^{2}+\mu_{3}^{2})/\mu^{2}$)
 for  VRE, Mw and iso regimes. $\nu_{\peak}$ for 
the VRE regime decreases faster than that for the Mw and iso regimes as $\bmu$ 
orientation changes from case 1 ($\mu_{1}=\mu/\sqrt{3}$)  to case 2 ($\mu_{1}=0$), but
 the decrease is within $10\%$. }
\label{Omepeak_mu}
\end{figure}

Figure \ref{Omepeak_mu} shows the variation of $\nu_{\peak}$ ({\it upper panel}) and $j_{\peak}$
  ({\it lower panel}) of spinning dust spectrum with $\mu_{\perp}^{2}/\mu^{2}$. It 
can be seen that for the Mw and iso $f(\theta)$, $\nu_{\peak}$ changes rather 
slowly with $\mu_{\perp}^{2}/\mu^{2}$, with somewhat larger variation found 
in the VRE regime. 
The peak emissivity for all regimes exhibits similar trend that
decreases with increasing $\mu_{\perp}^{2}/\mu^{2}$, though such a decrease
 is only $\sim 10\%$ when $\bmu$ orientation changes from case 1 
($\mu_{1}=\mu/\sqrt{3}$) to case 2 ($\mu_{1}=0$).

\subsubsection{Variation of distribution of electric dipole moment}
Our standard model (model A, see Table 1) assumed grains of a given size $a$ to
 have three values of dipole moment, $\beta=\beta_{0},~\beta_{0}/2$ and 
$2\beta_{0}$
 with  $\beta_{0}=0.4~\D$, appearing in Equation (\ref{mu2}).

\begin{figure}
\includegraphics[width=0.5\textwidth]{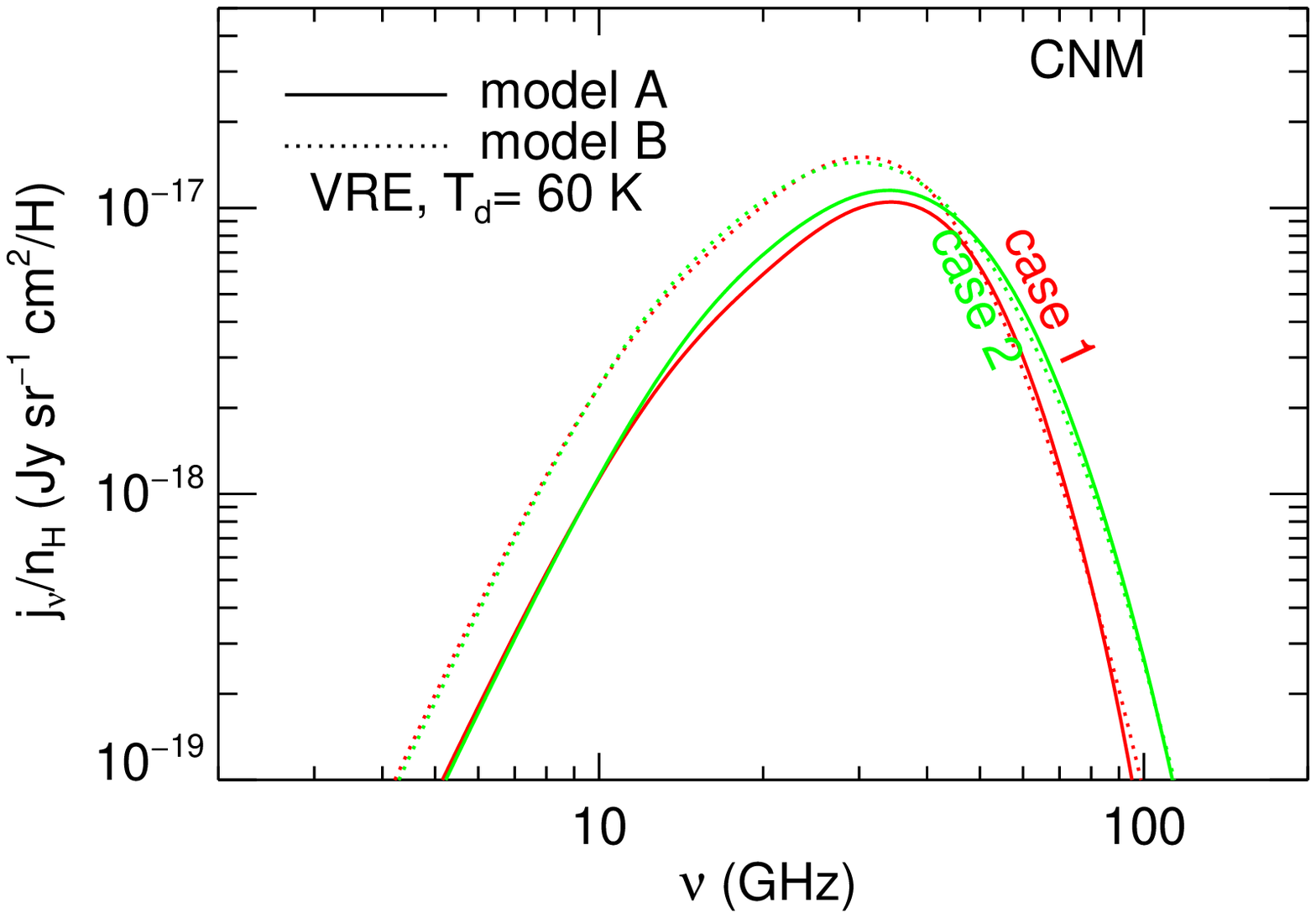}
\includegraphics[width=0.5\textwidth]{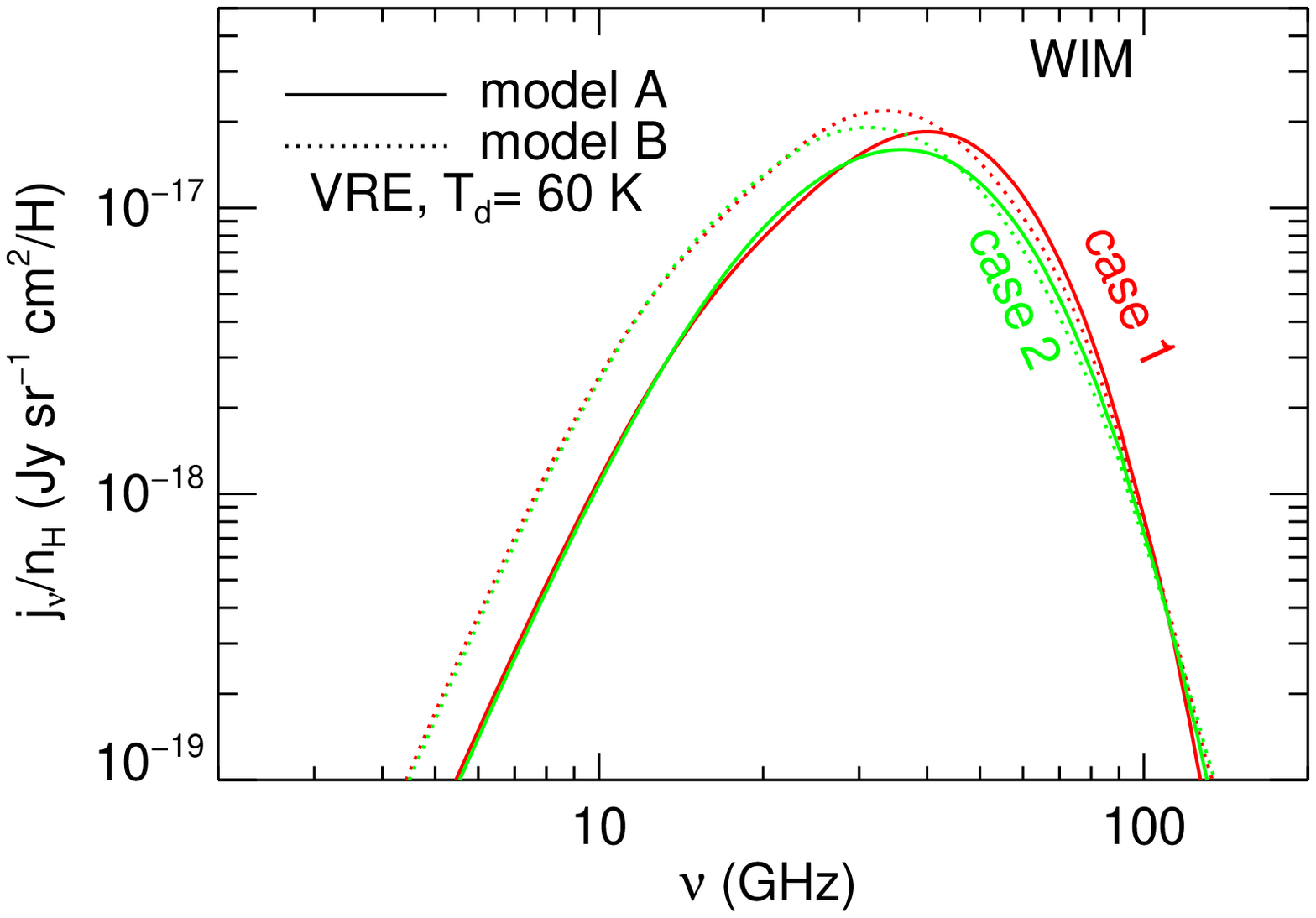}
\caption{Emissivity per H for the CNM ({\it upper}) and WIM ({\it lower}) for model A 
and B of dipole moment distribution and for case 1 ($\mu_{1}=\mu/\sqrt{3}$) and case 2 ($\mu_{1}=0$) of $\bmu$ 
orientation and for the VRE ($T_{\d}=60~\K$) regime.}
\label{jnu_beta}
\end{figure}

Here we examine the possibility of considerable variation in the dipole moment 
per grain, by considering a model (model B) in which $40 \%$ of grains have 
dipole moment parameter $\beta=\beta_{0}$; $15 \%$ have $\beta=2\beta_{0}$ and
 $\beta_{0}/2$; $10\%$ have $\beta=4\beta_{0}$ and $\beta_{0}/4$;
 and $5\%$ have $\beta=8\beta_{0}$ and $\beta_{0}/8$. 
Emissivity per H for this model is compared with the typical model A in Figure
 \ref{jnu_beta}.

 Figure \ref{jnu_beta} shows emission spectra for model A
 and B in cases 1 ($\mu_{1}=\mu/\sqrt{3}$) and 2 ($\mu_{1}=0$) of $\bmu$ orientation, for the CNM and WIM. 
Results for the VRE ($T_{\d}=60$ K) regime are presented. 
The peak frequency from model B is lower than for model A, 
but model B has a higher peak emissivity. The larger value of $\langle \beta^{2}\rangle
=3.12\beta_{0}^{2}$ in model B leads to stronger electric dipole emission but also
 stronger rotational damping.

\subsection{Variation of values of dipole moment and gas density}\label{jnu_den}

Here we assume that all physical parameters are constant, 
including gas temperature and dust temperature. Only the value of characteristic 
dipole  moment $\beta_{0}$ and gas density $n_{\H}$ are varied in calculations.
 We run LE simulations for $16$ values of $\beta_{0}$ from $0.05-3.2$ D, and 
32 values of $n_{\H}$ from 
$10^{-2}-1\cm^{-3}$ for the WIM and from $1-100 \cm^{-3}$ 
for the CNM. 
For a given value of $\beta_{0}$, the gaseous rotational damping 
time and the damping and excitation coefficient from infrared emission 
vary with the gas density $n_{\H}$ as 
\bea
\tau_{\H}&=&\tau_{\H}(\overline{n}_{\H})\frac{\overline{n}_{\H}}{n_{\H}},\\
F_{\IR}&=&F_{\IR}(\overline{n}_{\H})\frac{\overline{n}_{\H}}{n_{\H}},
~~~G_{\IR}=G_{\IR}(\overline{n}_{\H})\frac{\overline{n}_{\H}}{n_{\H}},
\ena
where $\overline{n}_{\H}$ is the typical gas density given in Table \ref{ISM}.
Other rotational damping  and excitation coefficients are independent of 
gas density. 

 The obtained distribution function for grain angular momentum is 
used to calculate spinning dust emissivity as functions of $\beta_{0}$ and 
$n_{\H}$ (see Sec. \ref{jnu-disk}).

\begin{figure*}
\includegraphics[width=0.5\textwidth]{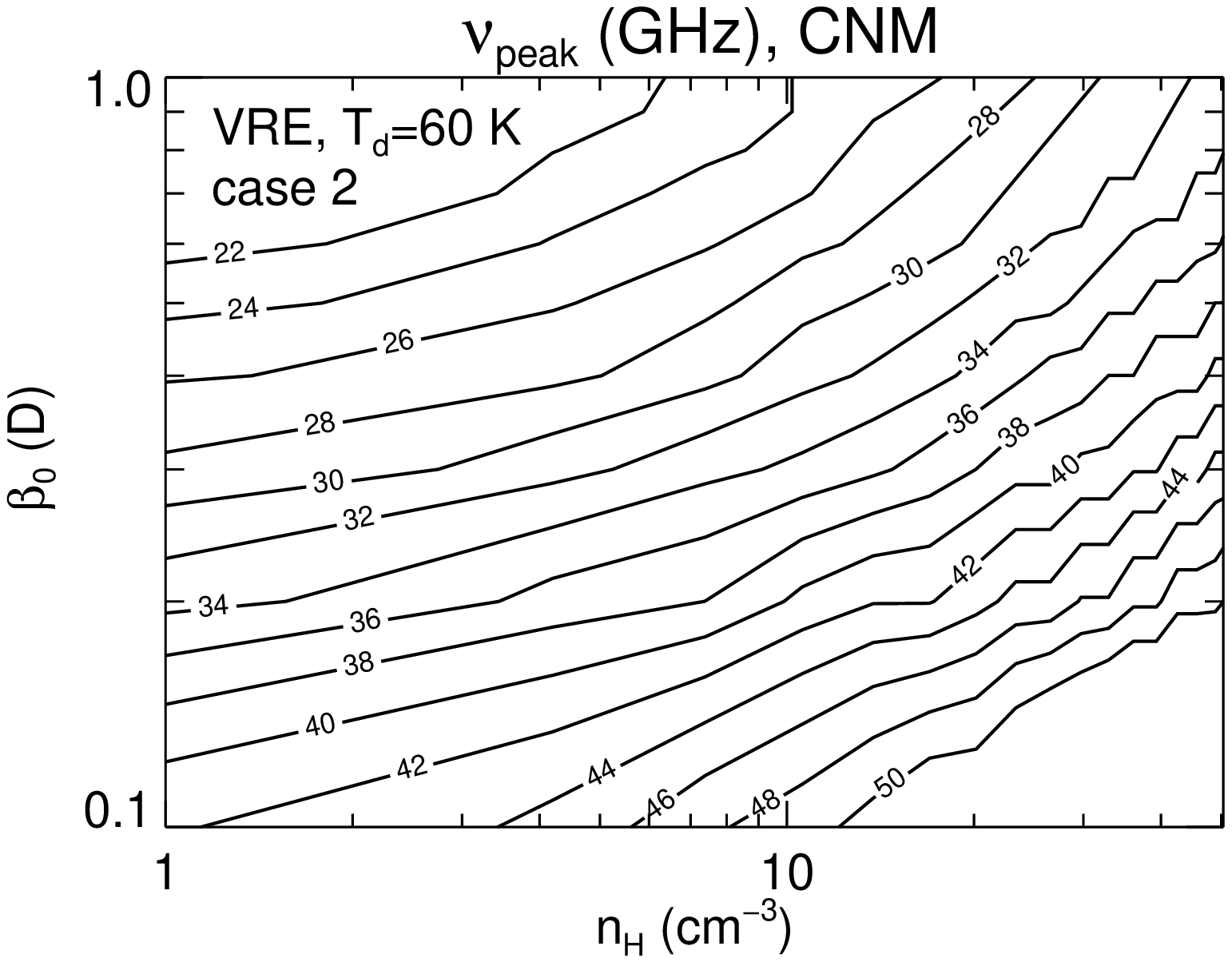}
\includegraphics[width=0.5\textwidth]{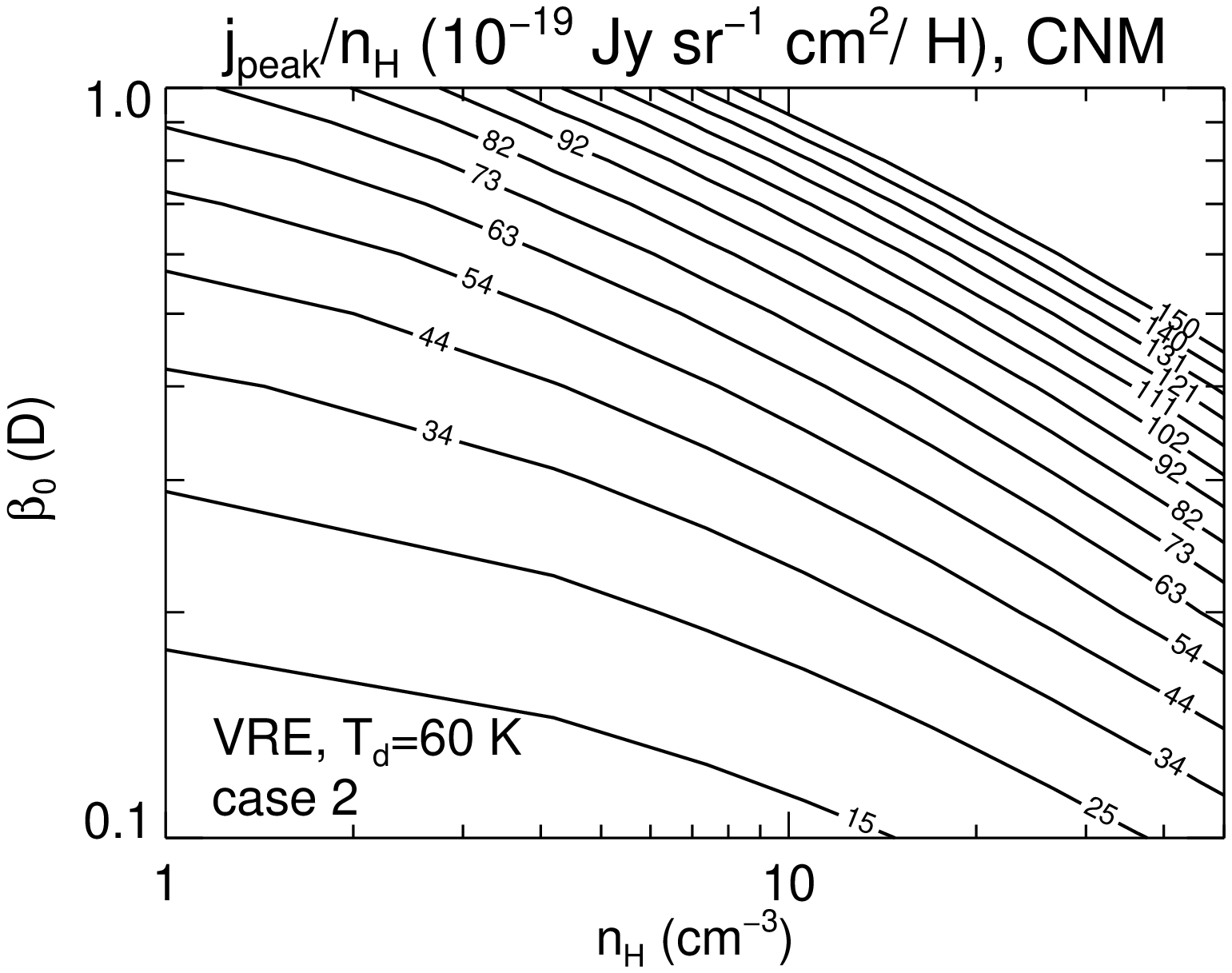}
\includegraphics[width=0.5\textwidth]{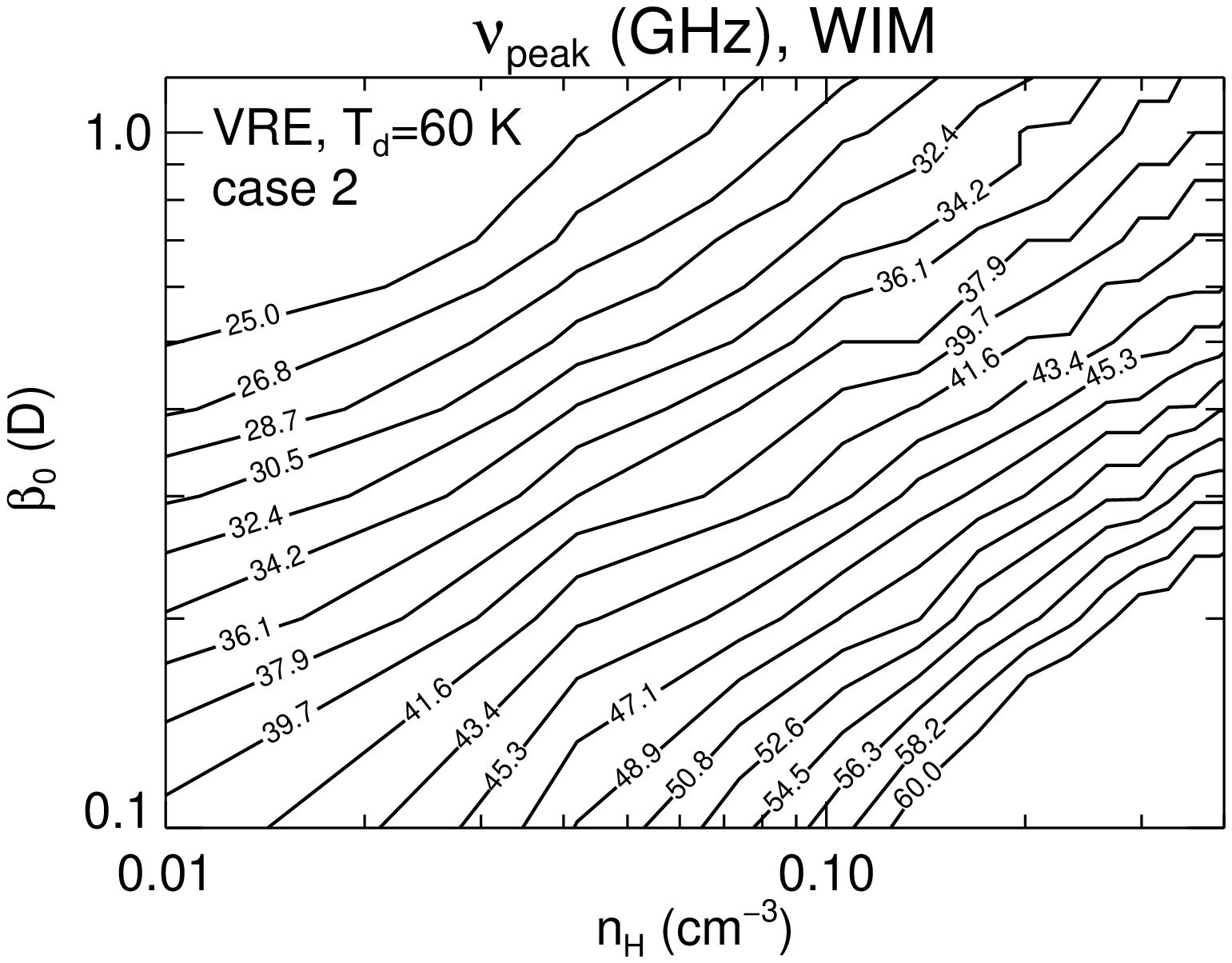}
\includegraphics[width=0.5\textwidth]{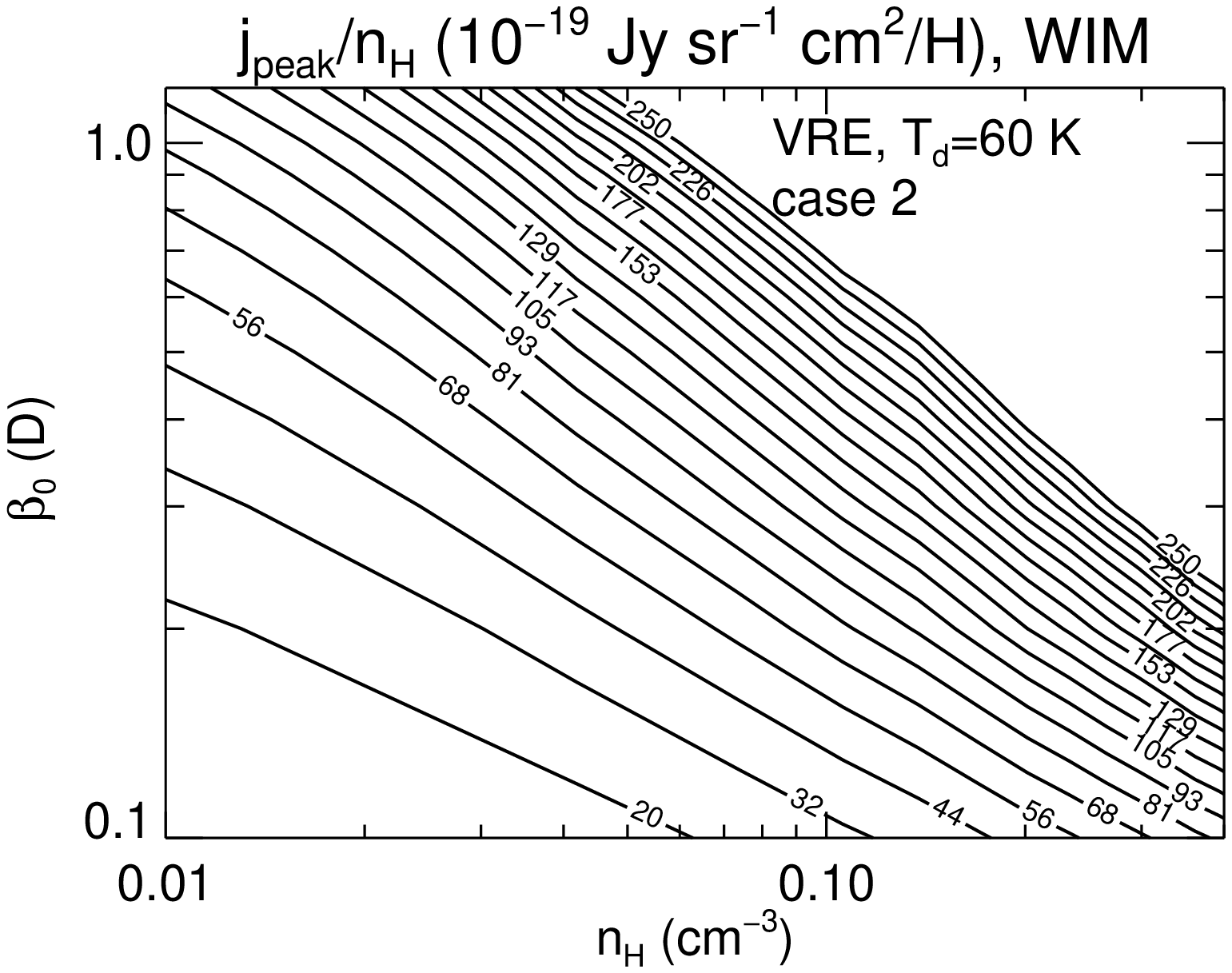}
\caption{Contour of peak frequency and peak emissivity in the plane of 
$n_{\H}$ and $\beta_{0}$ for the CNM ({\it upper}) and the WIM ({\it lower}),
 for the VRE ($T_{\d}= 60$ K) regime. Case 2 ($\mu_{1}=0$) of $\bmu$ orientation
 is considered. Peak frequency 
increases with $n_{\H}$ but decreases with $\beta_{0}$. Emissivity 
increases with both $n_{\H}$ and $\beta_{0}$.}
\label{nupk_contour}
\end{figure*}

Figure \ref{nupk_contour} shows the contour of peak frequency $\nu_{\peak}$ and peak emissivity
$j_{\peak}$  
in the plane of $n_{\H}, \beta_{0}$ for case 2 ($\mu_{1}=0$) of $\bmu$ orientation
in the CNM and  WIM and for the VRE regime with $T_{\d}=60~\K$. 
 For a given $\beta_{0}$, both $\nu_{\peak}$ and $j_{\peak}$ increase with $n_{\H}$ 
 due to the increase of collisional excitation with $n_{\H}$.
For a given $n_{\H}$, as $\beta_{0}$ increases, $\nu_{\peak}$ decreases, 
but $j_{\peak}$ increases. That is because as $\beta_{0}$
 increases, electric dipole damping rate increases, which results in lower
rotational rate of grains. Meanwhile, the emissivity increases with $\beta_{0}$ 
as $\beta_{0}^{2}$. 

 The results ($\nu_{\peak}$ and $j_{\peak}$) for the Mw regime are slightly larger 
than those for the former one due to the lack of internal relaxation.

From Figure \ref{nupk_contour} it can be seen that the amplitude of variation 
of $\nu_{\peak}$ and $j_{\peak}$ as functions of $n_{\H}$ and $\beta_{0}$ is 
very large. Thus, for a fixed grain size distribution, $n_{\H}$ and $\beta_{0}$ 
are the most important parameters 
in characterizing the spinning dust emission.
 
\section{Effect of Turbulence on Spinning Dust Spectrum}

Turbulence is present at all scales of astrophysical systems (see Armstrong et al.
1995; Chepurnov \& Lazarian 2010). The effects of turbulence on many 
astrophysical phenomena, e.g. star formation, cosmic rays transport and
acceleration, have been widely studied. In the context of electric dipole 
emission from spinning dust, compressible turbulence will produce density variations
 that can affect grain charging, rotational excitation, and damping. 
We expect enhancement of spinning dust emissivity per particle 
from denser clumps, with an increase in the emission integrated along 
a line of sight. This issue is quantified in
the following. For simplicity, we disregard the effect of turbulence on
 the grain charging.

\subsection{Numerical Simulations of MHD turbulence}

Gas density fluctuations in a turbulent medium is generated from Magneto-hydromagnetic 
(MHD) simulations. The sonic and 
Alfvenic Mach number
are defined as usual $M_{\rm s}=\langle\frac{\delta V}{c_{s}}\rangle$, 
$M_{\rm A}=\langle\frac{\delta V}{v_{\rm A}}\rangle$ with $\delta V$ are amplitude for 
injection velocity, $c_{s}$ and $v_{\rm A}=B_{0}/\sqrt{4\pi\rho}$ are sound 
speed and Alfvenic speed. 

The simulations are performed by solving the set of non-ideal MHD equations, 
in conservative form:

\begin{equation}
\frac{\partial \rho}{\partial t} + \mathbf{\nabla} \cdot (\rho{\bf v}) = 0,
\end{equation}

\begin{equation}
\frac{\partial \rho {\bf v}}{\partial t} + \mathbf{\nabla} \cdot 
\left[ \rho{\bf v v} + \left( p+\frac{B^2}{8 \pi} \right) {\bf I} - 
\frac{1}{4 \pi}{\bf B B} \right] = \rho{\bf f},
\end{equation}

\begin{equation}
\frac{\partial \mathbf{B}}{\partial t} - \mathbf{\nabla \times (v \times B)} = 0,
\end{equation}
\noindent
with $\mathbf{\nabla \cdot B} = 0$, where $\rho$, ${\bf v}$ and $p$ are the 
plasma density, velocity and pressure, respectively,
 ${\bf I}$ is the diagonal 
unit tensor, ${\bf B}$ is the magnetic 
field, and ${\bf f}$ represents the external acceleration source, responsible for 
the turbulence injection. Under this 
assumption, the set of equations is closed by an {\it isothermal equation of state}
 $p = c_{s}^2 \rho$. The equations are solved
 using a second-order-accurate and non-oscillatory MHD code 
described in detail in Cho \& Lazarian (2003).
Parameters for the models of MHD simulations
are shown in Table \ref{tab:mhd}, including the sonic Mach number $M_{\rm s}$ 
and Alfvenic Mach number $M_{\rm A}$. 
\begin{table}[htb]
\begin{center}
\caption{\label{tab:mhd}
         Models of $512^{3}$ MHD simulations}
\begin{tabular}{l c c c c c}
\hline\hline

Model &  $M_{\s}$ & $M_{\rm A}$ & $v_{A}/c_{s}$& $\langle \rho^{2}\rangle/\langle\rho\rangle^{2}$\cr
\hline
$1$ & 2.0 &0.7 & 2.9 &1.7\cr
$2$ & 4.4 &0.7 & 6.3 &2.1\cr
$3$ &  7.0&0.7 & 10  & 3.0\cr
\hline
\end{tabular}
\end{center}
\end{table}

\subsection{Results: influence of Gas Density Fluctuations}

\begin{figure}
\includegraphics[width=0.5\textwidth]{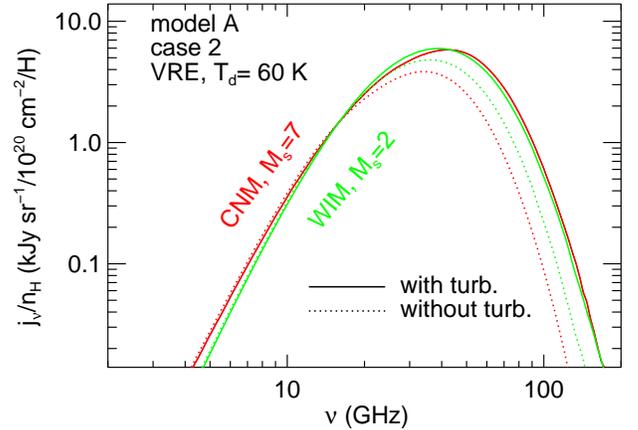}
\caption{Spinning dust emissivity per H 
in the presence of compressible turbulence 
with sonic Mach number $M_{\rm s}=2$ and $7$, compared to that from 
constant gas density $n_{\H}=\overline{n}_{\H}$ for the CNM (red) and WIM (green). 
The peak emissivity is increased, and the spectrum is shifted to 
higher frequency due to compressible turbulence. Case 2 ($\mu_{1}=0$) of $\bmu$ orientation
 is considered}
\label{jnu_M2M7}
\end{figure}

In a medium with density fluctuations, the effective emissivity is 
\bea
\langle j_{\nu}\rangle=\int_{0}^{1} f(x) j_{\nu}(x\langle\rho\rangle)dx,
\ena
where $f(x)dx$ is the fraction of the mass with $\rho/\langle\rho\rangle \in
(x,x+dx)$. We use compression distributions $f(x)$ obtained from MHD simulations
for $M_{\rm s}=2$ and $7$ to evaluate $\langle j_{\nu}\rangle$ for 
the WIM and CNM, respectively. 
We assume case 2 ($\mu_{1}=0$) of $\bmu$ orientation. It can be seen that the turbulent
compression increases the emissivity, and shifts the peak to higher $\nu_{\peak}$.

\begin{figure}
\includegraphics[width=0.5\textwidth]{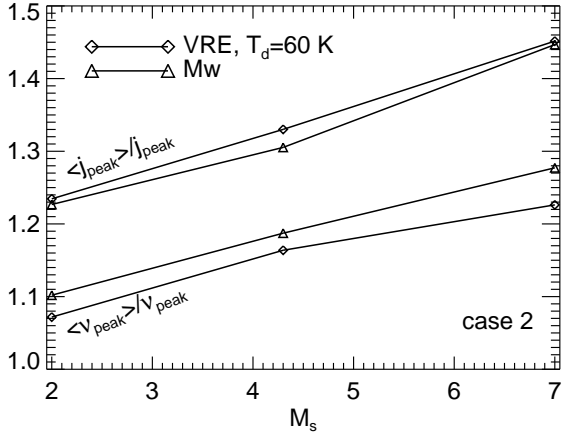}
\caption{Increase of peak emissivity and peak frequency due to compressible turbulence as functions 
of Mach sonic number M$_{\rm s}$. Results for both fast internal relaxation 
(VRE, $T_{\d}=60$ K) and without internal relaxation (Mw) are shown. 
Case 2 ($\mu_{1}=0$) of $\bmu$ orientation is considered.}
\label{deltanu_Ms}
\end{figure}

The increase of peak frequency and peak emission intensity, 
as a function
of $M_{\rm s}$ are shown in Figure \ref{deltanu_Ms}. 
$\langle j_{\peak}\rangle$ and $j_{\peak}$ are emissivity for the turbulent and uniform
media, respectively. $\langle\nu_{\peak}\rangle$ and $\nu_{\peak}$ are peak frequency 
for the former and latter. Results for
$M_{\rm s}=2$ and $7$ are obtained from the spinning dust spectra for 
the WIM and CNM, respectively. 
For $M_{\rm s}=4.4$, results for the CNM are chosen. It is shown that the 
emission intensity can be increased
 by factors from $\sim 1.2-1.4$, and peak frequency is increased by
factors from $1.1-1.2$ as $M_{s}$ increases from $2-7$. 

\subsection{Observational constraining turbulence effects}

The discussion of interstellar conditions adopted in DL98 was limited by 
idealized interstellar phases. It is now recognized that turbulence plays
an important role in shaping the interstellar medium.

The distribution of phases, for instance, CNM and WNM of the ISM
 at high latitudes can be obtained from absorption lines. Similarly, studying 
fluctuations of emission it is possible to constrain parameters of turbulence. 
In an idealized case of a single phase medium with fluctuations of density with
 a given characteristic size one can estimate the value of the 3D fluctuation by
 studying the 2D fluctuations of column density. A more sophisticated techniques
 of obtaining sonic Mach numbers\footnote{It may be seen that Alfven Mach 
numbers have subdominant effect on the distribution of densities 
(see  Kowal, Lazarian \& Beresnyak 2008). Thus in our study (see Table 3) we did not vary the 
Alfven Mach number.} have been developed recently (see  Kowal et al. 2008; Esquivel \& Lazarian 2010; Burkhard et al. 2009,
 2010). In particular, Burkhart et al. (2010), using just column density 
fluctuations of the SMC, obtained a distribution of Mach numbers corresponding 
to the independent measurements obtained using Doppler shifts and absorption data. With
 such an input, it is feasible to quantify the effect of turbulence in actual 
observational studies of spinning dust emission.  

\section{Galactic CMB foreground components and fitting model}

\subsection{Galactic CMB foreground components}

The Galactic CMB foreground consists of four 
components: synchrotron emission ($I_{\nu}^{\rm syn}$), free-free emission ($I_{\nu}^{\rm ff}$), 
thermal dust emission ($I_{\nu}^{\rm td}$), and spinning 
dust  emission ($I_{\nu}^{\sd}$):
\bea
I_{\nu}=I_{\nu}^{\rm syn}+I_{\nu}^{\rm td}+I_{\nu}^{\rm ff}+I_{\nu}^{\sd}.
\ena

The synchrotron emission consists of soft and hard components. The soft
component  has the spectrum $I_{\nu}^{\rm syn}\sim \nu^{-\beta}$ with
$\beta=0.7-1.2$, and is dominant below $1 \GHz$. The soft
synchrotron is produced from electrons accelerated by supernova shocks that 
spiral about
the Galactic magnetic field (Davies, Watson \& Gutierrez 1996). The hard 
component, also named ``Haze'', is present around the Galactic center,
and its origin is still unknown (Finkbeiner 2004). The free-free emission 
arising from ion-electron scattering has a flatter spectral index 
$I_{\nu}^{\rm ff}\sim \nu^{-\beta}$ with $\beta\approx 0.1$, and is an important 
CMB foreground in the range $10-100$ GHz.
The third component due to thermal dust emission (vibrational emission of dust grains) 
dominates  above 100 GHz. Finally, the spinning dust component 
dominates the range $20-60$ GHz.

Dobler et al. (2009) subtracted the synchrotron contribution using the Haslam et al. (1982) 408 MHz map plus a model for the ``haze'' component. The remaining
emission was decomposed into two components: (1) a ``thermal-dust-correlated''
spectrum correlated with the SFD dust map (basically, 100 $\mu$m) and (2) an
H$\alpha$-correlated spectrum correlated with observed (reddening-corrected)
H$\alpha$.  Both components included an ``anomalous emission'' component, 
peaking around 30--40 GHz, that is attributable to spinning dust.

The spinning dust intensity is given by 
 \bea
 I_{\nu}^{\sd}=\int j_{\nu} dl,\label{eq:Inu}
\ena
where $j_{\nu}$ is the spinning dust emissivity, and the integral is taken 
along a line of sight. Below, we constrain the physical parameters of spinning dust by fitting 
the model of foreground emissions to H$\alpha$-correlated and thermal
 dust-correlated emission spectra. 
 
 PAHs are likely to be irregular,
but we do not attempt to determine the degree of irregularity in the present 
paper.
Instead, we calculate  spinning dust spectra for disk-like grains, and 
account for the effect of the grain shape irregularity by scaling the emission to
account for irregularity. 
Consider a simple model of the grain irregular shape,  with the ratio of 
semi-axes $\eta=b_{3}/b_{2}=1.2$ (see \S 4.3). Then, the model peak frequency and peak emissivity
are scaled with those from disk-like grains as follows:
\bea
\nu_{\peak}^{\rm mod}=\nu_{\peak}(\disk)\times \frac{\nu_{\peak}(\eta)}{\nu_{\peak}(\disk)},\\
j_{\peak}^{\rm mod}=j_{\peak}(\disk)\times \frac{j_{\peak}(\eta)}{j_{\peak}(\disk)},
\ena
where we take $\nu_{\peak}(\eta)/\nu_{\peak}(\disk)$ and 
$j_{\peak}(\eta)/j_{\peak}(\disk)$ from Figures 5 and 6. 

\subsection{Fitting to $\H{\alpha}$-correlated emission}
 The H${\alpha}$ intensity is approximately given by (Draine 2011)
\bea
I_{\H\alpha}&=&0.361T_{4}^{-0.942-0.031\ln T_{4}}\frac{\rm EM}{\cm^{-6} 
\pc} \mbox{~R},\label{Ialpha}
\ena
where $T_{4}=T_{e}/(10^{4}~\K)$, $T_{e}$ is the electron temperature, 
and EM is the emission measure.

\begin{figure}
\includegraphics[width=0.48\textwidth]{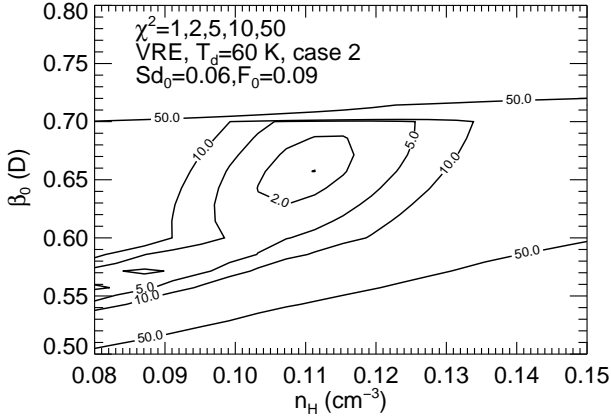}
\caption{Fitting three-component model to H$\alpha$-correlated spectrum: 
the $\chi^2$ contours in the plane of $n_{\H}$ and $\beta_{0}$
 for VRE ($T_{\d}=60~\K$) regime and case 2 ($\mu_{1}=0$) of $\bmu$. 
The parameters at which $\chi^{2}$ has minimum are shown. }
\label{chi2_contour}
\end{figure}
 \begin{figure}
\includegraphics[width=0.48\textwidth]{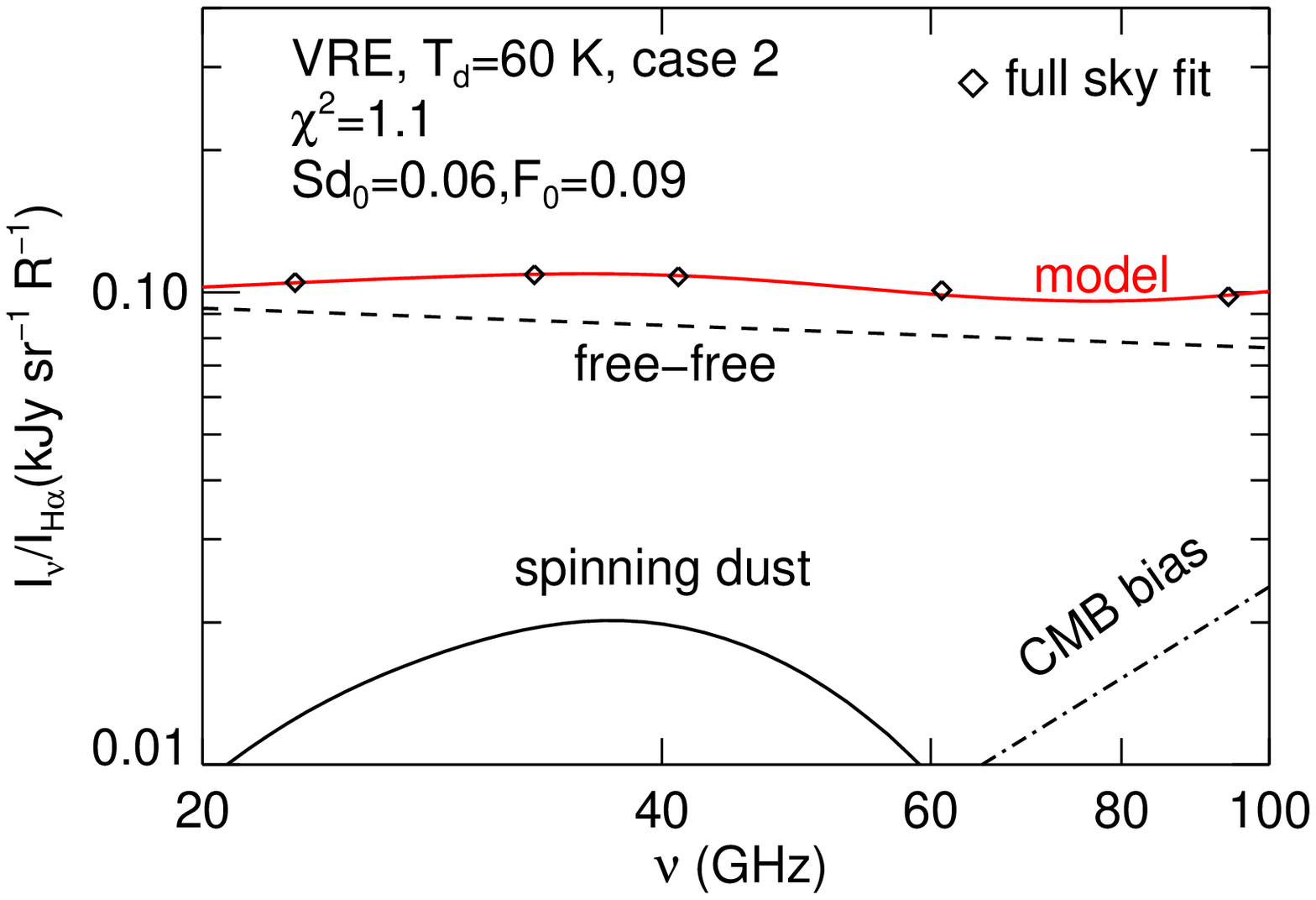}
\includegraphics[width=0.48\textwidth]{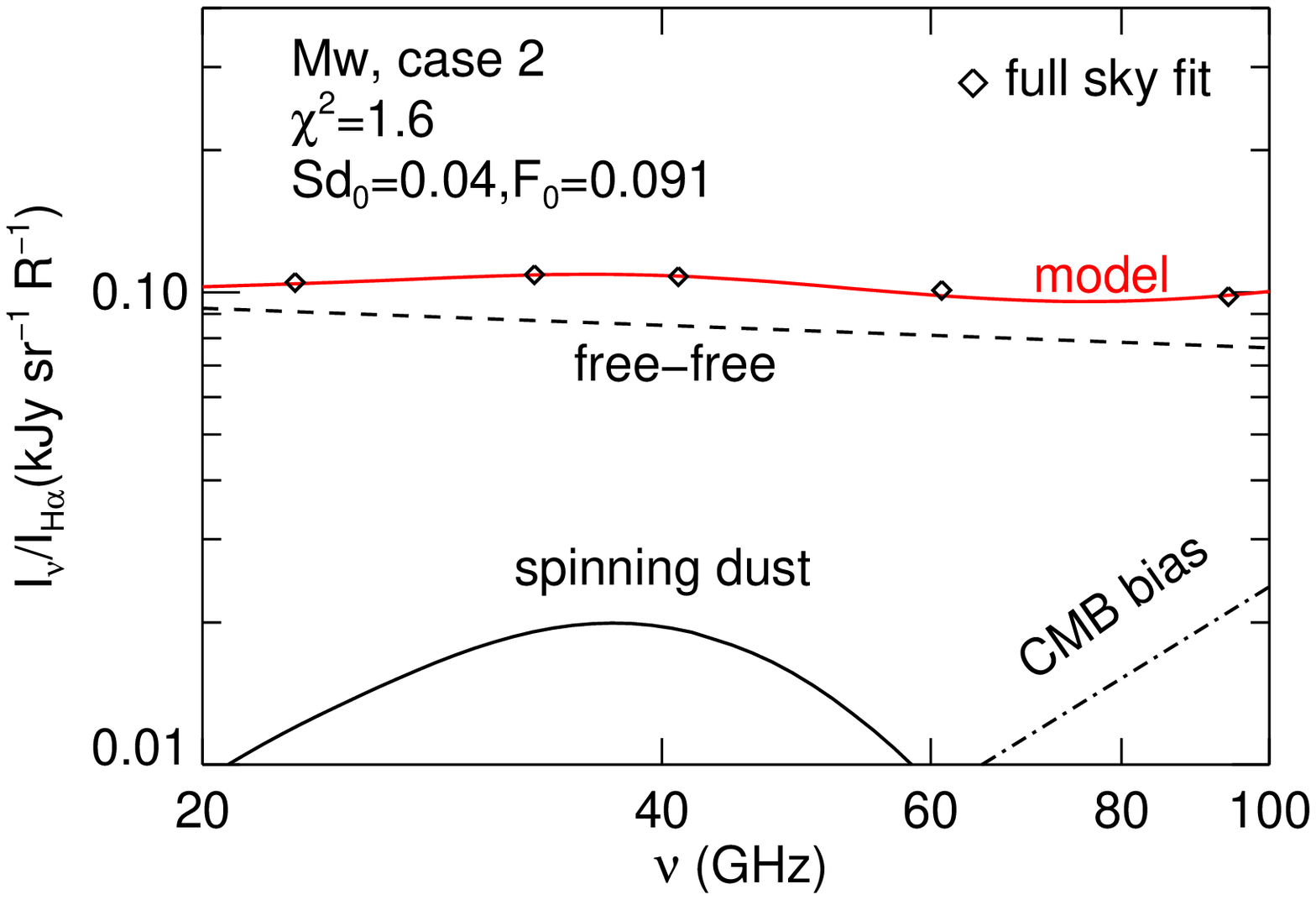}
\caption{A three-component fit model (solid line) to H${\alpha}$-correlated 
spectrum using CMB5 data (diamonds) from Dobler et al. (2009) 
for VRE ({\it upper}) and Mw ({\it lower}) regimes using case 2 ($\mu_{1}=0$) of $\bmu$.
Three components of free-free emission, spinning dust emission
 and CMB bias are shown separately. As shown are best fits parameters 
and $\chi^2$.}
\label{Halpha_fit}
\end{figure}

Dobler et al. (2009) have determined the spectrum $I^{\rm obs}$ of 
the H$\alpha$-correlated emission in the {\it WMAP} five-year data. The thermal
dust contribution has been almost completely removed as part of the SFD-
correlated component, and synchrotron emission has similarly been removed by
its correlation with the Haslam et al. (1982) 408 MHz map. Within the frequency
range considered here, the residual thermal dust emission
and synchrotron emission are assumed negligible.

Thus, as in Dobler et al. (2009), we fit the H$\alpha$-correlated spectrum with three components:
free-free emission, spinning dust emission, and CMB cross-correlation bias.
We present the ratio of intensity to H$\alpha$ intensity. The free-free emission
is given by (Draine 2011)
\bea
\frac{I_{\nu}^{\rm ff}}{I_{\H\alpha}}=F_{0}\left(\frac{\nu}{23 \GHz}\right)^{-0.12}~ {\rm kJy ~sr}^{-1} {\rm R}^{-1}.\label{Iff_Ialpha}
\ena
where 
\bea
F_{0}=0.197T_{4}^{0.614+0.942\ln T_{4}}.\label{eq:F0}
\ena

The CMB bias is defined as $\partial B_{\nu}(T)/\partial T|_{T=2.7~ \K}\propto \nu^{2}$, and 
we obtain
\bea
\frac{I_{\nu}^{\rm cmb}}{I_{\H\alpha}}=C_{0}\left(\frac{\nu}
{23 \GHz}\right)^{2}\frac{1}{{\rm plc}(\nu)}~~ {\rm kJy~ sr}^{-1} {\rm R}^{-1},
\ena
 where $C_{0}$ is a free parameter, and plc is the Planck correction factor to convert thermodynamic temperature $\Delta T$ to antenna temperature at frequency $\nu$.

The spinning dust component is taken as the emission from the WIM. We 
obtain
\bea
\frac{I_{\nu}^{\sd}}{I_{\H\alpha}}&=&Sd_{0}\times\left[\frac{I_{\nu}^{\sd}}{I_{\H\alpha}}\right]_{\WIM},~~~\label{Isd_Ialpha}
\ena
where $\left[I_{\nu}^{\sd}/I_{\H\alpha}\right]_{\WIM}$ is calculated for our model for the
WIM with a standard PAH abundance, and the fitting parameter  $Sd_{0}$  allows for the
possibility that the PAH abundance in the WIM may differ from our standard value. 

Using Equation (\ref{Ialpha}) for the above equation, we obtain the spinning dust intensity per H$\alpha$
\bea
\left[\frac{I_{\nu}^{\sd}}{I_{\H\alpha}}\right]_{\WIM}&=&\left(\frac{j_{\nu}}{n_{\H}}\right)_{\WIM}\times\left(\frac{n_{\H}}{j_{\H\alpha}}\right)_{\WIM},\nonumber\\
&=&\left(\frac{j_{\nu}}{n_{\H}}\right)_{\WIM}\times\left(\frac{\cm^{-5}/n_{\H}}{(n_{e}/n_{\H})^{2}}\right)\nonumber\\
&&\times\frac{8.55\times 10^{18}T_{4}^{0.942+0.031\ln T_{4}}}{\rm R}~{\rm kJy~ sr}^{-1}.~~~~~~~~\label{Isd_p}
\ena

The model emission intensity per H$\alpha$ is then
\bea
\frac{I_{\nu}^{\rm mod}}{I_{\H\alpha}}&=&\Biggl\{F_{0}\left(\frac{\nu}{23 \GHz}\right)^{-0.12}+
C_{0}\left(\frac{\nu}
{23 \GHz}\right)^{2}\frac{1}{{\rm plc}(\nu)}\Biggr\}~\frac{{\rm kJy~ sr}^{-1}}{{\rm R}}\nonumber\\
&&+Sd_{0}\times\left[\frac{I_{\nu}^{\sd}}{I_{\H\alpha}}\right]_{\WIM}.\label{eq:Imod}
\ena

For the spinning dust model, we consider a range of gas density $n_{\H}\in [0.01,1]
\cm^{-3}$ and
 the value of dipole moment $\beta_{0}\in [0.1,1]$ D. 
For each ($n_{\H}$, $\beta_{0}$), we compute  $\left(j_{\nu}/n_{\H}\right)_{\WIM}$. We  then vary the values of $F_{0}, Sd_{0}$ and $C_{0}$.
From a given value $F_{0}$ we derive the gas temperature
$T_{4}$ using Equation (\ref{eq:F0}). When $T_{4}$ and $\left(j_{\nu}/n_{\H}\right)_{\WIM}$ 
 are known, we obtain $\left[I_{\nu}^{\sd}/I_{\H\alpha}\right]_{\WIM}$ from Equation (\ref{Isd_p}).

The fitting process is performed by minimizing the $\chi^{2}$ function
\bea
\chi^2=\sum_{\nu}\frac{|I^{\rm mod}_{\nu}-I^{\rm obs}_{\nu}|^{2}}
{\sigma_{\nu}^{2}},\label{chisq}
\ena
where $I_{\nu}^{\rm obs}$ is the measured H$\alpha$-correlated emission intensity, and
 $\sigma_{\nu}$ is the mean noise per observation at the frequency band $\nu$
(see Hinshaw et al. 2007; Dobler et al. 2009). 

In general, $\chi^2$ is a function of three amplitude parameters $F_{0}, C_{0}, 
Sd_{0}$, and 
two physical parameters $n_{\H}$ and $\beta_{0}$.
Therefore, the fitting proceeds by minimizing $\chi^2$ over five parameters.

Figure \ref{chi2_contour} shows $\chi^2$ contours as functions of
$\beta_{0}$ and $n_{\H}$ for case 2 ($\mu_{1}=0$) and for VRE ($T_{\d}=60~\K$),
for the values of $Sd_{0}$ and $F_{0}$  for which
 $\chi^{2}$ is minimized.  
 We can see that the distribution of $\chi^2$ 
is localized and centered around the standard values in the plane $n_{\H}-\beta_{0}$.

Figure \ref{Halpha_fit} shows the fit of the three-component model to the H$\alpha$-correlated
 foreground spectrum for VRE and Mw regimes for case 2 of $\bmu$.
 Both regimes provide a good fit to the data.

Table \ref{tab:bestfit_WIM} shows best fit parameters and $\chi^2$ for different
 cases of $\bmu$ orientation and different regimes of internal relaxation. 
We can see that both orientations of $\bmu$ (case 1 and case 2) can produce a good fit 
with low $\chi^{2}$, but case 2 exhibits a relatively better fit. The best fit gas 
density is in the range $0.08- 0.15\cm^{-3}$. In case 1, the best fit dipole moment  for 
the VRE and Mw regimes is $\beta_{0}\sim 0.76$ and $1.0$ D. In case 2, the best
 fit requires $\beta_{0}\sim 0.65$ D and $0.84$ D for VRE and Mw respectively.  
Case 1 requires higher $\beta_{0}$ to reproduce the observations than case 2
because the effect of grain shape irregularity is more important for case 1. 
Similarly, the Mw regime requires higher $\beta_{0}$ than VRE because the 
effect of irregularity is stronger for the former case.

From Table \ref{tab:bestfit_WIM}, it can be seen that the best fit value
 $Sd_{0}\sim 0.04-0.06$ is significantly lower than
  the value $Sd_{0}=0.3$ in Dobler et al. (2009). Our lower $Sd_{0}$  
  stems from the higher value of $\beta_{0}$ 
  required,  from the increase of emissivity in the HDL10 model
  compared to the DL98 model used in Dobler et al. (2009), and the further (modest)
  increase in emissivity for irregular grains.
  As a result, a higher depletion of small PAHs is required to obtain a good fit. 
  
  In addition, from best fit values $F_{0}$ in Table \ref{tab:bestfit_WIM} we
  derive the gas temperature $T_{e}\sim 2700-3500$ K as found by Dobler 
  et al. (2009). This is lower than the
  typical temperature $T_{e}=8000$ K usually assumed for the WIM. 
  Dong \& Draine (2011)
   proposed a model of three components that can explain the low gas temperature in
    the WIM.

\begin{table}[htb]
\begin{center}
\caption{\label{tab:bestfit_WIM}
         Best fit parameters and $\chi^2$ for H$\alpha$-correlated emission}
\begin{tabular}{l c c c c c c c c}
\hline
\hline
$\bmu$ & $f_{\rm int}(\theta)$ & $n_{\H}~({\rm cm}^{-3})$ &$\beta_{0}~(\D)$  
& $Sd_{0}$ & $F_{0}$ & $C_{0}$& $\chi^2$ \cr
\hline
\hline
case~1 & VRE  &0.15 & 0.76  &0.04 &0.09&0.0012 &1.9 \cr
       & Mw         &0.08& 1.0  &0.06 &0.1&0.001 & 4.5\cr
\hline
case~2 & VRE  &0.11 & 0.65  &0.06  &0.09 &0.0012&1.1 \cr
       & Mw          & 0.14 & 0.84  &0.04 &0.09 &0.0012&1.6 \cr
       \cr
\hline\hline
\end{tabular}
\end{center}
\end{table}

\subsection{Fitting to thermal dust-correlated emission}

In addition to the H$\alpha$-correlated emission, the foreground induces a dust-correlated emission spectrum with a usual thermal dust emission component falling from $94$ to $40$ GHz, and another component  rising from $40$ to 
$23$ GHz (see Bennett et al. 2003).
 The latter component is consistent with the spinning dust emission (DL98ab; 
de Oliveira et 
al. 2004; HDL10). Although this peak frequency around $22$ GHz is consistent with 
prediction by the DL98 model, it is 
 lower than the prediction by the improved model of HDL10, using the same 
parameters for the CNM as in DL98b. 

As in Dobler et al. (2009), we fit the thermal dust-correlated emission spectrum
 with a three-component model including spinning dust, thermal dust 
and CMB bias.\footnote{Free-free emission is not important for the thermal 
dust-correlated emission spectrum.} 

Dobler et al. (1999) determined the spectrum $I_{\nu}/T_{94\GHz}({\rm FDS})$ for the contribution
$I_{\nu}$ to the intensity from the thermal-dust-correlated emission, where
$T_{94\GHz}({\rm FDS})$ is the antenna temperature at 94 GHz predicted for the 
FDS dust model (Finkbeiner, Davis \& Schlegel 1999).
 
Planck collaboration (2011c) showed that the thermal dust emission can be 
approximated by
\bea
{I_{\nu}^{\rm td}}=\epsilon(\nu)B_{\nu}(T_{\rm td})N_{\H},\label{eq:Itd}
\ena
where  $T_{\rm td}=17.6~ \K$ and $\epsilon({\nu})=0.92\times 10^{-25}(\lambda/250~\mu \mtxt)^{-1.8}\cm^{-2}\H^{-1}$.

We model the  spinning dust contribution as
\bea
{I_{\nu}^{\sd}}=Sd_{0}\left(\frac{j_{\nu}}{n_{\H}}\right)_{\CNM}.
\ena
Thus, we seek to fit the Dobler et al. (2009) spectrum by 
\bea
\frac{I_{\nu}^{\rm mod}}{T_{94\GHz}({\rm FDS})}&=&\Biggl\{ \left[\left(\frac{\nu}{94\GHz}\right)^{3.8}
+ Sd_{0}\frac{\left(j_{\nu}/n_{\H}\right)_{\CNM}}{[\epsilon(\nu)B_{\nu}(T_{\rm td})]_{\nu=94\GHz}}\right] \nonumber\\
&&+ C_{0}\left(\frac{\nu}{94\GHz}\right)^{2}\Biggr\}~ {\rm kJy}\sr^{-1}\mtxt\K^{-1},~
\ena
where $A, ~Sd_{0}$ and $C_{0}$ are adjustable parameters. For each trial $n_{\H}$
and $\beta_{0}$, we adjust $A,~Sd_{0}$ and $C_{0}$ to minimize $\chi^{2}$. We
consider different possible values of $n_{\H}$ and $\beta_{0}$.

Best fit parameters and $\chi^2$ for different situations are 
shown in Table \ref{tab:bestfit_CNM}.
In case 1 ($\mu_{1}=\mu/\sqrt{3}$), the best fit corresponds to $n_{\H} = 6.0-6.5 \cm^{-3}$,  
and $\beta_{0}\sim 0.94-0.97~\D$ for  VRE and Mw regimes.
In case 2 ($\mu_{1}=0$), the best fit corresponds to $n_{\H}= 7.0-7.5\cm^{-3}$
and $\beta_{0}\sim 0.95~\D$ for VRE and  Mw regimes.
 The significance of fitting is low in both cases 1 and 2. Case 2 fits slightly 
 better ($\chi^{2}=30-35$) than case 1 ($\chi^{2}=54-56$).

\begin{figure}
\includegraphics[width=0.48\textwidth]{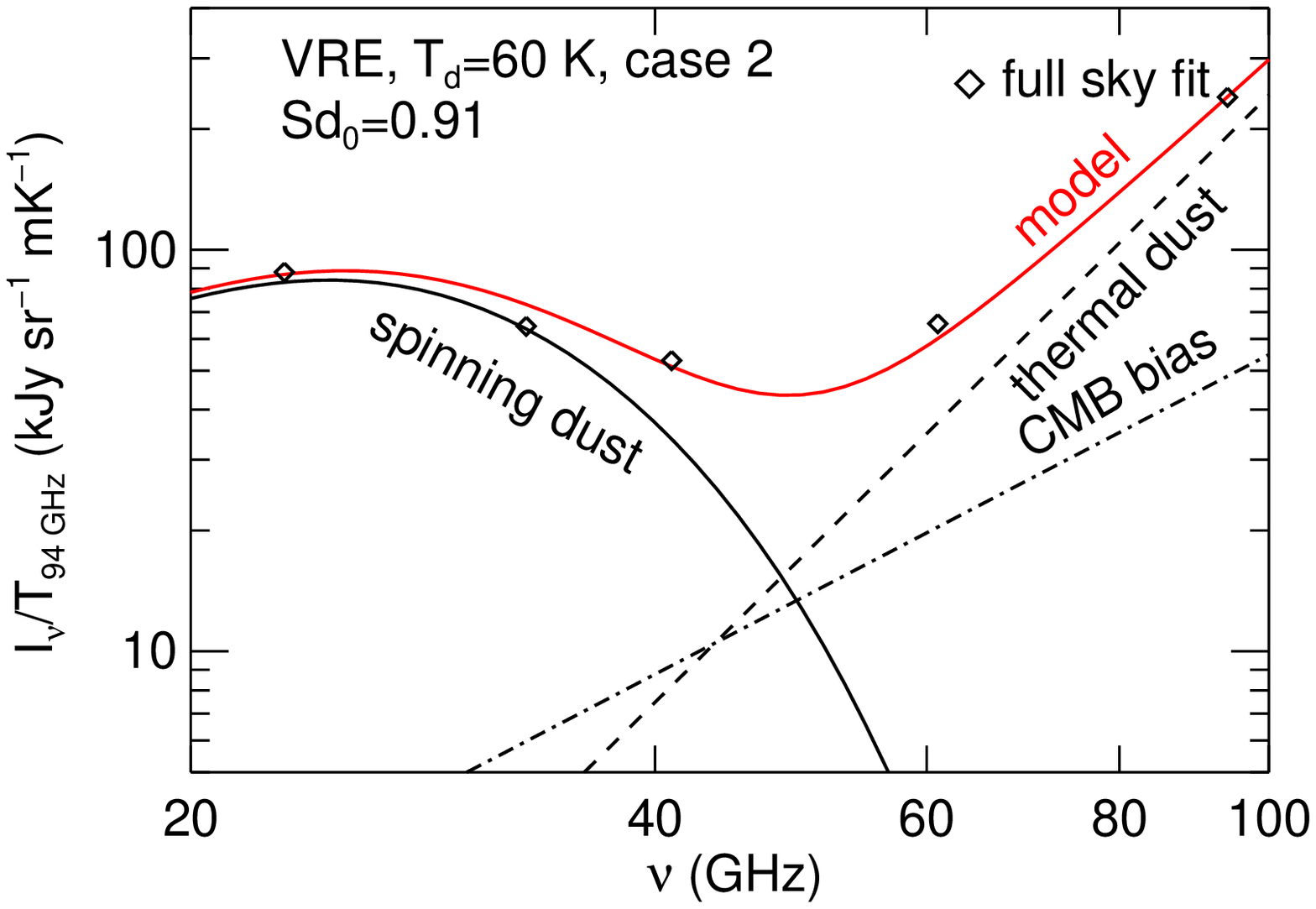}
\includegraphics[width=0.48\textwidth]{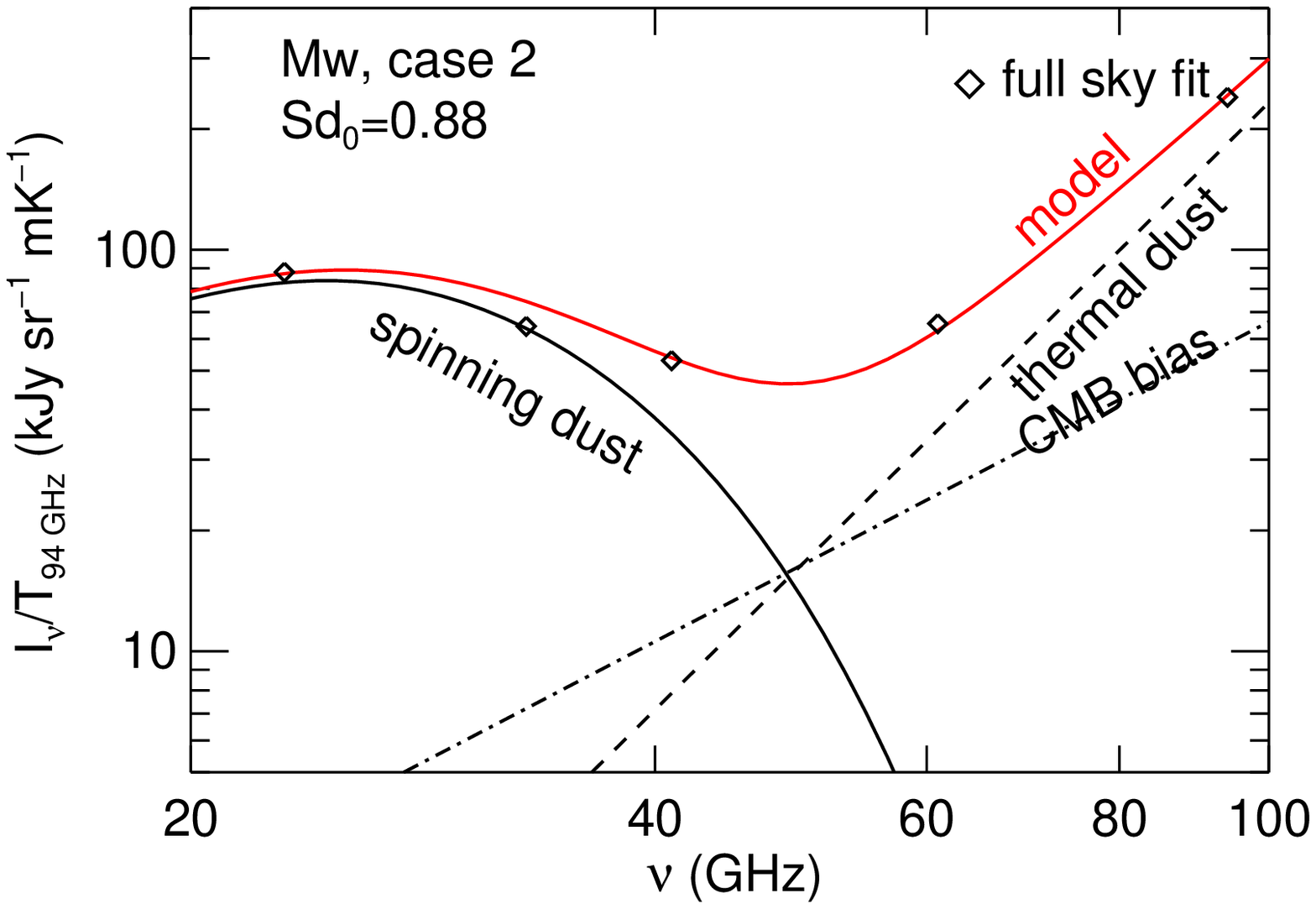}
\caption{A three-component fit model (solid line) to thermal dust-correlated 
 emission spectrum (diamonds) 
for VRE ($T_{\d}=60~\K$, {\it upper}) and Mw ({\it lower}) regimes using case 2 ($\mu_{1}=0$) of
 $\bmu$. Three
components of spinning dust emission, thermal dust emission
 and CMB bias are shown separately but solid, dashed and dashed-dotted lines. As shown is the best fit value of $Sd_{0}$.}
\label{thermal_fit}
\end{figure}

 Figure \ref{thermal_fit} shows the best-fit to the thermal dust-correlated spectrum 
 obtained using case 2 models, for both VRE and Mw regimes. 

We see that the fitting to thermal dust-correlated
emission has large $\chi^2$. The reason  
is that the curvature of the model spectrum is larger than that of the observed
 spectrum for frequencies $22-44$ GHz. 

\begin{table}[htb]
\begin{center}
\caption{\label{tab:bestfit_CNM}
         Best fit parameters and $\chi^2$ for thermal dust-correlated emission}
\begin{tabular}{l c c c c c c c c}
\hline
\hline
$\bmu$ & $f_{\rm int}(\theta)$ & $n_{\H}({\rm cm}^{-3})$ &$\beta_{0}(\D)$  
& $Sd_{0}$ & $A$ & $C_{0}$& $\chi^2$ \cr
\hline
\hline
case~1 & VRE & 6.5 & 0.97 &0.74 &184.3 & $4.5$ &54 \cr
             & Mw  & 6.0 & 0.94 &0.63 &176.1 & $4.7$ &56\cr
\hline
case~2 & VRE & 7.5 & 0.95 & 0.91 &192.4 &$ 2.9$  &30 \cr

             & Mw  & 7.0 & 0.96 & 0.88 &184.2 &$ 3.5 $&35 \cr
             \cr
\hline\hline
\end{tabular}
\end{center}
\end{table}

\subsection{Classical vs. quantum treatment}

Ysard \& Verstraete (2010, hereafter YV10) questioned the validity of classical 
mechanics in the DL98 model for spinning dust emission. YV10 put forward a
 quantum-mechanical formalism for treating the rotational emission. 
 However, they only calculated 
spinning dust emission for the rotational transition $\Delta J=-1$
 and $\Delta K=0$, which is induced by the oscillation of the dipole 
component along the grain symmetry axis.

In our view, a quantum description of spinning dust is unnecessary, because, 
as shown in DL98b for spherical grain, the angular quantum number 
\bea
J\equiv\frac{I_{1}\omega}{\hbar}\approx
 72 \left(\frac{N_{\rm C}}{20}\right)^{5/6}\left(\frac{T_{\rm rot}}{100~\K}\right)^{1/2},
\ena
 which shows that even smallest PAHs have angular  quantum number $J \gg 1$, 
therefore, the classical treatment should be valid. 

\begin{figure}
\includegraphics[width=0.5\textwidth]{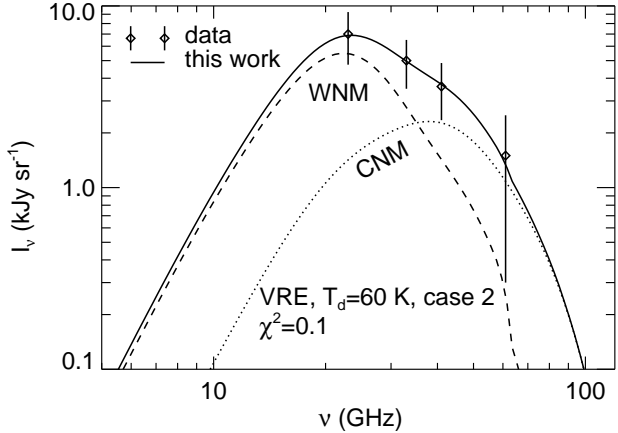}
\caption{Comparison of our study with the anomalous spectrum extracted 
from the observation data in YV10.
 The solid line is the total emissivity resulting from the CNM 
and WNM with $N_{\H}({\rm CNM})=3.5\times 10^{20}\cm^{-2}$ and $N_{\H}({\rm WNM})=1.4\times 10^{21}\cm^{-2}$. Squared symbols show observation data. All grains have
 the characteristic electric dipole moment $\beta=\beta_{0}=0.3~\D$. Case 2
($\mu_{1}=0$) of $\bmu$ is considered.}
\label{jnu_fit}
\end{figure}

Miville-Desch{\^e}nes et al. (2008) made an attempt to separate the spinning 
dust component from the Galactic foreground components (mostly synchrotron
 and anomalous emission) using
 both the {\it WMAP} intensity and polarization data. The inferred spinning dust 
spectrum is presented in Figure \ref{jnu_fit} (diamond symbols). 

Ysard, Miville-Desch{\^e}nes \& Verstraete (2010) presented a fit to observation 
data for spinning dust, which is extracted from WMAP data for
regions with latitude $b=22.4^{\circ}$  (the spinning dust 
spectrum extracted from WMAP data using the quantum mechanical approach
from YV10). 
They assumed a
lower cut-off of the grain size corresponding to the number of carbon atom
 $N_{\min}=24$ 
and $48$ for the CNM and WNM, and dipole moment $\beta_{0}=0.3~ \D$. 
Converting to the grain size using the usual relationship 
 $a=10 (N_{\C}/410)^{1/3}\Angstrom$, we obtain $a_{\min}\sim 3.88$ and
 $4.89\Angstrom$, respectively. 
However, since $a_{\min}$ chosen for the WNM is much larger than that for 
the CNM, the peak frequency of its emission spectrum is expected to be much 
lower than that for the CNM, because the peak frequency decreases with 
$a_{\min}$ increasing 
(see \S 5.2). This disagrees with results shown in Figure 5 of Ysard et al. (2010),
 where the peak frequency of the WNM ($\nu_{\peak}\sim 45$ GHz) is much larger 
than that for the CNM ($\nu_{\peak}\sim 22$ GHz). 

To see how well the improved model of spinning dust in HDL10 fits to observation data
 for selected regions in Ysard et al. (2010),
 we assume the same spinning dust parameters as in Ysard et al. (i.e. gas density, 
dipole moment $\beta_{0}$ and $a_{\min}$). 

The total emission intensity from the CNM and WNM with column density $N
_{\H}$(CNM) and $N_{\H}$(WNM) is given by
\bea
I_{\nu}^{\rm mod}=N_{\H}({\rm CNM})\left(\frac{j_{\nu}^{\sd}}{n_{\H}}\right)_{\rm CNM} +N_{\H}({\rm WNM}) \left(\frac{j_{\nu}^{\sd}}{n_{\H}}\right)_{\rm WNM},~~~\nonumber
\ena
where $\left[{j_{\nu}^{\sd}}/{n_{\H}}\right]$ is the spinning dust emissivity per H
 calculated for the CNM and WNM, respectively (see Table 2). 
The fitting proceeds by minimizing $\chi^2$ (Eq.  \ref{chisq})  with  $N_{\H}$(CNM)
 and $N_{\H}$(WNM) as free parameters. 
We found that the best fit is achieved at $N_{\H}({\rm CNM})=3.5\times 10^{20}\cm^{-2}$
and $N_{\H}({\rm WNM})=1.4\times 10^{21}\cm^{-2}$. 
The best fit is shown in Figure \ref{jnu_fit} and $\chi^{2}$ is also shown.
 We can see that our improved model can reproduce the observational data
 for high latitude regions. 

\section{Discussion}

\subsection{Relative importance of effects}

The present paper extends further the improved
model of spinning dust emission in HDL10 by including the following effects: 
irregular grain shape, dust temperature fluctuations due to transient heating, 
orientation of dipole moment, and fluctuations of gas density
due to compressible turbulence.

\subsubsection{Power Spectrum of Grains of Irregular Shape}

First of all, we find that a torque-free rotating irregular grain radiates
at multiple frequency modes with angular frequency 
$\omega_{m_{i}}$ and $\omega_{n_{i}}$ depending on a dimensionless
parameter $q$, the ratio of the 
rotational energy to the rotational energy along the axis of major inertia. 

In case 1 ($\mu_{1}=\mu/\sqrt{3}$) of $\bmu$ orientation, the mode $\omega_{m_{-1}}$ dominates the 
dipole emission for $q$ close to $1$ (when $\ba_{1}$ makes a small angle
with $\bJ$). As $q$ increases, the mode $\omega_{m_{0}}$ increases accordingly, 
 and become dominant.
Physically, the mode $\omega_{m=0}$ arises from the precession of the dipole
moment component parallel to $\ba_{1}$ axis about the angular momentum. Its emission
power increases with the angle $\theta$ between $\ba_{1}$ and $\bJ$, 
and is maximum when $\ba_{1}$ is perpendicular to $\bJ$, i.e.,
$\theta=\pi/2$ or $q=q_{\sp}$.

In case 2 ($\mu_{1}=0$, i.e. no dipole component along $\ba_{1}$), the mode $m=0$ does not 
exist when the grain rotates about $\ba_{1}$ for $q<q_{\sp}$. For 
$q>q_{\sp}$, the irregular grain rotates about $\ba_{3}$, therefore, the 
dipole moment component along $\ba_{3}$ induces the frequency mode
 $m=0$.

In both case 1 and case 2, the modes $m=\pm 1$, which result from the oscillation
 of dipole moments along $\ba_{2}$ and $\ba_{3}$, always exist, and 
have power increasing with $q$ up to $q=q_{\sp}$, then decrease in importance 
as $q$ increases further.

\subsubsection{Increase of spinning dust emission with the degree of grain 
shape irregularity}

For both case 1 ($\mu_{1}=\mu/\sqrt{3}$) and 2 ($\mu_{1}=0$) of $\bmu$ orientation, we 
find that the peak frequency 
 increases rapidly with increasing degree of grain shape irregularity $\eta$ 
 (see \S 4.3). The 
underlying reason is that the deformation of the grain from the disk-like shape 
allows the 
grain to rotate along the axis of minimum moment of inertia $\ba_{3}$, 
in addition to rotation about $\ba_{1}$ and $\ba_{2}$ axes. For a given angular
 momentum, the angular velocity along $\ba_{3}$ is largest, so that
the emission frequency of the irregular grain is in general higher than that
 of the disk-like grain with the same mass. 

Peak emissivity also increases with the grain shape irregularity for case 1.
But for case 2,  $j_{\peak}$ decreases with $\eta$ for $T_{\vib}\le 60~\K$, 
and starts to increase with $\eta$ as $T_{\vib}$ increases.

When the internal thermal fluctuations are rather weak ($T_{J}>T_{\vib}$) so that the
 grain tends to rotate with the minimum rotational energy, 
$\theta$ and $q$ 
deviate slightly from $\theta=0$ and $q=1$, and
 the rotational emission spectrum of irregular grains differs only slightly
from that of disk-like grains. 
In contrast, for strong thermal fluctuations ($T_{J}\ll T_{\vib}$), the thermal fluctuations 
 enable grains to spend an important fraction of time
rotating with large $q$. Because the total power emission increases with
 $q$, the higher probability of 
rotating with larger $q$ results in larger 
emissivity and peak frequency than disk-like grains.

\subsubsection{Orientation of dipole moment}

We investigated the effects of changing the orientation of the electric dipole
moment $\bmu$. We found (see Fig. 9) that as the orientation is varied from having
a component parallel to $\ba_{1}$ (case 1, $\mu_{1}=\mu/\sqrt{3}$) to being entirely perpendicular to 
$\ba_{1}$ (case 2, $\mu_{1}=0$), the peak frequency $\nu_{\peak}$ and the peak emissivity 
both decrease by $\sim 10\%$. Given the sensitivity of the spectrum to other
variables (mass distribution, $|\mu|$, grain shape, gas density, gas temperature),
it does not seem possible to infer the orientation of $\bmu$ from observations of
the spectrum.

\subsubsection{Density Fluctuation due to Compressible Turbulence}

We identified interstellar turbulence as another factor that influences
emission spectrum.
The spinning dust emissivity is determined by two principal processes: collisional
 and radiative damping and excitation. Local compression due to turbulence 
increases the collisional excitation, that results in 
increased spinning dust emissivity. We found that the emission is increased by a factor 
from $1.2-1.4$ as the sonic Mach number increases from $M_{\s}=2-7$.

\subsection{Constraints from Spinning dust Emission}

Recent studies showed that the correspondence of the DL98 model to 
observations can be improved by adjusting the parameters of the model. 
For instance, the five-year ({\it WMAP}) data showed a broad 
bump with frequency at $\sim40$ GHz in the H$\alpha$-correlated emission
(Dobler \& Finkbeiner 2008; Dobler et al. 2009). Using H$\alpha$ as a tracer 
of the WIM,
they showed that this bump can be explained using the DL98 model 
by varying either dipole moment or gas density of the WIM. 

 By fitting the 3 components of synchrotron, free-free emission and spinning 
dust emission to H$\alpha$-correlated spectrum, Dobler et al. (2009) found that 
the spinning dust model with $\beta_{0}=0.15$ D and density 
$n_{\H}=0.15\cm^{-3}$ could reproduce the observed spectrum, with 
$\nu_{\peak}=37\GHz$. 
Our best fit values from the improved model correspond to
 $n_{\H}=0.08-0.15\cm^{-3}$ and $\beta_{0}$ in the range of $0.6-1.0$ D. 

In addition to the H$\alpha$-correlated emission, WMAP data show a thermal dust-
correlated spectrum that declines from $23$ GHz to $40$ GHz.
This peak at low frequency is lower than the HDL10 model prediction
using the standard parameters for the CNM. 

Our results show that the thermal dust-correlated data can be fitted 
with spinning dust emission from the CNM of density $n_{\H}\sim 6-8\cm^{-3}$
 and $\beta_{0}\sim 0.95-1.0~\D$. Similar to Dobler et al. (2009),
 we also found that the best-fit model did not provide a very good fit 
(high $\chi^{2}$). The reason for that is the rotational spectrum is steeper than
the observation data in frequency range $20-50$ GHz.  
Dobler et al. (2009) suggested that the superposition of spinning dust spectra 
from different ISM phases along a sight-line would produce a flatter spectral slope. 
With our results for irregular grains, one can see that by averaging
 the rotational spectrum over various degree of irregularity $\eta$ with the 
 fraction of irregular grains decreasing with $\eta$ (see \S 4),  the obtained
 spectrum becomes shallower, that can improve the fit. 
 
\subsection{Range of applicability of the model of spinning dust emission}

The model of spinning dust emission has been used to interpret the anomalous
microwave emission in the general ISM (e.g. Finkbeiner 2004; Dobler \& Finkbeiner
 2008; Gold et al. 2009, 2011; Planck Collaboration 2011c), in star forming regions
 in the nearby galaxy NGC 6946 (Scaife et al. 2010;
Murphy et al. 2010) and in the Persus and Ophiuchus clouds (Cassasus et al. 2008; 
Tibbs et al. 2010; Planck Collaboration 2011a). Early Planck results have been
interpreted as showing an emission excess from spinning dust in the Magellanic
Clouds (Bot et al. 2010; Planck Collaboration 2011b). 

This paper together with HDL10 presents a comprehensive model 
of spinning dust, accounting for non-disk-like (``irregular'') grains, new 
electric dipole emission modes from torque-free rotation of irregular grains, 
and investigating the effects of variation in grain dipole moment, 
gas density, and different regimes of vibrational-rotational mode coupling. 
We believe that apart from being an important CMB foreground, the spinning dust
 spectrum can become an important diagnostic tool to constrain physical properties of 
astrophysical dust (e.g. size distribution, shape, electric dipole moment and 
gas density) in various environments.

\section{Summary}

The model of spinning dust emission is further extended by accounting for 
 effects of irregular grain shape, fluctuations of dust temperature, 
 and effects of ISM turbulence. We consider both regimes of fast internal 
relaxation and without internal relaxation. 
Our main results are as follows.

1. Small grains of irregular shape radiate in general at multiple harmonic 
frequency modes. The rotational emission shifts to higher frequency as the 
degree of grain shape irregularity
 increases, but the spectral profile remains similar. The effect of the 
 grain shape irregularity is more important for higher dust temperature or 
 stronger internal thermal fluctuations. Depending on the irregularity parameter
 $\eta$, 
peak frequency and peak emissivity can be increased by a factor of up to $1.4$, 
relative to disk-likes grains of the same mass (see Figs. 5 and 6) for 
case 1 ($\mu_{1}=\mu/\sqrt{3}$).

2. Fluctuations of dust temperature $T_{\vib}$ also increase the rotational
emissivity relative to the emissivity for grains of a steady low temperature.

3. Fluctuations of gas density and gas pressure due to compressible turbulence
 enhance both emission and peak frequency of spinning dust spectrum compared
to that in uniform media. 
An increase in emission by a factor from $1.2-1.4$ is expected as the sonic 
Mach number $M_{\s}$ increases from $2-7$.

4. Spinning dust parameters (e.g., gas density and dipole moment) are constrained 
by fitting the improved model to {\it WMAP}  cross-correlation foreground spectra, 
including H$\alpha$ and thermal dust-correlated spectra. We find a reduced PAH 
abundance in the WIM ($Sd_{0}\approx 0.05$) with dipole moment parameter
$\beta_{0}\approx 0.7$ D. For the thermal-dust-correlated emission, we find a normal
PAH abundance ($Sd_{0}\approx 0.9$) and $\beta_{0}\approx 0.95$ D.

5. Our improved model also provides a good fit to WMAP data for 
selected regions at high latitude ($b=22.4^{\circ}$) obtained by Ysard et al. (2010).

\acknowledgements TH and AL acknowledge the support of the Center for
Magnetic Self-Organization and the NASA grant NNX11AD32G. TH acknowledges
 the support from Ranger Supercomputing Center. BTD acknowledges research support
 from NSF grant AST 1008570.

\appendix

\section{A. Electric Dipole emission from an irregular grain}
\subsection{A1. Torque-free motion}
The effective size $a$ of an irregular grain with volume $V$ is defined as the radius
of a sphere with the same volume $V$, i.e., 
\bea
a=\left(3V/4\pi \right)^{1/3}.\label{eq:aef}
\ena

In general, an irregular grain can be characterized by a triaxial ellipsoid with moments 
of inertia $I_{1}, I_{2}$ and $I_{3}$ around three principal  axes 
${\bf a}_{1}, {\bf a}_{2}$ and ${\bf a}_{3}$, respectively. 
Define dimensionless parameters $\alpha_{i}$ so that the moments of inertia
are written as
\bea
I_{j}=\alpha_{j}I_{\rm sph},
\ena
where $I_{\rm sph}=\left(8\pi/15\right) \rho a^{5}$ is the moment of inertia of the 
equivalent sphere of radius $a$, and $\rho$ is the mass density of the grain.

For a torque-free rotating
 grain, its angular momentum $\bJ$ is conserved
, while the angular velocity ${\bomega}$ nutates and wobbles with respect to $\bJ$. 
We can term the wobbling associated with the irregularity in 
the grain shape {\it irregular wobbling}, to avoid confusion with thermal
 wobbling (also thermal fluctuations) induced by the Barnett relaxation 
(Purcell 1979) and nuclear relaxation (Lazarian \& Draine 1999b).

A detailed description of the  torque-free motion for an asymmetric top in terms 
of Euler angles $\theta, \phi$ and $\psi$ (see Fig. \ref{euler}) can 
be found in classical textbooks (e.g. Landau \& Lifshitz 1976; see also WD03), 
and a brief summary is given below.
 
Consider an ellipsoid with three principal axes ${\bf a}_{1}, {\bf a}_{2},
{\bf a}_{3}$ and moments of inertia $I_{1}> I_{2}> I_{3}$. 

Let define a dimensionless quantity
\bea
k^{2}=\frac{(I_{2}-I_{3})(q-1)}{(I_{1}-I_{2})(1-I_{3}q/I_{1})},\label{eq:k2}
\ena
where $q={2I_{1}E_{\rm rot}}/{J^{2}}$ is the ratio of total kinetic energy to 
the rotational energy along the axis of major inertia $\ba_{1}$.

({\it a}) For $q<I_{1}/I_{2}, k^{2}<1$, the solution of Euler equations is
\bea
\omega_{1}&=\pm\frac{J}{I_{1}}\left(\frac{I_{1}-I_{3}q}{I_{1}-I_{3}}\right)^{1/2} \mbox{dn}
(\tau),\label{omep1}\\
\omega_{2}&=-\frac{J}{I_{2}}\left(\frac{I_{2}(q-1)}{I_{1}-I_{2}}\right)^{1/2} \mbox{sn}
(\tau),\label{omep2}\\
\omega_{3}&=\pm\frac{J}{I_{3}}\left(\frac{I_{3}(q-1)}{I_{1}-I_{3}}\right)^{1/2} \mbox{cn}
(\tau)\label{omep3}
\ena
where cn, sn and dn are  hyperbolic trigonometric functions, and $\tau$ is
given by
\bea
\tau\equiv tJ\left[\frac{(I_{1}-I_{2})(1-I_{3}q/I_{1})}{I_{1}I_{2}I_{3}}\right]^{1/2},
\ena
and the sign $\pm$ in $\omega_{1}$ and $\omega_{3}$ are taken the same. We denote
 the rotation with $+$ and $-$ sign as positive and negative rotation state. 
 For $q<I_{1}/I_{2}$, the grain mostly rotate about the axis of major inertia 
$\ba_{1}$, while it rotates about $\ba_{3}$ for $q>I_{1}/I_{2}$.

The rotation period around the axis of major inertia $\ba_{1}$ is 
\bea
P_{\tau}=4F(\pi/2,k^{2}),
\ena
where $F$ is the elliptic integral defined by
\bea
F(\epsilon, m)=\int_{0}^{\epsilon} d\theta (1-m \sin^{2}\theta)^{-1/2}.
\ena

({\it b}) For $q> I_{1}/I_{2}$, angular velocities are given by
\bea
\omega_{1}&=\pm\frac{J}{I_{1}}\left(\frac{I_{1}-I_{3}q}{I_{1}-I_{3}}\right)^{1/2} \mbox{cn}
(\tau),\label{omen1}\\
\omega_{2}&=-\frac{J}{I_{2}}\left(\frac{I_{2}(1-I_{3}q)}{I_{2}-I_{3}}\right)^{1/2} \mbox{sn}
(\tau),\label{omen2}\\
\omega_{3}&=\pm\frac{J}{I_{3}}\left(\frac{I_{3}(q-1)}{I_{1}-I_{3}}\right)^{1/2} \mbox{dn}
(\tau),\label{omen3}
\ena
where 
\bea
\tau\equiv tJ\left[\frac{(I_{2}-I_{3})(q-1)}{I_{1}I_{2}I_{3}}\right]^{1/2}.
\ena
Rotation period for this case is given by
\bea
P_{\tau}=4F(\pi/2,k^{-2}).
\ena

({\it c}) For $q\sim I_{1}/I_{3}$, Equation (\ref{eq:k2}) shows that 
$k^{2}\rightarrow \infty$,
  the rotation of the grain is about the axis near $\ba_{3}$.
Thus, $\omega_{3}\approx J/I_{3}$, and  $\omega_{1}\sim \omega_{2}\sim 0$.
From Euler equations, we obtain
\bea
\frac{I_{1}\d\omega_{1}}{dt}=\omega_{2}\omega_{3}\left(I_{2}-I_{3}\right),\\
\frac{I_{2}\d\omega_{2}}{dt}=\omega_{3}\omega_{1}\left(I_{3}-I_{1}\right).
\ena
Substituting $\omega_{3}=\Omega_{0}=J/I_{3}$, the equations are rewritten as
\bea
\frac{dJ_{1}}{dt}=\Omega_{0}J_{2}\left(1-\frac{I_{3}}{I_{2}}\right),\label{eq:dj1dt}\\
\frac{dJ_{2}}{dt}=-\Omega_{0}J_{1}\left(1-\frac{I_{3}}{I_{1}}\right),\label{eq:dj2dt}
\ena
where $J_{k}=I_{k}\omega_{k}$.
Taking the first derivative of Equation (\ref{eq:dj1dt}), and using 
(\ref{eq:dj2dt}), we obtain  solutions for $J_{1}$ and $J_{2}$:
\bea
J_{1}&=&A\cos\omega t,\\
J_{2}&=&\frac{A}{\omega}\left(1-\frac{I_{3}}{I_{1}}\right)\sin\omega t
=\frac{A}{\left(1-{I_{3}}/{I_{2}}\right)^{1/2}}\left(1-\frac{I_{3}}{I_{1}}
\right)^{1/2}\sin\omega t,
\ena
where $\omega^{2}=\Omega_{0}^{2}\left(1-\frac{I_{3}}{I_{2}}\right) 
\left(1-\frac{I_{3}}{I_{1}}\right)$,
 and $A$ is a constant of integration.
Denote $A/\left(1-\frac{I_{3}}{I_{2}}\right)^{1/2}=\alpha J$ with $\alpha$ 
is a small parameter, then 
\bea
J_{1}=\alpha J\left(1-\frac{I_{3}}{I_{2}}\right)^{1/2}\cos\omega t,\\
J_{2}=\alpha J\left(1-\frac{I_{3}}{I_{1}}\right)^{1/2}\sin\omega t.
\ena

The value of $\alpha$ is found by using the relation
\bea
q=\frac{2I_{1}E}{J^{2}},\label{eq:q}\\
E=\frac{J_{1}^{2}}{2I_{1}}+\frac{J_{3}^{2}}{2I_{3}},\label{eq:E}
\ena
where we have use the fact that at $t=0$, $J_{2}=0$. Substituting $J_{1}$
 and $J_{3}$ in Equation (\ref{eq:E}) and plugging it into (\ref{eq:q}), we obtain
\bea
q=\alpha^{2}\left(\frac{I_{3}}{I_{2}}-1\right)+\frac{I_{1}}{I_{3}},
\ena
Hence,
\bea
\alpha=\left(\frac{I_{1}/I_{3}-q}{1-I_{3}/I_{2}}\right)^{1/2}
\ena
As $q=I_{1}/I_{3}$, then $\alpha=0$, i.e., $J_{1}=J_{2}=0$ and $J_{3}=J$.

When the angular velocity components are known, we can infer the orientation
of the grain axes in the inertial coordinate system using Euler angles:
\bea
\cos\theta=\frac{I_{1}\omega_{1}}{J},~~~ \tan\psi=\frac{I_{2}\omega_{2}}{I_{3}\omega_{3}}.\label{euler_angle}
\ena

\subsection{A2. Flip states}
For a given $\bJ$, there are two sets of solution ($\pm$ sign) 
for $\omega_{i}$ of the Euler motion equations (see Eqs \ref{omep1}-\ref{omep3}).
It can be seen that for $q<q_{\sp}$, two rotation states $\pm$ correspond to 
$\omega_{1}>0$ and $\omega_{1}<0$, i.e., $\ba_{1}.\bJ>0$ and $\ba_{1}.\bJ<0$. We
define these rotation states as positive flip state and negative flip state 
(also WD03; Hoang \& Lazarian 2008). For 
$q>q_{\sp}$, the similar situation occurs with $\omega_{3}$, and there are 
positive and negative flip states with respect to $\ba_{3}$. 

\subsection{A3. Electric Dipole Emission for Irregular Grain}

Let us consider the general case where the dipole is fixed in the
grain body, and given by 
\bea
\bmu=\mu_{1}\ba_{1}+\mu_{2}\ba_{2}+\mu_{3}\ba_{3}.\label{eq:mu}
\ena
where $\mu_{i}$ are components of electric dipoles along three principal axes.
In Paper I we disregarded the third term in Equation (\ref{eq:mu}) 
because of grain's axisymmetry. 

 In the inertial coordinate system $\xhat\yhat\zhat$ (see Fig. \ref{euler}),
 $\ba_{1}, \ba_{2}$ and $\ba_{3}$ 
are described as
\bea
\ba_{1}&=&\sin\phi\sin\theta \xhat-\cos\phi\sin\theta \yhat+\cos\theta \zhat,\\
\ba_{2}&=&(\cos\phi\cos\psi-\sin\phi\sin\psi\cos\theta) \xhat+(\sin\phi\cos\psi+
\cos\phi\sin\psi\cos\theta)\yhat+ \sin\psi\sin\theta \zhat,\\
\ba_{3}&=&-(\cos\phi\sin\psi+\sin\phi\cos\psi\cos\theta) \xhat+(-\sin\phi\sin\psi
+\cos\phi\cos\psi\cos\theta)\yhat+ \cos\psi\sin\theta \zhat,~
\ena
where $\phi,\psi$ and $\theta$ are Euler angles.

Complex motion of the grain principal axes with respect to $\bJ$ results in an 
acceleration for dipole moment in the inertial coordinate system $\xhat\yhat\zhat$:
\bea
\ddot{\bmu}=\mu_{1}\ddot{\ba}_{1}+\mu_{2}\ddot{\ba}_{2}+\mu_{3}\ddot{\ba_{3}},
\label{dotmua}
\ena
where $\ddot{\ba}_{1}$ and $\ddot{\ba}_{2}$ are given by
\bea
\ddot{\ba}_{1}&=&\left[-(\dot{\phi}^{2}+\dot\theta^{2})\sin\phi\sin\theta+
\dot\theta\dot\phi\cos\phi\cos\theta\right]\xhat\nonumber\\
&&+\left[\dot{\phi}^{2}\sin\theta\cos\phi+\dot\theta^{2}\cos\phi\sin\theta+
\dot\theta\dot\phi\sin\phi\cos\theta\right]\yhat-\dot\theta^{2}\cos\theta\zhat,
\label{dota1}\\
\ddot{\ba}_{2}&=&\left[-(\dot{\phi}^{2}+\dot{\psi}^{2})(\cos\phi\cos\psi-
\sin\phi\sin\psi\cos\theta)-2\dot\phi\dot\psi(-\sin\phi\sin\psi+
\cos\phi\cos\psi\cos\theta)\right]\xhat\nonumber\\
&&+\left[\dot\theta^{2}\sin\phi\sin\psi\cos\theta+\dot\theta\dot\phi\cos\phi
\sin\psi\sin\theta+\dot\theta\dot\psi\sin\phi\cos\psi\sin\theta\right]\xhat\nonumber\\
&&+\left[-(\dot{\phi}^{2}+\dot{\psi}^{2})(\sin\phi\cos\psi+\cos\phi\sin\psi\cos\theta)
-2\dot\phi\dot\psi(\cos\phi\sin\psi+\sin\phi\cos\psi\cos\theta)\right]\yhat\nonumber\\
&&+\left[-\dot\theta^{2}\cos\phi\sin\psi\cos\theta+\dot\theta\dot\phi\sin\phi\sin\psi
\sin\theta-\dot\theta\dot\psi\cos\phi\cos\psi\sin\theta\right]\yhat\nonumber\\
&&+\left[-({\dot\psi}^{2}+\dot\theta^{2})\sin\psi\sin\theta+\dot\theta\dot\psi
\cos\psi\cos\theta\right] \zhat,\label{dota2}\\
\ddot{\ba}_{3}&=&\left[(\dot{\phi}^{2}+\dot{\psi}^{2})(\cos\phi\sin\psi+
\sin\phi\cos\psi\cos\theta)+2\dot\phi\dot\psi(\sin\phi\cos\psi+\cos\phi\sin\psi
\cos\theta)\right]\xhat\nonumber\\
&&+\left[\dot\theta^{2}\sin\phi\cos\psi\cos\theta+\dot\theta\dot\phi\cos\phi
\cos\psi\sin\theta-\dot\theta\dot\psi\sin\phi\sin\psi\sin\theta\right]\xhat
\nonumber\\
&&+\left[(\dot{\phi}^{2}+\dot{\psi}^{2})(\sin\phi\sin\psi-\cos\phi\cos\psi\cos\theta)
-2\dot\phi\dot\psi(\cos\phi\cos\psi-\sin\phi\sin\psi\cos\theta)\right]\yhat\nonumber\\
&&+\left[-\dot\theta^{2}\cos\phi\cos\psi\cos\theta+\dot\theta\dot\phi\sin\phi
\cos\psi\sin\theta+\dot\theta\dot\psi\cos\phi\sin\psi\sin\theta\right]\yhat\nonumber\\
&&+\left[-({\dot\psi}^{2}+\dot\theta^{2})\cos\psi\sin\theta-\dot\theta\dot\psi
\sin\psi\cos\theta\right] \zhat,\label{dota3}
\ena

The precession and rotation rates $\dot\phi$ and $\dot\psi$ are 
related to the angular velocity components as follows (Landau \& Lifshitz 1976):
\bea
\omega_{1}=\dot\phi\cos\theta+\dot\psi,\\
\omega_{2}=\dot\phi\sin\theta\sin\psi+\dot\theta\cos\psi,\\
\omega_{3}=\dot\phi\sin\theta\cos\psi-\dot\theta\sin\psi.
\ena
By solving equations, we obtain
\bea
\dot\theta=\omega_{2}\cos\psi-\omega_{3}\sin\psi,\\
\dot\phi=\frac{\omega_{2}\sin\psi+\omega_{3}\cos\psi}{\sin\theta}={J}
\frac{I_{2}\omega_{2}^{2}+I_{3}\omega_{3}^{2}}{I_{2}^{2}\omega_{2}^{2}+I_{3}^{2}
\omega_{3}^{2}},\\
\dot\psi=\omega_{1}-\dot\phi\cos\theta,~~\label{omepa}
\ena

Plugging into Equation (\ref{dotmua})  with the usage of Equations (\ref{dota1})
 and (\ref{dota2}) we obtain the acceleration components as functions of time. 
Performing Fourier transform for these components gives us the spectrum of electric 
dipole emission (see Fig. \ref{fft}).

The dipole emission power of this torque-free rotating grain can be 
obtained by averaging the $\ddot\mu^{2}$ over time:
\bea
P_{\ed}=\frac{2}{3c^{3}}\langle \ddot\bmu^{2}\rangle
\equiv \frac{1}{T}\int_{0}^{T}\frac{2}{3c^{3}}\ddot\bmu^{2} dt
\ena

\section{B. Electric dipole damping for disk-like grain}

In the grain body system $\ba_{1}\ba_{2}\ba_{3}$, the dipole moment is given by 
Equation (\ref{eq:mu}).
The orientation of axes $\ba_{i}$ in an inertial system are determined by Euler 
angles (see Fig. \ref{euler}).
The increase of grain's angular momentum over time arising from the acceleration
 of dipole emission is then
\bea
\frac{d{\bJ}}{dt}=-\frac{2}{3c^{3}}[\dot{\bmu}\times \ddot{\bmu}],\label{dldt}
\ena

Using  Equations (\ref{eq:mu}) and (\ref{dotmua}) for (\ref{dldt}) and averaging
 it over $\phi$ and $\psi$ from 0 to $2\pi$ due to torque-free motion, the non-
vanished component is given by
\bea
\frac{d{J_{z}}}{dt}=-\frac{2}{3c^{3}}\frac{J^{3}}{I_{\|}^{3}}\left[\frac{\mu_{\perp}^{2}}{2}
\{\cos^{4}\theta(h^{3}-3h+2)+\cos^{2}\theta(3h-2h^{3})+h^{3}\}+
\mu_{1}^{2}h^{3}\sin^{2}\theta\right],\label{dlzdt}
\ena
where we assumed $\mu_{2}^{2}=\mu_{3}^{3}=\mu_{\perp}^{2}/2$. The 
components  $dJ_{x}/dt$ and $dJ_{y}/dt$ are averaged out to zero.

For case 1 with $\mu_{\|}:\mu_{\perp}=2:3$, i.e., 
$\mu_{1}^{2}=\mu_{2}^{2}=\mu_{3}^{2}=\mu^{2}/3$, 
we obtain (similar to HDL10):
\bea
\frac{dJ_{z}}{dt}=-\frac{2\mu^{2}}{9c^{3}}\frac{J^{3}}{I_{\|}^{3}}
\left[\cos^{4}\theta(h^{3}-3h+2)+\cos^{2}
\theta(-2h^{3}+3h)+h^{3}(1+\sin^{2}\theta)\right]
\ena

For case 2 with $\mu_{\|}:\mu_{\perp}=0:1$, i.e., $\mu_{1}=0$ and 
$\mu_{2}^{2}=\mu_{3}^{2}=\mu^{2}/2$,
\bea
\frac{dJ_{z}}{dt}=-\frac{\mu^{2}}{3c^{3}}\frac{J^{3}}{I_{\|}^{3}}
\left[\cos^{4}\theta(h^{3}-3h+2)+\cos^{2}
\theta(-2h^{3}+3h)+h^{3}\right].
\ena

In dimensionless variables, we have
\bea
\frac{dJ_{z}'}{dt'}=-\frac{2}{3}\frac{J_{z}^{'3}}{\tau'_{\ed,\eff}},
\ena
where $\tau'_{\ed,\eff}=\tau_{\ed,\eff}/\tau_{\H,\|}$ with
\bea
\tau_{\ed,\eff}&=&\tau_{\ed,\|}\times\frac{2}{\cos^{4}\theta(h^{3}-3h+2)+\cos^{2}
\theta(-2h^{3}+3h)+h^{3}(1+\sin^{2}\theta)},\\
\tau_{\ed,\|}&=&\frac{3I_{\|}^{2}c^{3}}{4\kB T_{\gas}\mu^{2}},
\ena
for case 1, and
\bea
\tau_{\ed,\eff}&=&\tau_{\ed,\|}\times\frac{2}{\cos^{2}\theta(-2h^{3}+3h)+
\cos^{4}\theta(h^{3}-3h+2)+h^{3}},\\
\tau_{\ed,\|}&=&\frac{I_{\|}^{2}c^{3}}{2\kB T_{\gas}\mu^{2}},
\ena
for case 2.

For $\theta=0$, then $\tau_{\ed,\eff}=\tau_{\ed,\|}$.

\end{document}